\documentclass[a4paper,11pt]{article}
\pdfoutput=1
\usepackage{feynmp-auto}
\usepackage[centertags]{amsmath}
\usepackage{amsfonts} 
\usepackage{amssymb} 
\usepackage{amsthm}
\allowdisplaybreaks[1]
\usepackage{graphicx}
\usepackage{bm}
\usepackage{epsfig,amssymb,amsmath,psfrag,subfigure,eurosym}
\usepackage{float}
\usepackage{jheppub}
\newcommand{\eq}[1]{Eq.~(\ref{#1})}
\newcommand{\comm}[2]{\left[#1,#2\right]}
\newcommand{\ii}{\mathrm{i}}
\newcommand{\rme}{\mathrm{e}}

\newcommand{\vev}[1]{\langle #1 \rangle}

\newcommand{\tr}{\mathrm{tr}\,}

\newcommand{\cA}{{\mathcal{A}}}

\newcommand{\cE}{{\mathcal{E}}}

\newcommand{\cG}{{\mathcal{G}}}

\newcommand{\cN}{{\mathcal{N}}}
\newcommand{\cO}{{\mathcal{O}}}

\newcommand{\cS}{{{S}}}

\newcommand{\cW}{{\mathcal{W}}}
\newcommand{\cY}{{\mathcal{Y}}}
\newcommand{\cZ}{{\mathcal{Z}}}
\newcommand\re[1]{(\ref{#1})}

\newcommand{\one}{{\rm 1\kern -.9mm l}}

\newcommand{\vphi}{\bm{\varphi}}

\newcommand{\nordg}[1]{:\! #1 \!:_g}
\newcommand{\nord}[1]{:\! #1 \!:}
\newcommand{\eee}{\mathrm{e}}

\def\XXint#1#2#3{{\setbox0=\hbox{$#1{#2#3}{\int}$}
\vcenter{\hbox{$#2#3$}}\kern-.5\wd0}}

\newcommand{\beqas}{\begin{eqnarray*}}
\newcommand{\eeqas}{\end{eqnarray*}}
\def\eeq{\end{equation}}
\def\be{\begin{equation}}
\def\ee{\end{equation}}
\def\bea{\begin{eqnarray}}
\def\eea{\end{eqnarray}}
\def\eq{\begin{equation}}
\def\eqe{\end{equation}}
\def\eqa{\begin{eqnarray}}
\def\eqae{\end{eqnarray}}

\makeatletter
\let\save@mathaccent\mathaccent
\newcommand*\if@single[3]{%
  \setbox0\hbox{${\mathaccent"0362{#1}}^H$}%
  \setbox2\hbox{${\mathaccent"0362{\kern0pt#1}}^H$}%
  \ifdim\ht0=\ht2 #3\else #2\fi
  }
\newcommand*\rel@kern[1]{\kern#1\dimexpr\macc@kerna}
\newcommand*\widebar[1]{\@ifnextchar^{{\wide@bar{#1}{0}}}{\wide@bar{#1}{1}}}
\newcommand*\wide@bar[2]{\if@single{#1}{\wide@bar@{#1}{#2}{1}}{\wide@bar@{#1}{#2}{2}}}
\newcommand*\wide@bar@[3]{%
  \begingroup
  \def\mathaccent##1##2{%
    \let\mathaccent\save@mathaccent
    \if#32 \let\macc@nucleus\first@char \fi
    \setbox\z@\hbox{$\macc@style{\macc@nucleus}_{}$}%
    \setbox\tw@\hbox{$\macc@style{\macc@nucleus}{}_{}$}%
    \dimen@\wd\tw@
    \advance\dimen@-\wd\z@
    \divide\dimen@ 3
    \@tempdima\wd\tw@
    \advance\@tempdima-\scriptspace
    \divide\@tempdima 10
    \advance\dimen@-\@tempdima
    \ifdim\dimen@>\z@ \dimen@0pt\fi
    \rel@kern{0.6}\kern-\dimen@
    \if#31
      \overline{\rel@kern{-0.6}\kern\dimen@\macc@nucleus\rel@kern{0.4}\kern\dimen@}%
      \advance\dimen@0.4\dimexpr\macc@kerna
      \let\final@kern#2%
      \ifdim\dimen@<\z@ \let\final@kern1\fi
      \if\final@kern1 \kern-\dimen@\fi
    \else
      \overline{\rel@kern{-0.6}\kern\dimen@#1}%
    \fi
  }%
  \macc@depth\@ne
  \let\math@bgroup\@empty \let\math@egroup\macc@set@skewchar
  \mathsurround\z@ \frozen@everymath{\mathgroup\macc@group\relax}%
  \macc@set@skewchar\relax
  \let\mathaccentV\macc@nested@a
  \if#31
    \macc@nested@a\relax111{#1}%
  \else
    \def\gobble@till@marker##1\endmarker{}%
    \futurelet\first@char\gobble@till@marker#1\endmarker
    \ifcat\noexpand\first@char A\else
      \def\first@char{}%
    \fi
    \macc@nested@a\relax111{\first@char}%
  \fi
  \endgroup
}
\makeatother

\title{\boldmath Two-point correlators in non-conformal $\mathcal{N}=2$ gauge theories}
\author[a]{M. Bill\`o,}
\affiliation[a]{Universit\`a di Torino, Dipartimento di Fisica
and I.\,N.\,F.\,N. - sezione di Torino,\\
Via P. Giuria 1, I-10125 Torino, Italy\\}
\emailAdd{billo@to.infn.it}

\author[b]{F.~Fucito,}
\affiliation[b]{I.\,N.\,F.\,N. - sezione di Roma Tor Vergata\\
Via della Ricerca Scientifica, I-00133 Roma, Italy\\}
\emailAdd{fucito@roma2.infn.it}

\author[c]{G.P.~Korchemsky,}
\affiliation[c]{Institut de Physique Th\'eorique\footnote{Unit\'e Mixte de Recherche 3681 du CNRS}, Universit\'e Paris Saclay, CNRS, CEA, 91191 Gif-sur-Yvette\\}
\emailAdd{gregory.korchemsky@ipht.fr}

\author[d]{A.~Lerda\,}
\affiliation[d]{Universit\`a del Piemonte Orientale, Dipartimento di Scienze e Innovazione Tecnologica\\
and I.\,N.\,F.\,N. - sezione di Torino,
Via P. Giuria 1, I-10125 Torino, Italy\\}
\emailAdd{lerda@to.infn.it}

\author[b]{and J.~F.~Morales\,}
\emailAdd{morales@roma2.infn.it}

\abstract{We study the two-point correlation functions of 
chiral/anti-chiral operators in $\mathcal{N}=2$ supersymmetric Yang-Mills 
theories on $\mathbb{R}^4$ with gauge group  SU($N$) and $N_f$ massless 
hypermultiplets in the fundamental representation. 
We compute them in perturbation theory, using dimensional regularization 
up to two loops, and show that field-theory observables built out of  
dimensionless ratios of two-point renormalized correlators on $\mathbb{R}^4$ are 
in perfect agreement with the same quantities computed using localization 
on the four-sphere, even in the non-conformal case $N_f\not=2N$.
}

\dedicated{In memoriam of our collaborator and friend Yassen Stanislavov Stanev}
\preprint{ROM2F/2018/02}

\begin{document}
\maketitle
\flushbottom

\section{Introduction}
\label{sec:intro}

Supersymmetric Yang-Mills theories (SYM) have long been considered as an ideal play-ground to get exact results in Quantum Field Theory. In recent times, exact formulae for special observables in theories with  extended supersymmetries have been found. 
In four-dimensional theories with maximal ${\cN}=4$ supersymmetry,
the exact resummation of the infinite series of perturbative corrections to
the expectation value of circular Wilson loops, also in presence of chiral operators, has been performed \cite{Erickson:2000af,Berenstein:1998ij,Drukker:2000rr,Semenoff:2001xp,Fucito:2015ofa}.
These results are based on the counting of the relevant rainbow-like Feynman diagrams 
by means of a matrix model. The introduction of this matrix model has been considered 
\emph{ad hoc} until it was shown \cite{Pestun:2007rz} that localization for the ${\cN}=2^*$ theory on 
the four-sphere $S^4$, after having performed the ${\cN}=4$ limit, predicts its existence. Moreover, 
localization provides a non-perturbative formula for the circular Wilson loop in a general 
${\cN}=2$ theory which takes into account both perturbative and non-perturbative, instanton and 
anti-instanton, corrections in an interacting matrix model. A two-loop test of this formula against 
perturbation theory was presented in \cite{Andree:2010na} in the case of superconformal QCD. 

It is natural to ask whether localization on $S^4$ can be used to compute non-trivial quantities other than the Wilson loop expectation value. In  
\cite{Baggio:2014sna,Baggio:2015vxa,Gerchkovitz:2016gxx,Baggio:2016skg,Rodriguez-Gomez:2016ijh,Rodriguez-Gomez:2016cem,Baggio:2014ioa} it has been proposed that two-point correlators  between chiral and anti-chiral operators in a superconformal ${\cN}=2$ theory on $\mathbb R^4$
can be computed from the partition function of the theory on the four-sphere with chiral and anti-chiral insertions at the north and south pole respectively; localization expresses this partition function as a matrix model. In a conformal $\cN=2$ theory, two-point correlators between a chiral operator 
$O_{\vec n}={\rm tr} (\varphi^{n_1}) {\rm tr} (\varphi^{n_2}) \cdots\,$, where $\vec n= (n_1,n_2,\dots)$ and $\varphi$ is the complex
scalar of the gauge vector multiplet, and an anti-chiral operator 
$\widebar O_{\vec m}$ made out of the complex conjugate 
field $\widebar{\varphi}$, take the form  
\begin{equation}
\label{2pt-CFT}
\big\langle O_{\vec n}(x) \,\widebar O_{\vec m}(0)\big\rangle
=  \frac{G_{\vec{n}, \vec{m}} (g_0) }{(4 \pi^2 x^2)^ n} \,\delta_{n,m}\,,
\end{equation}
where $n=\sum_i n_i$ and $m=\sum_j m_j$ are the {scaling} dimensions of the two operators.
$G_{\vec{n}, \vec{m}} (g_0)$ is a non-trivial function of the coupling constant $g_0$, but bears no dependence on the distance $x$ since chiral and anti-chiral operators are protected in conformal ${\cN}=2$ theory. 
The two-point functions on a four-sphere also take the form (\ref{2pt-CFT}) but
with $x^2$ being the chordal distance on $S^4$. The function 
$G_{\vec{n}, \vec{m}} (g_0)$ is the same on the sphere and in flat space, and 
it is given by a two-point function in a matrix model obtained from localization.  

Explicit tests of the match between the field theory and the matrix model descriptions of the correlator (\ref{2pt-CFT}) have been performed up to two loops for low-dimensional operators in SU(2) and SU(3) gauge theories with $N_f=4$ and $N_f=6$ matter hypermultiplets. The results where extended in \cite{Billo:2017glv} to generic chiral operators in a superconformal  SU($N$) theory with $N_f=2N$.  Also the one-point functions of chiral operators in presence of a circular Wilson loop can be expressed in terms of the matrix model obtained via localization of the Wilson 
loop on $S^4$, as checked up to two loops in \cite{Billo:2018oog}. 

It is of obvious importance to investigate to what extent the matrix model description of the 
correlation function persists {in $\cN=2$ theory} in non-conformal set-ups. A first step in this 
direction was carried out in \cite{Billo:2017glv} for a SU$(N)$ theory with 
$N_f$ flavors, where suitable operators were chosen in such a way that  their two-point correlators vanish in perturbation theory up to a given loop order, leaving a finite contribution at the next leading 
loop order. In such a situation, a perfect match between the perturbation theory and localization was 
shown for an arbitrary rank and any number of flavors at two and three loops. This strongly hints that 
chiral/anti-chiral correlators are related to the $S^4$ matrix model also beyond the conformal case. 

Considering generic chiral/anti-chiral correlation functions away from the conformal 
point $N_f= 2N$, we encounter important differences. 
For $N_f\neq 2N$, the gauge coupling and the operators
are not anymore protected from quantum corrections and have to 
be renormalized to account for the ultraviolet (UV) divergences.  
As a consequence, the two-point correlation functions of renormalized operators $O^R_{\vec n}$ and ${\widebar O}^R_{\vec m}$ depend on the renormalization 
scale $\mu$ and are no longer forced to have just a power-like dependence on the distance $x$ as in (\ref{2pt-CFT}). Instead, they take the general form
\begin{equation}
\label{2pt-nonCFT}
\big\langle O^R_{\vec n}(x) \,{\widebar O}^R_{\vec m}(0)\big\rangle
= \frac{G_{\vec n, \vec m}^R (g,\nu)}{(4\pi^2 x^2)^n}\,\delta_{n,m}~, 
\end{equation}
where $g= g(\mu)$ is the renormalized coupling and the dimensionless quantity
\begin{equation}
\label{defnu}
\nu= 2 + \gamma_{\text{E}} + \ln \pi \mu^2 x^2
\end{equation}
parametrizes the distance separation. 
Here $\gamma_{\rm E}$ denotes the Euler-Mascheroni constant. 
In contrast to the conformal case, the function $G_{\vec n, \vec m}^R(g,\nu)$ depends 
non-trivially on the distance $x^2$ through the quantity $\nu$. 
Such a dependence cannot be obtained from localization on $S^4$, 
because in this case the operators are inserted at the opposite poles of the four-sphere 
and the distance between them is fixed in terms of the sphere radius.  
Still our results show that, up to two loops, the $\nu$-dependence is very simple 
and can be put in the factorized form  
\begin{equation}
G_{\vec n, \vec m}^R (g,\nu)  = \frac{G_{\vec n, \vec m}^R (g, 0)}{\big(1+\frac{1}{2}\beta_0\,g^2\nu\big)^n} \delta_{nm}
+ O(g^6)  \,,
\label{gfactorn}
\end{equation}
where $\beta_0$ is the expansion coefficient of the exact one-loop $\beta$-function of the theory, namely
\begin{equation}
\label{defbetaintro}
\beta_0 = \frac{N_f-2N}{8\pi^2}~.
\end{equation}
Remarkably, the $\nu$-dependent prefactor in (\ref{gfactorn}) depends only 
on the scaling dimension $n$ and not on the details of the operators. As a consequence, 
up to two loops at least, ratios of correlators of the same scaling dimension 
are actually $\nu$-independent and can be compared directly against the matrix 
model results. 
We show that the field theory results for such observables are indeed in perfect agreement 
with the predictions from the localization matrix model.  This is consistent with the fact 
that all Feynman diagrams contributing to the ratios are finite in four dimensions, and 
finite loop integrals on  $S^4$ and $\mathbb{R}^4$ yield the same result after 
replacing propagators in flat space by those on the four-sphere, as we will show 
in Section~\ref{secn:sphere}.

Furthermore, one can ask whether the results for the renormalized correlators at 
a given renormalization scale can be directly matched against those coming 
from localization. 
At one loop, we show that this is indeed the case if we choose 
$\mu^2 x^2=\rme^{\gamma_{\text{E}}}/\pi$ in the minimal 
subtraction scheme. Moreover, by considering the field theory 
on $S^D$ and evaluating the relevant one-loop integrals in dimensional regularization, we 
find that, apart from the obvious replacement of propagators, not only the divergent parts 
but also the finite parts agree with the results on $\mathbb{R}^D$ for $D\to 4$. 
While the agreement of the divergent part is expected, since the divergences are 
sensible to short distances and do not distinguish between the sphere and flat space, 
the agreement of the finite part is neither expected nor guaranteed {\it{a priori}}, 
but nevertheless it holds.        
At two loops, we find that the matrix model results reproduce the field theory ones,
up to a term proportional to $(2N-N_f)$ and to the dimension $n$ of the operators. 
This suggests that the difference could be interpreted as a conformal anomaly which, in 
non-conformal theories, affects the correlation functions in going from the 
four-sphere to the flat space.  

In this paper we keep the numbers of colors $N$ and flavors $N_f$ arbitrary and 
compute the two-point correlators for a general choice of chiral/anti-chiral operators. 
On the field theory side we do this at the two-loop level.
We summarize our findings in Section~\ref{secn:pert} and give a detailed account 
of the Feynman diagram computations in Appendices~\ref{app:loop_integrals} 
and \ref{app:Feynman diagrams}.
To keep track of the various particles exchanged in the loops we use a superfield 
formalism. The loop integrals are evaluated using the integration methods pioneered 
in \cite{Chetyrkin:1981qh} (see also \cite{Grozin:2005yg} for a review). 
In Section~\ref{secn:rencorr} we compute the renormalized correlators and 
their anomalous dimensions.
Our results suggest that the anomalous dimensions $\gamma_{\vec n,0}$ of the chiral 
operators are one-loop exact and are given by the simple formula 
$\gamma_{\vec n,0}= \frac{n}{2}\,\beta_0$. 
In Section~\ref{secn:loc} we discuss the computation of the correlators on the matrix 
model side, building on the techniques described in \cite{Billo:2017glv}, and compare the 
results with those previously obtained from the field theory side. 
To facilitate the comparison, we show that, up to two loops, 
the localization matrix model can be re-expressed as a complex matrix model encoding 
the color factors and the combinatorics of the Feynman diagrams that contribute to the 
chiral/anti-chiral correlators. 
Finally, in Section~\ref{secn:sphere} we discuss the field theory calculation of the 
two-point correlators on the four-sphere $S^4$, and in Section~\ref{secn:concl} 
we present our conclusions. Several technical details for such calculations are
provided in Appendix~\ref{app:feynsphere}.

\section{Two-point correlators from perturbation theory}
\label{secn:pert}

We consider a $\mathcal{N}=2$ SYM theory with gauge group SU($N$) and 
$N_f$ hypermultiplets in the fundamental representation.  For $N_f=2N$ the theory is 
conformally invariant also at the quantum level.
We denote by $\varphi(x)$ the complex scalar field of the $\cN=2$ vector multiplet which,
in $\cN=1$ notation, is the lowest component of a chiral superfield $\Phi$. 
In this theory a basis of chiral operators can be given in terms of the multi-trace operators
\begin{equation}
\label{On}
O_{\vec n}(x) = \tr\left(\varphi^{n_1}(x) \right)\, \tr\left(\varphi^{n_2}(x)\right) \, 
\ldots \,\tr\left(\varphi^{n_\ell}(x)\right)\,,
\end{equation}
where $\vec n = (n_1,n_2,\ldots,n_\ell)$. 
The scaling (bare) dimension of $O_{\vec n}(x)$ is 
\begin{equation}
n = \sum_{k=1}^\ell n_k ~.
\end{equation}
We expand the scalar field $\varphi(x) = \varphi^a(x)\, T^a$ over the SU$(N)$ generators 
$T^a$ ($a = 1,\ldots, N ^2-1)$ in the fundamental representation, normalized as
\begin{equation}
\label{tadef}
\tr (T^a T^b) = \frac 12 \,\delta_{ab}\,.
\end{equation}
In terms of the components $\varphi^a(x)$, the operators (\ref{On}) become
\begin{equation}
O_{\vec n}(x) = R_{\vec n}^{\,a_1\ldots a_n}\, \varphi^{a_1}(x)\ldots \varphi^{a_n}(x)\,,
\label{Ona}
\end{equation}
where $R_{\vec n}^{a_1\ldots a_n}$ is a completely symmetric tensor %
\footnote{Explicitly,
	\begin{equation*}
	R_{\vec{n}}^{\,a_1\dots a_n} 
	= \tr \big(T^{(a_1}\cdots T^{a_{n_1}}\big)~
	\tr \big(T^{a_{n_1+1}}\cdots T^{a_{n_1+n_2}}\big)\ldots
	\tr \big(T^{a_{n_1 + \ldots + n_{\ell-1}+1}}\cdots T^{a_n)}\big)
	\end{equation*}
	where the indices are symmetrized with strength 1. \label{footnote:R}}. 
In an analogous way, we define 
the anti-chiral operators $\widebar{O}_{\vec n}(x)$ using
the complex conjugate field $\widebar{\varphi}(x)$ instead of $\varphi(x)$.

We are interested in computing the two-point correlation functions
\begin{equation}
\big\langle\,O_{\vec n}(x) \,\widebar{O}_{\vec m}(0)\,\big\rangle
\label{OnOm}
\end{equation} 
in non-conformal $\mathcal{N}=2$ theories using standard perturbative techniques. We perform our calculations at the origin of moduli space, where the scalar fields have vanishing vacuum expectation values. This is a preferred point in the sense that here the
breaking of conformal invariance occurs only at the quantum level, as a consequence of the dimensional transmutation phenomenon.
Therefore, this is the natural place in which to test whether the matrix model approach based on localization agrees with the standard perturbative field-theory 
calculations also in the non-conformal case.
Unlike the $\mathcal{N}=4$ theory where the correlators (\ref{OnOm}) 
are exact at tree-level, 
in $\mathcal{N}=2$ theories they receive quantum corrections, starting from one loop for
$N_f\not=2N$ and from two loops in the conformal case $N_f=2N$. 
Moreover, in the non-conformal theories, the loop integrals are UV divergent, in general, 
and need to be regularized. Here, we use the dimensional regularization taking 
the space-time dimension to be $D=4-2\epsilon$. As a consequence, the bare gauge 
coupling constant, $g_0$, becomes dimensionful.

In general, the bare two-point functions (\ref{OnOm}) take the form 
\begin{equation}
\label{formOO}
\big \langle \,O_{\vec n}(x) \, \widebar{O}_{\vec m}(0)\,\big\rangle
= \Delta^n(x)  \, G_{\vec n,\vec m}(g_0,\epsilon,x) \, \delta_{nm} 
\end{equation}
where $n= m$ is the common scaling dimension of the two operators, and 
\begin{equation}
\Delta(x) =  \int \frac{d^D k}{(2\pi)^D} \frac{\rme^{\ii k\cdot x}}{k^2}  = 
\frac{\Gamma(1-\epsilon)}{4\pi\,(\pi x^2)^{1-\epsilon}}~.
\label{Delta}
\end{equation}
is the massless scalar propagator in $D$-dimensions. The correlator 
$G_{\vec n,\vec m}(g_0,\epsilon,x)$ can be computed at weak coupling as an 
expansion in powers of $g_0^2$. 
We refer to \cite{Billo:2017glv} for details on the Feynman 
rules that are needed to perform this calculation; they are summarized for convenience in 
Appendix~\ref{app:Feynman diagrams}.

\begin{figure}[H]
	\vspace{0.2cm}
	\begin{center}
		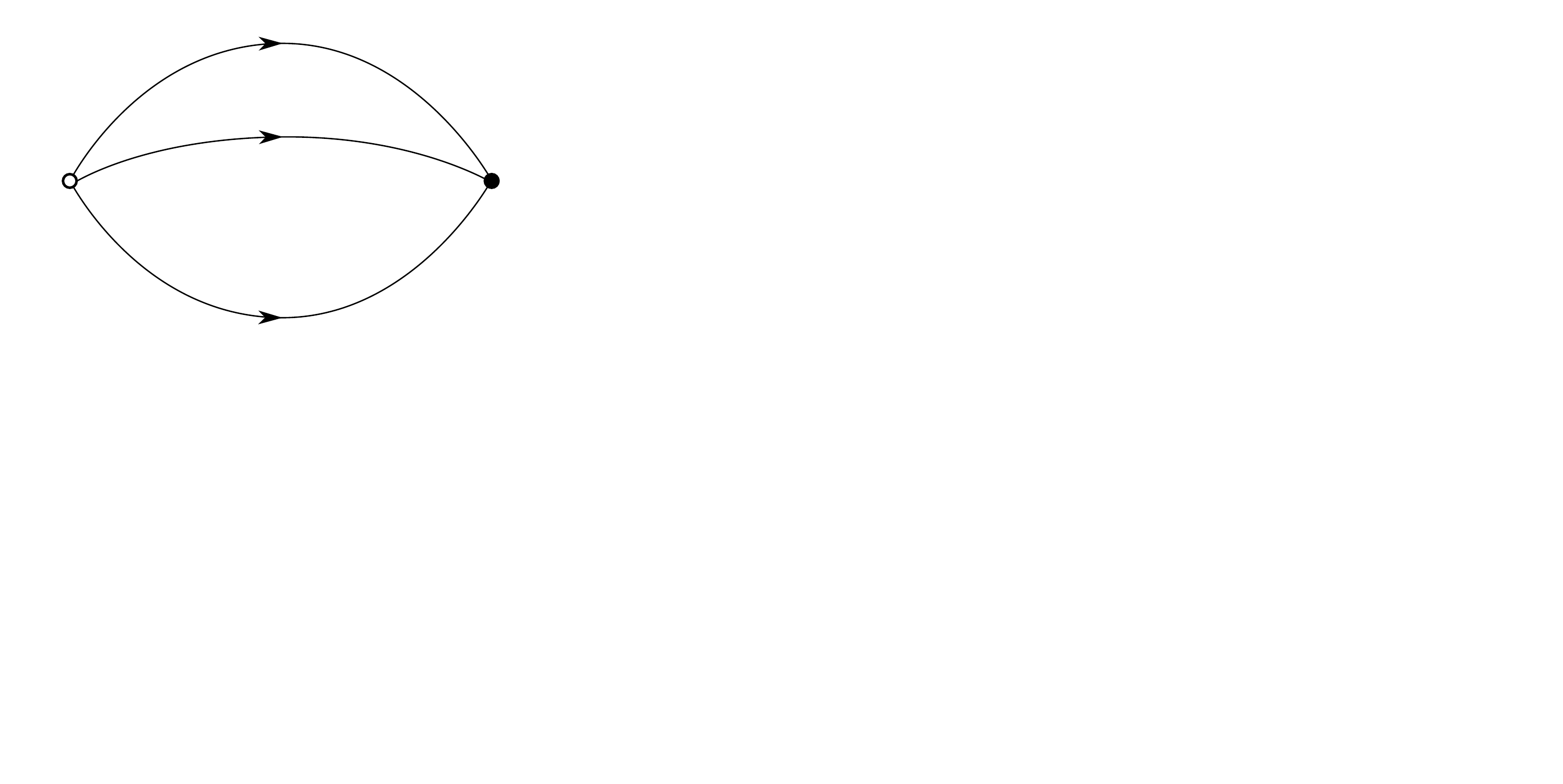
	\end{center}
	\vspace{-0.2cm}
\caption{The Feynman diagrams contributing to the two-point correlators. White and black 
dots stand for the operators $O_{\vec n}$ and $\bar{O}_{\vec{m}}$, respectively. Lines with 
arrows denote free scalar propagators. The effective vertex $v_{k,\ell}$ represents the sum 
of irreducible diagrams with $k$ external legs at $\ell$ loops.}
\label{fig:diagrams}
\end{figure}

The diagrams which contribute to the two-point functions (\ref{OnOm}) up to order 
$g_0^4$ are schematically represented in Fig.~\ref{fig:diagrams}.
\noindent
The diagram $(a)$ is the tree-level contribution, the diagram $(b)$ is 
the one-loop correction, while the other four diagrams represent the two-loop part.
The blobs labeled by $v_{k,\ell}$ stand for the sum of all irreducible 
diagrams of order $g_0^{2\ell}$ with $k$ external lines - half of them 
connected to the chiral fields $\varphi$ of $O_{\vec n}$, 
half to the anti-chiral ones of $\widebar{O}_{\vec m}$.

A convenient way to organize the computation of these diagrams 
is to consider the $\mathcal{N}=4$ theory, remove all contributions 
from Feynman diagrams involving loops of the adjoint hypermultiplet 
(which we call $H$) and add those with loops of the fundamental matter multiplets (which 
we call $Q$ and $\widetilde{Q}$) \cite{Andree:2010na}. Since the two-point correlators
in $\mathcal{N}=4$ theory are exact at tree-level, we can write
\begin{equation}
\begin{aligned}
G_{\vec n,\vec m}\,&=\,G_{\vec n,\vec m}\big|_{\mathcal{N}=4}
-G_{\vec n,\vec m}\big|_{H}+G_{\vec n,\vec m}\big|_{Q,\widetilde{Q}} \phantom{\Big|}\\
\,&=\, G_{\vec n,\vec m}\big|_{\mathrm{tree}}
-G_{\vec n,\vec m}\big|_{H}+G_{\vec n,\vec m}\big|_{Q,\widetilde{Q}} \phantom{\Big|}\,,
\end{aligned}
\label{ampldec}
\end{equation}
where, in an obvious notation, 
$G_{\vec n,\vec m}\big|_{H}$ stands for all diagrams in the $\mathcal{N}=4$ 
theory with the adjoint hypermultiplet $H$ circulating in the loops, 
and $G_{\vec n,\vec m}\big|_{Q,\widetilde{Q}}$ stands for
the same diagrams in the $\mathcal{N}=2$ theory  with loops of 
fundamental matter multiplets $Q$ and $\widetilde{Q}$. In the following, we will 
sometimes refer to this method as ``performing the computation in the 
difference theory''. We stress that in the difference theory one should take into account 
only diagrams involving loops of the adjoint hypermultiplet or loops of the fundamental 
ones, but not both.

\subsection{Tree-level}
The tree-level contribution to the correlator (\ref{formOO}) comes from 
the diagram in Fig.~\ref{fig:diagrams}$(a)$. 
To obtain its explicit expression,  one contracts the 
fields $\varphi$ in $O_{\vec n}$ with the fields $\widebar{\varphi}$ 
in $\widebar{O}_{\vec m}$ by means of a free scalar propagator 
\begin{equation}
\big\langle\,\varphi^a(x)\,\widebar{\varphi}^{\,b}(0)\,\big\rangle
= \parbox[c]{60mm}{\hspace*{-20mm}
\begingroup%
  \makeatletter%
  \providecommand\color[2][]{%
    \errmessage{(Inkscape) Color is used for the text in Inkscape, but the package 'color.sty' is not loaded}%
    \renewcommand\color[2][]{}%
  }%
  \providecommand\transparent[1]{%
    \errmessage{(Inkscape) Transparency is used (non-zero) for the text in Inkscape, but the package 'transparent.sty' is not loaded}%
    \renewcommand\transparent[1]{}%
  }%
  \providecommand\rotatebox[2]{#2}%
  \newcommand*\fsize{\dimexpr\f@size pt\relax}%
  \newcommand*\lineheight[1]{\fontsize{\fsize}{#1\fsize}\selectfont}%
  \ifx\svgwidth\undefined%
    \setlength{\unitlength}{250bp}%
    \ifx\svgscale\undefined%
      \relax%
    \else%
      \setlength{\unitlength}{\unitlength * \real{\svgscale}}%
    \fi%
  \else%
    \setlength{\unitlength}{\svgwidth}%
  \fi%
  \global\let\svgwidth\undefined%
  \global\let\svgscale\undefined%
  \makeatother%
  \begin{picture}(1,0.08574668)%
    \lineheight{1}%
    \setlength\tabcolsep{0pt}%
    \put(0,0){\includegraphics[width=\unitlength,page=1]{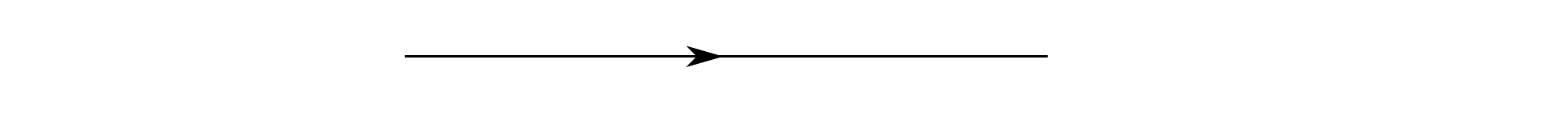}}%
    \put(0.24719949,0.06965537){ \makebox(0,0)[lt]{\lineheight{1.25}\smash{\begin{tabular}[t]{l}$a$\end{tabular}}}}%
    \put(0.63655864,0.06965537){ \makebox(0,0)[lt]{\lineheight{1.25}\smash{\begin{tabular}[t]{l}$b$\end{tabular}}}}%
    \put(0.24719949,0.00311278){ \makebox(0,0)[lt]{\lineheight{1.25}\smash{\begin{tabular}[t]{l}$x$\end{tabular}}}}%
    \put(0.63655864,0.00311278){ \makebox(0,0)[lt]{\lineheight{1.25}\smash{\begin{tabular}[t]{l}$0$\end{tabular}}}}%
    \put(0.70994262,0.04179632){ \makebox(0,0)[lt]{\lineheight{1.25}\smash{\begin{tabular}[t]{l}$\!\!= \,\Delta(x)\,\delta^{ab}$ \end{tabular}}}}%
    \put(0,0){\includegraphics[width=\unitlength,page=2]{propphi.pdf}}%
  \end{picture}%
\endgroup%
}~.
\label{propagator-tree}
\end{equation}
In this way one finds that the correlator
$\big \langle O_{\vec n}(x) \, \widebar{O}_{\vec m}(0)\big\rangle$ at tree-level 
takes the form \re{formOO} with
\begin{equation}
\label{Gtree}
G_{\vec n,\vec m}\,\big|_{\text{tree}} 
= n!\, R_{\vec n}^{\,a_1\ldots a_n}\, R_{\vec m}^{\,a_1 \ldots a_n}
\end{equation}
being a constant that is determined by the color structure of the two operators. 
For example, for the first operators of even dimension, one finds  \cite{Billo:2017glv} 
\begin{equation}\label{Trees}
\begin{aligned}
G_{(2),(2)}\big|_{\text{tree}}\,&=\frac{N^2-1}{2}~,\\
G_{(2,2),(2,2)}\big|_{\text{tree}}\,&=\frac{N^4-1}{2}~,\\
G_{(4),(2,2)}\big|_{\text{tree}}\,&=\frac{(N^2-1)(2N^2-3)}{2N}~,\\
G_{(4),(4)}\big|_{\text{tree}}\,&=\frac{(N^2-1)(N^4-6N^2+18)}{4N^2}~.
\end{aligned}
\end{equation}
Explicit expressions can be easily found also for operators with higher dimension.

\subsection{One-loop diagrams}

\begin{figure}[H]
	\vspace{0.4cm}
	\begin{center}
		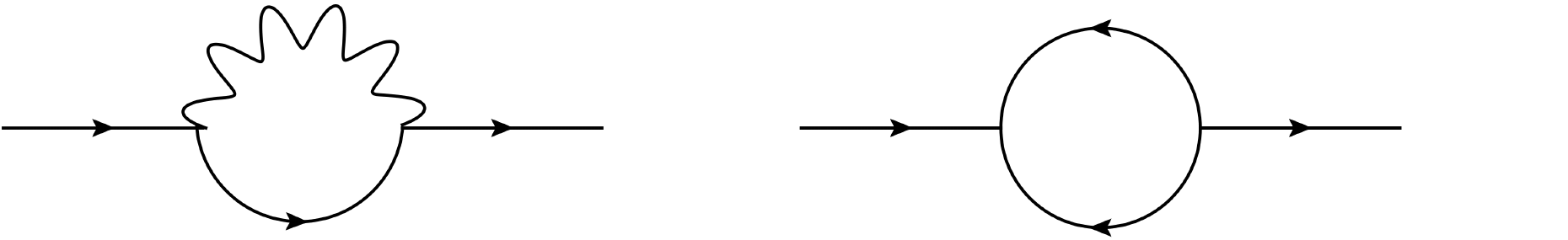
	\end{center}
	\vspace{-0.2cm}
	\caption{The vanishing of the one-loop propagators of the three adjoint 
		scalars $\Phi_I$ in the $\cN=4$ theory.
		The wavy line corresponds to the vector superfield. In the second diagram, 
		the three-point vertices are proportional to the totally anti-symmetric 
		tensor $\epsilon_{IJK}$.}
	\label{fig:pphi23-1l}
\end{figure}

We first observe that in the $\cN=4$ theory there are no one-loop corrections 
to the propagators of  the adjoint scalars $\Phi_I$ ($I=1,2,3$).  At one loop, this is 
schematically represented in Fig.~\ref{fig:pphi23-1l}. 
\noindent
This implies that there is no one-loop correction to the propagator of the adjoint 
hypermultiplet $H$ in the difference theory.

\begin{figure}[H]
	\vspace{0.4cm}
	\begin{center}
\begingroup%
  \makeatletter%
  \providecommand\color[2][]{%
    \errmessage{(Inkscape) Color is used for the text in Inkscape, but the package 'color.sty' is not loaded}%
    \renewcommand\color[2][]{}%
  }%
  \providecommand\transparent[1]{%
    \errmessage{(Inkscape) Transparency is used (non-zero) for the text in Inkscape, but the package 'transparent.sty' is not loaded}%
    \renewcommand\transparent[1]{}%
  }%
  \providecommand\rotatebox[2]{#2}%
  \newcommand*\fsize{\dimexpr\f@size pt\relax}%
  \newcommand*\lineheight[1]{\fontsize{\fsize}{#1\fsize}\selectfont}%
  \ifx\svgwidth\undefined%
    \setlength{\unitlength}{250bp}%
    \ifx\svgscale\undefined%
      \relax%
    \else%
      \setlength{\unitlength}{\unitlength * \real{\svgscale}}%
    \fi%
  \else%
    \setlength{\unitlength}{\svgwidth}%
  \fi%
  \global\let\svgwidth\undefined%
  \global\let\svgscale\undefined%
  \makeatother%
  \begin{picture}(1,0.15934586)%
    \lineheight{1}%
    \setlength\tabcolsep{0pt}%
    \put(0,0){\includegraphics[width=\unitlength,page=1]{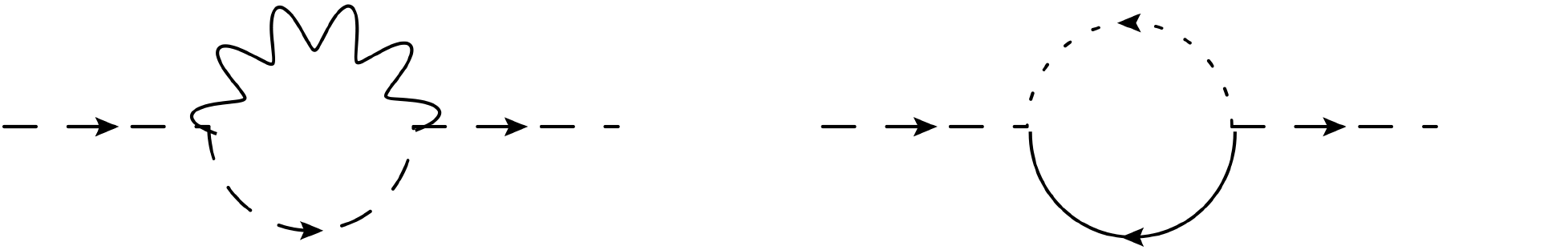}}%
    \put(0.43792458,0.07319104){\color[rgb]{0,0,0}\makebox(0,0)[lt]{\lineheight{1.25}\smash{\begin{tabular}[t]{l}\textbf{$+$}\end{tabular}}}}%
    \put(0.95269553,0.07335798){\color[rgb]{0,0,0}\makebox(0,0)[lt]{\lineheight{1.25}\smash{\begin{tabular}[t]{l}\textbf{$=\,0$}\end{tabular}}}}%
    \put(-0.00108718,0.09488059){\color[rgb]{0,0,0}\makebox(0,0)[lt]{\lineheight{1.25}\smash{\begin{tabular}[t]{l}\textbf{$u$}\end{tabular}}}}%
    \put(0.33300174,0.09488059){\color[rgb]{0,0,0}\makebox(0,0)[lt]{\lineheight{1.25}\smash{\begin{tabular}[t]{l}\textbf{$v$}\end{tabular}}}}%
    \put(0.519527,0.09488059){\color[rgb]{0,0,0}\makebox(0,0)[lt]{\lineheight{1.25}\smash{\begin{tabular}[t]{l}\textbf{$u$}\end{tabular}}}}%
    \put(0.89257755,0.09488059){\color[rgb]{0,0,0}\makebox(0,0)[lt]{\lineheight{1.25}\smash{\begin{tabular}[t]{l}\textbf{$v$}\end{tabular}}}}%
  \end{picture}%
\endgroup%

	\end{center}
	\vspace{-0.2cm}
	\caption{The vanishing of the one-loop propagator of the $Q$ superfield 
		represented by a dashed line. In the second diagram, the continuous internal line 
		represents the $\Phi_1=\Phi$ superfield of the $\cN=2$ theory, while the dotted 
		line represents the $\widetilde{Q}$ superfield. 
		The indices $u$ and $v$ belong to 
		the fundamental representation of SU$(N)$.
		The same happens if the role of $Q$ and $\widetilde{Q}$ 
		is exchanged.}
	\label{fig:pQ1l}
\end{figure}

The one-loop correction to the propagator of the fundamental matter superfields 
$Q$ and $\widetilde{Q}$ vanishes as well. Indeed, as shown in Fig.~\ref{fig:pQ1l}, the 
contribution of this diagram is similar to the previous one upon replacing 
the generators in the adjoint with those in the fundamental representation,
so that the same cancellation mechanism at work for the adjoint 
scalars applies here as well.

\begin{figure}[H]
	\vspace{0.2cm}
	\begin{center}
		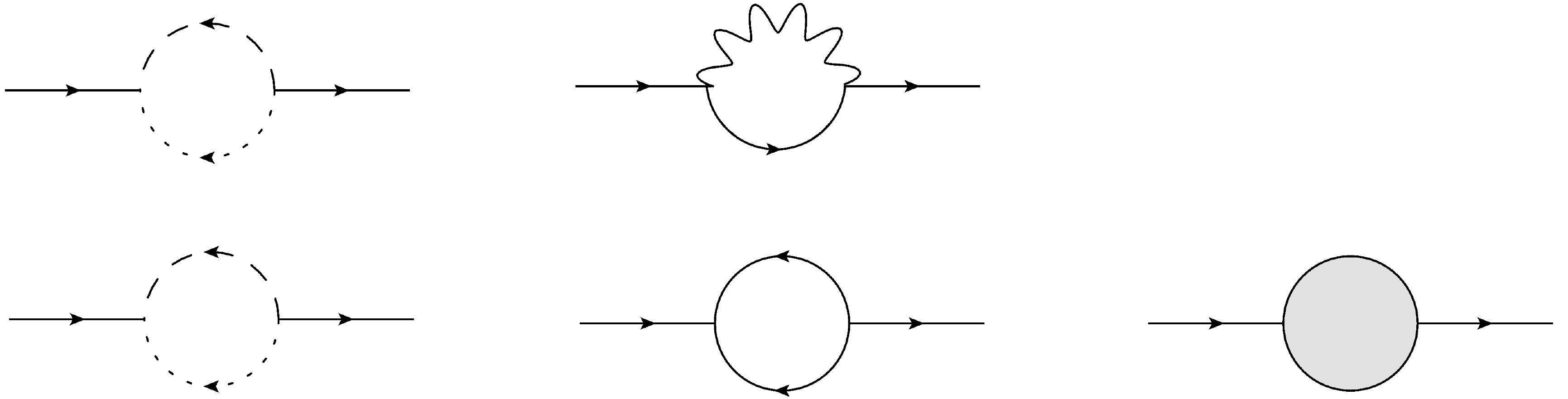
	\end{center}
	\vspace{-0.2cm}
	\caption{The one-loop correction to the scalar propagator. 
	In the second line we have used the relation shown in Fig.~\ref{fig:pphi23-1l} for 
	$I=1$ to replace the loop diagram with the vector propagator with the one with 
	a scalar loop.}
	\label{fig:onelooppphi-iniz}
\end{figure}

To find the contribution of the diagram shown in Fig.~\ref{fig:diagrams}$(b)$,
we need to compute the one-loop correction to the 
propagator of the scalar field in the $\mathcal{N}=2$ theory.
This is represented in Fig.~\ref{fig:onelooppphi-iniz}, where in the second
line we have used the $\mathcal{N}=4$ result of Fig.~\ref{fig:pphi23-1l} for $I=1$,
to replace the diagram with the vector propagator by the diagram with a scalar loop.
\noindent
Explicitly computing these superdiagrams (see Appendix \ref{app:Feynman diagrams} 
for details), we find
\begin{equation}
{\parbox[c]{.25\textwidth}{\includegraphics[width = .25\textwidth]{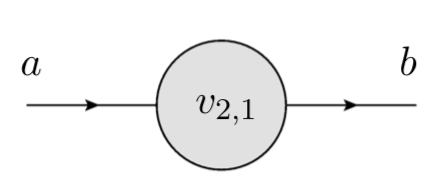}}}
\equiv ~v_{2,1}\, \Delta(x)\, \delta^{ab}\,,
\label{oneloopa2}
\end{equation}
where
\begin{equation}
v_{2,1}
= \frac{g_0^2}{8 \pi^2}\, (2N - N_f)\, \frac{(\pi x^2)^\epsilon\,\Gamma(1-\epsilon)}{2\epsilon(1-2\epsilon)}
~.
\label{v1loop}
\end{equation}
The correction (\ref{oneloopa2}) can be of course inserted in any 
of the $n$ propagators connecting $O_{\vec{n}}$ and $\widebar{O}_{\vec{m}}$, 
so that the one-loop contribution to the two-point correlator \re{formOO} corresponding 
to Fig.~\ref{fig:diagrams}$(b)$ is 
\begin{equation}
\label{onelres}
G_{\vec n,\vec m}\,\big|_{\text{1-loop}} \,=\, n\,v_{2,1}
\, G_{\vec n,\vec m}\,\big|_{\text{tree}}~.
\end{equation}

\subsection{Two-loop diagrams}

At order $g_0^4$ there are several diagrams that contribute to the correlator 
(\ref{formOO}). They are schematically represented by the last four diagrams, 
from $(c)$ to $(f)$, of Fig.~\ref{fig:diagrams}. A detailed derivation of the various 
contributions and the evaluation of the corresponding loop integrals can be found 
in Appendix \ref{app:Feynman diagrams}. Here we simply summarize the results
for the building blocks of each of these diagrams.

\subsubsection{$v_{2,1}^2\,$- contributions}

The two-loop reducible contributions proportional to $v_{2,1}^2$ arise from 
two insertions of the one-loop effective interaction vertex (\ref{oneloopa2}). 
These can occur either on two different scalar propagators connecting the operators 
$O_{\vec{n}}$ and $\widebar{O}_{\vec{m}}$, or on a single propagator.
These two possibilities correspond, respectively, to the diagrams $(c)$ and $(d)$
of Fig.~\ref{fig:diagrams}.

\begin{figure}[H]
	\vspace{0.4cm}
	\begin{center}
		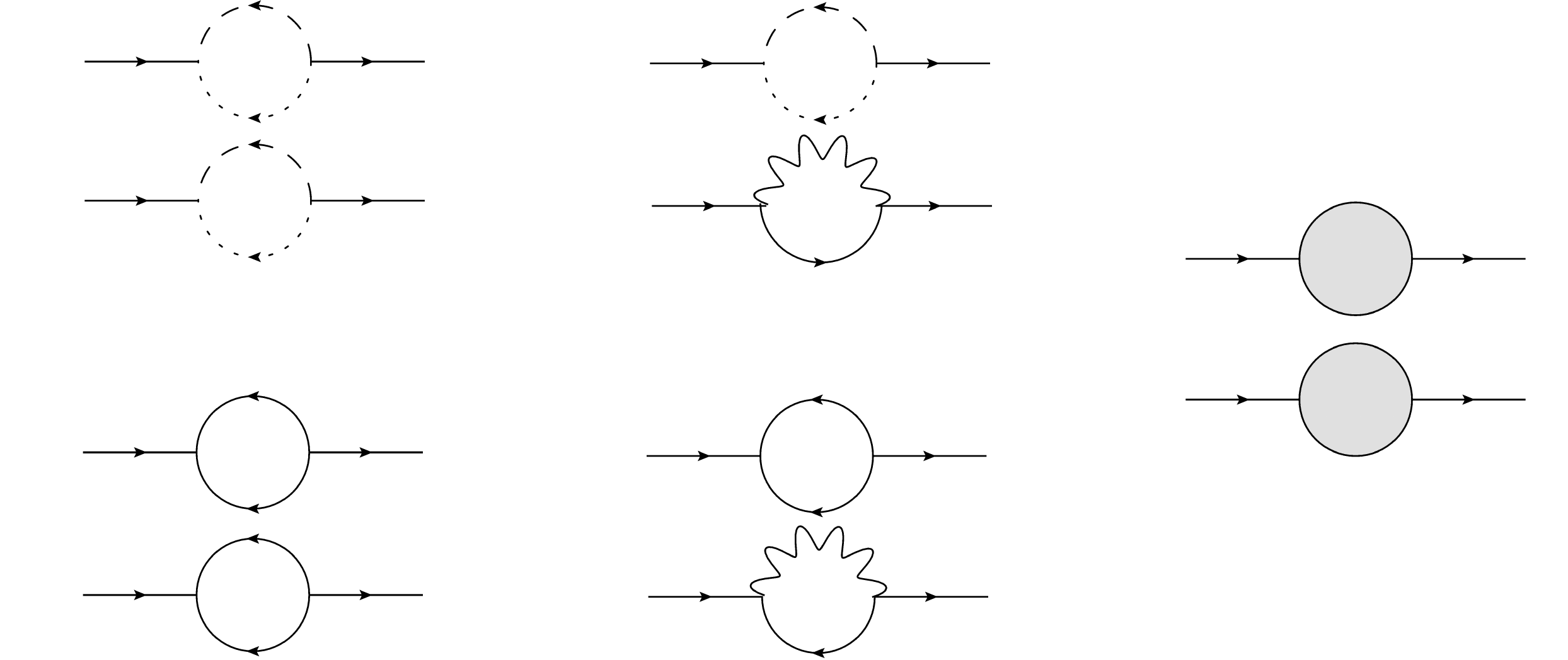
	\end{center}
	\vspace{-0.2cm}
	\caption{The simultaneous corrections to two scalar propagators in the difference 
	theory can be expressed in terms of the one-loop correction. The factor of $2$ in front 
	of the diagrams with the vector field propagator is a multiplicity factor.}
	\label{fig:2p1l}
\end{figure}

\begin{figure}[H]
	\vspace{0.4cm}
	\begin{center}
		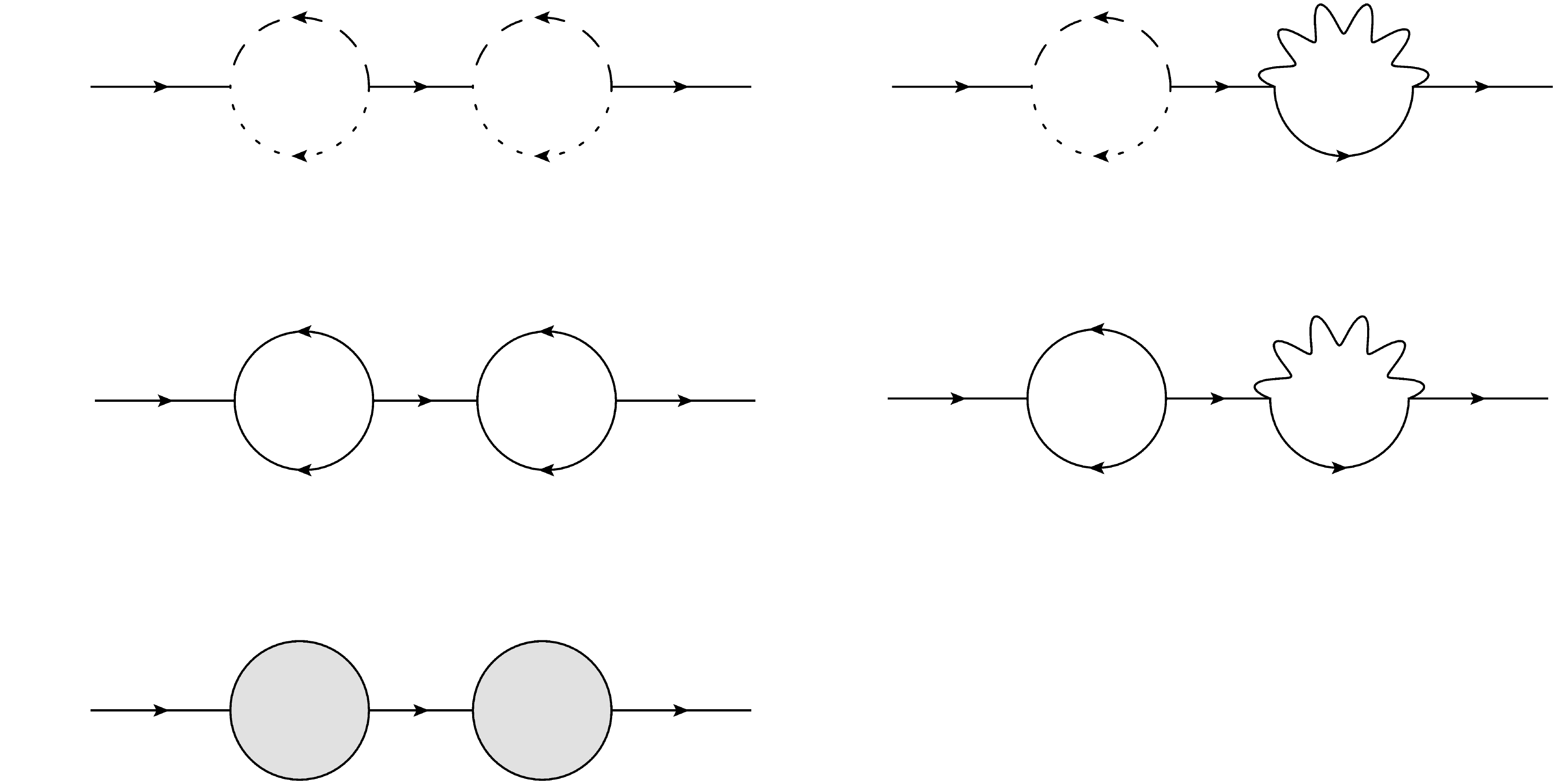
	\end{center}
	\vspace{-0.2cm}
	\caption{In the difference theory, the reducible diagrams that correct the $\varphi$ 
	propagator at two loops can be expressed in terms of the one-loop contribution.}
	\label{fig:pphi2l}
\end{figure}

In the difference theory, the diagram $(c)$ contains as building blocks the diagrams 
represented in the left-hand side of Fig.~\ref{fig:2p1l}.
\noindent
If we exploit the identity of Fig.~\ref{fig:pphi23-1l} to replace the diagrams
with the vector field propagator in favor of the ones with a scalar loop, we can easily 
realize that the diagrams in the left-hand side Fig.~\ref{fig:2p1l}
precisely reconstruct the square represented in the right-hand side. 
Using (\ref{oneloopa2}), and taking into account the appropriate multiplicity 
factor of the diagrams, this gives
\begin{equation}
\label{2p1lres}
{\parbox[c]{.25\textwidth}{ \includegraphics[width = .25\textwidth]{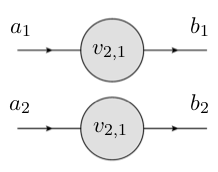}}}
~= ~\frac 12\, v_{2,1}^2\, \Delta^2(x)\,\delta^{a_1 b_1}\delta^{a_2 b_2}~.
\end{equation}
Let us now consider the contribution corresponding to the two-loop diagram in
Fig.~\ref{fig:diagrams}$(d)$.
In this case, the insertion of two one-loop corrections on the same scalar propagator
leads to the diagrams displayed in the first two lines  of Fig.~\ref{fig:pphi2l}.
\noindent
If we use again the identity of Fig.~\ref{fig:pphi23-1l} to replace the diagrams
containing the vector field propagator with 
those with a scalar loop, 
we reconstruct the square of the one-loop correction, as shown in the 
last line of Fig.~\ref{fig:pphi2l}. 
Evaluating explicitly the loop integrals in this case, we obtain that the result 
can be written as the square of the one-loop up to terms of order 
$\epsilon$ (see Appendix~\ref{app:Feynman diagrams} for details), namely
\begin{equation}
{\parbox[c]{.35\textwidth}{ \includegraphics[width = .35\textwidth]{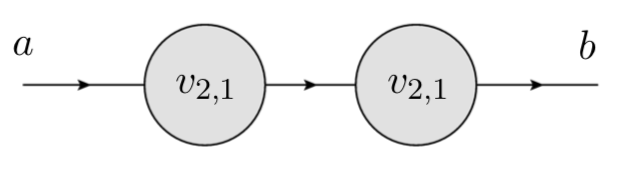}}}    
\approx  ~v_{2,1}^2\, \Delta(x)\, \delta^{ab} ~.
\label{2loopv212}
\end{equation}  
Here and in the following, we use the approximate symbol $\,\approx\,$ for equations 
that hold up to terms vanishing in the limit $\epsilon \to 0$.

\subsubsection{$v_{2,2}\,$- and $v_{4,2}\,$- contributions}

Let us now consider the two-loop irreducible corrections to the scalar propagator
which appear in Fig.~\ref{fig:diagrams}$(e)$. 
A first class of contributions arises from correcting one of the internal 
lines in the one-loop diagrams of Fig.~\ref{fig:onelooppphi-iniz} by using 
the one-loop propagator of the matter superfields 
$Q$ and $\widetilde{Q}$, or of the adjoint  hypermultiplet $H$. However, 
as we have seen before, these contributions vanish.

\begin{figure}[H]
	\vspace{0.4cm}
	\begin{center}
		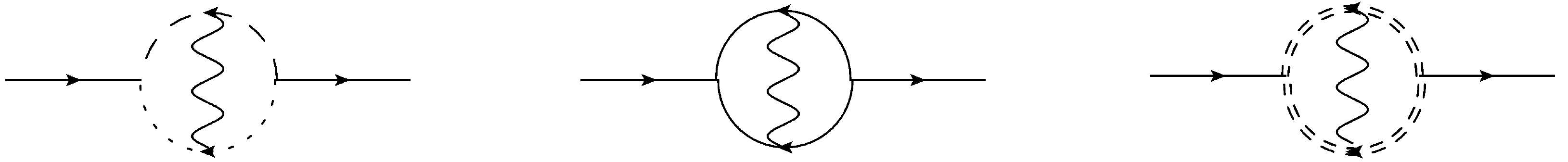
	\end{center}
	\vspace{-0.2cm}
	\caption{A class of diagrams that correct the scalar propagator at two loops. The dashed 
	double-line notation in the right-hand side is a convenient way to
	represent this contribution in the difference theory.}
	\label{newfig}
\end{figure}

Other terms that correct the scalar propagator at two loops in the difference theory 
are those represented in Fig.~\ref{newfig}.
\noindent
Here we have introduced the dashed double-line notation as a convenient way to
represent the difference between the loop with fundamental flavors and the loop with
the adjoint hypermultiplet. Actually there are other three classes of irreducible diagrams
that correct the scalar propagator at two loops. In Fig.~\ref{fig:pphi2l-irred} we 
have drawn all such diagrams, whose evaluation is presented in 
Appendix~\ref{app:Feynman diagrams} to which we refer for details.
\begin{figure}[H]
	\vspace{0.2cm}
	\begin{center}
		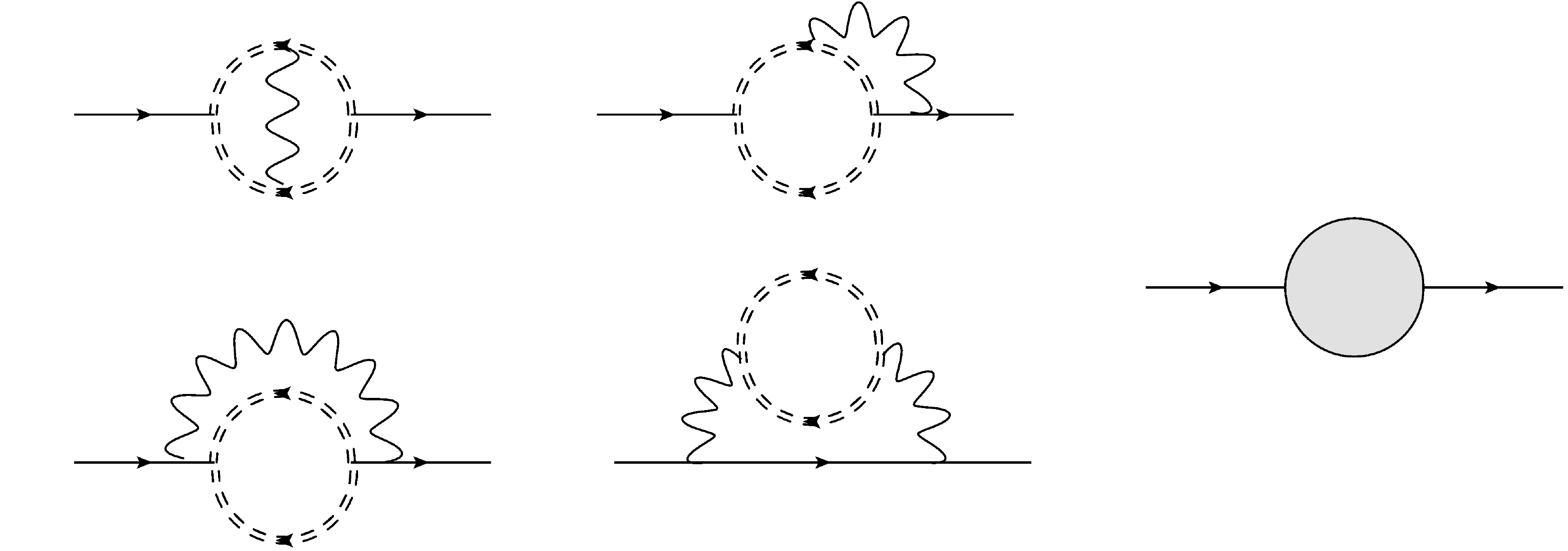
	\end{center}
	\vspace{-0.2cm}
	\caption{Irreducible two-loop diagrams that correct the $\varphi$ propagator in the 
	difference theory.}
	\label{fig:pphi2l-irred}
\end{figure}
\noindent
Summing all contributions, we find that the irreducible two-loop correction to 
the scalar propagator is
\begin{equation}
\label{2loop2}
{\parbox[c]{.25\textwidth}{ \includegraphics[width = .25\textwidth]{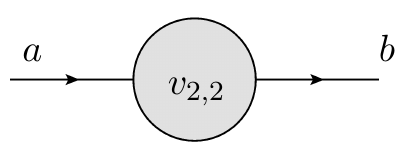}}} 
\equiv  ~ v_{2,2}\, \Delta(x)\, \delta^{ab}\,,
\end{equation}
where
\begin{equation}
v_{2,2}\approx -\Big(\frac{g_0^2}{8\pi^2}\Big)^2\left[3\,\zeta(3)
\Big(\frac{N_f}{2N}+N^2\Big)-N(2N-N_f)\frac{\Gamma^2(1-\epsilon)}{4\epsilon^2(1-2\epsilon)
	(1+\epsilon)}\right](\pi x^2)^{2\epsilon}~.
\label{v2}
\end{equation} 

In the difference theory, there are irreducible two-loop contributions that
involve two chiral and two anti-chiral fields and give rise to the diagram 
of Fig.~\ref{fig:diagrams}$(f)$ . 
These contributions can be further distinguished according to their overall color structure and
can be split into two terms which we denote as $v_{4,2}^{(A)}$ and $v_{4,2}^{(B)}$.

\begin{figure}[H]
	\vspace{0.4cm}
	\begin{center}
		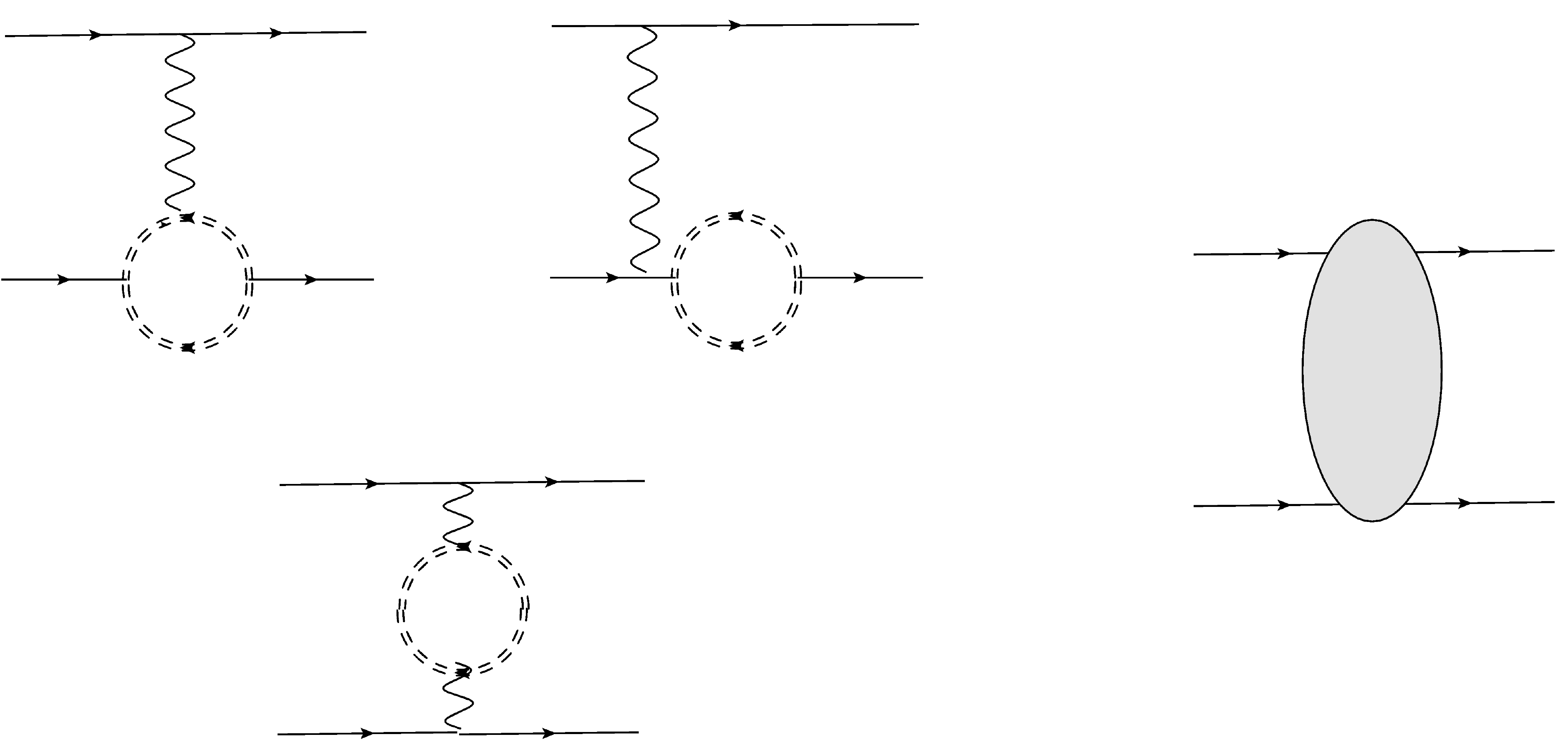
	\end{center}
	\vspace{-0.2cm}
	\caption{Irreducible two-loop diagrams in the difference theory that yield the $v_{4,2}^{(A)}$ contribution.}
	\label{fig:vh42diag}
\end{figure}

The diagrams yielding $v_{4,2}^{(A)}$ are drawn in Fig.~\ref{fig:vh42diag}, where again we have
used the dashed double-line notation to represent the difference between $Q$ and $H$ loops.
\noindent
They are explicitly evaluated in Appendix~\ref{app:Feynman diagrams}, and the final result is
\begin{equation}
\label{v42Aform}
{\parbox[c]{.25\textwidth}{ \includegraphics[width = .25\textwidth]{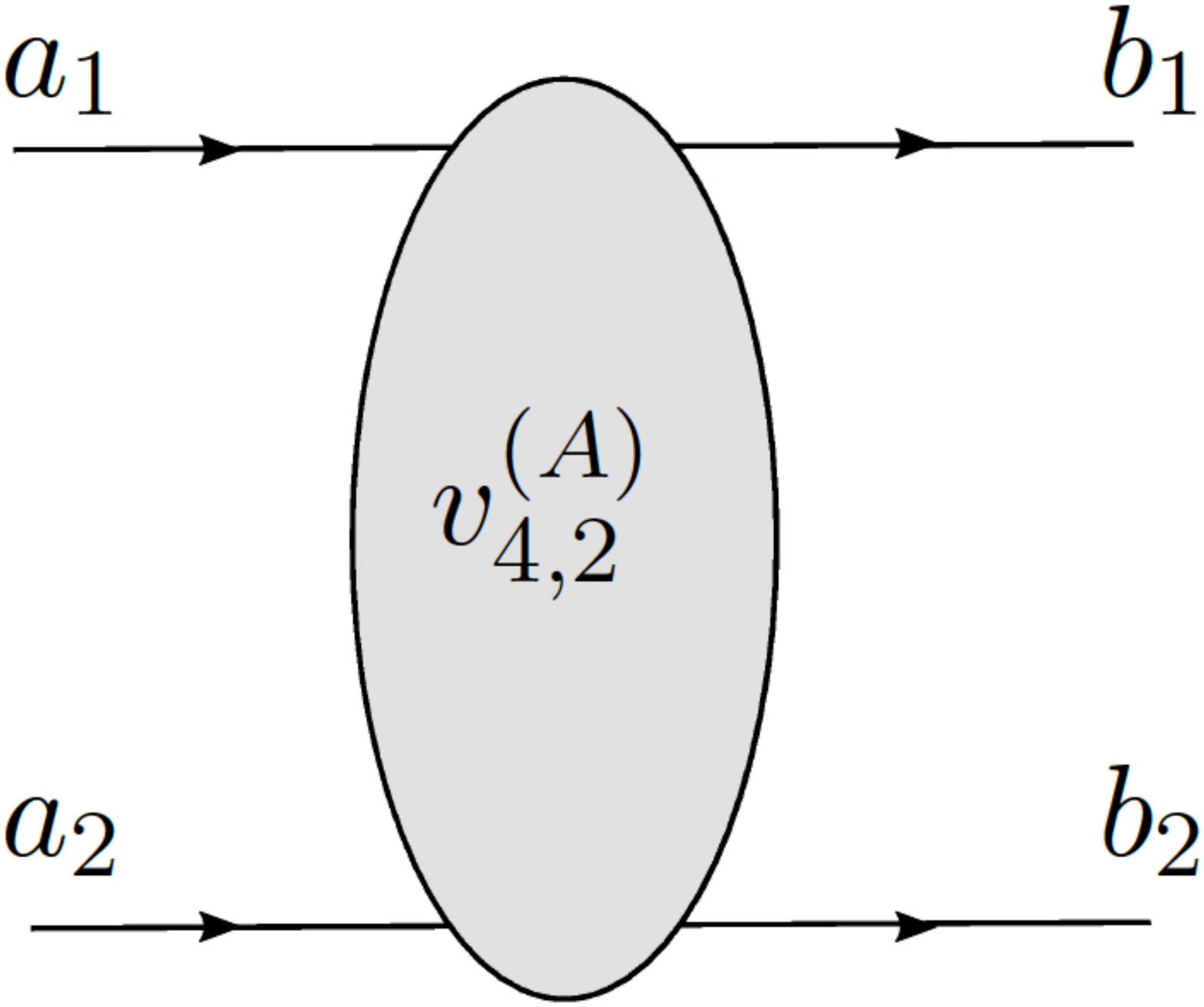}}} \equiv ~ v_{4,2}^{(A)} \,\Delta^2(x)  \, C_{4}^{(A)\,a_1 a_2 b_1 b_2}+ \cdots~,
\end{equation}	
where  
\begin{align}
\label{v42Ares}
{}& v_{4,2}^{(A)} \approx  \left(\frac{g_0^2}{8\pi^2}\right)^2  \,N\,(2N-N_f)
\left[\frac{21}{2} \zeta(3) + \frac{\Gamma^2(1-\epsilon)}{4\epsilon^2(1-2\epsilon)(1 + \epsilon)}\right]
(\pi x^2)^{2\epsilon}~,	
\\[3mm]
\label{C4Ais}
{}& C_{4}^{(A)\,a_1 a_2 b_1 b_2}
=  -\frac{1}{N}  \,f^{c\,a_1 b_1}\, f^{c\,a_2 b_2} ~,
\end{align}		
with $f^{abc}$ being the SU$(N)$ structure constants. In (\ref{v42Aform}) the ellipses 
stand for terms with color tensors that are anti-symmetric in $(a_1,a_2)$ and
$(b_1,b_2)$. 
Such terms do not contribute to the two-point correlation functions (\ref{OnOm}) because
they are contracted with the symmetric tensors $R_{\vec{n}}$
and $\widebar{R}_{\vec{m}}$ defined in (\ref{Ona}).

\begin{figure}[H]
	\vspace{0.4cm}
	\begin{center}
		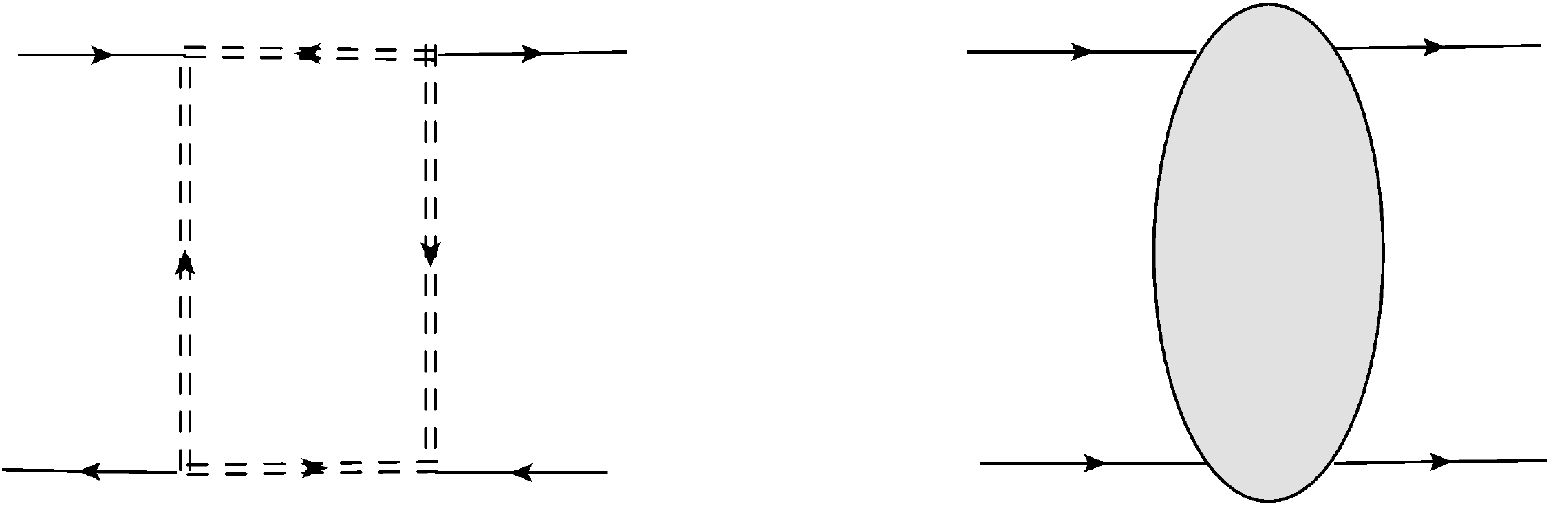
	\end{center}
	\vspace{-0.2cm}
	\caption{The irreducible two-loop diagram in the difference theory proportional to the colour tensor $C_{4}^{(B)\,a_1 a_2 b_1 b_2}$ defined in (\ref{C4}).}
	\label{fig:v42diag}
\end{figure}

The last two-loop diagram we have to consider is the one represented in Fig.~\ref{fig:v42diag}.
\noindent
This diagram was already computed in \cite{Billo:2017glv} and the result, which is also 
reviewed in Appendix~\ref{app:Feynman diagrams}, is
\begin{equation}
\label{2loop4}
{\parbox[c]{.22\textwidth}{ \includegraphics[width 
= .22\textwidth]{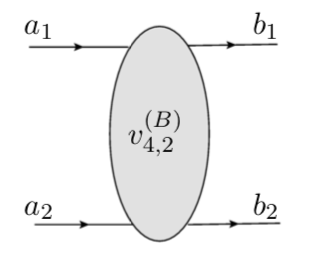}}} \equiv  ~ {v}_{4,2}^{(B)}\, \Delta^2(x) \,{C}_{4}^{(B)\,a_1a_2b_1b_2}
+\cdots~,
\end{equation}
where
\begin{align}
\label{v42is}
{}& v_{4,2}^{(B)}\approx \Big(\frac{g_0^2}{8\pi^2}\Big)^2 \,3\,\zeta(3) (\pi x^2)^{2\epsilon}~,
\\[3mm]\notag
{}& {C}_{4}^{(B)\,a_1a_2b_1b_2}= -(2N-N_f)\, \tr T^{a_1}T^{b_1}T^{a_2}T^{b_2}\,
\\
{}& ~~\qquad\qquad\qquad
-\,\frac{1}{2}\big(
\delta^{a_1b_1}\delta^{a_2 b_2}+\delta^{a_1a_2}\delta^{b_1 b_2}
+\delta^{a_1b_2}\delta^{a_2 b_1}\big)~.
\label{C4}
\end{align}
Again, the ellipses in (\ref{2loop4}) stand for terms with color tensors
which are anti-symmetric in $(a_1,a_2)$ and $(b_1,b_2)$ and, therefore, vanish when inserted
in the two-point correlation function.

\subsection{Effective vertices}

In order to later compare the results of perturbation theory to those of the matrix model, we find it convenient to introduce effective vertices following the ideas of
\cite{Billo:2017glv}. In particular, to obtain the color dependence of the two-point  function 
$G_{\vec{n},\vec{m}}$ we strip the $x$-dependence of the scalar fields and introduce 
the adjoint matrix $\vphi = \vphi^a T^a$ and its conjugate 
$\widebar{\vphi}=\widebar{\vphi}^aT^a$, such that
\begin{equation}
\label{treemc}
\big\langle\vphi^a\,\widebar{\vphi}^b\big\rangle=\delta^{ab}~,\qquad\qquad
\big\langle\vphi^a\,\vphi^b\big\rangle=\big\langle\widebar{\vphi}^a\,\widebar{\vphi}^b\big\rangle=0~. 
\end{equation}
We denote by $O_{\vec n}(\vphi)$ the operator obtained by replacing in (\ref{On}) the 
field $\varphi(x)$ with the constant matrix $\vphi$, and by $\widebar{O}_{\vec{m}}(\widebar{\vphi})$
the same with $\widebar{\varphi}(x)$ replaced with $\widebar{\vphi}$. 

With these definitions, it is straightforward to see that the tree-level correlator (\ref{Gtree}) 
can be written as
\begin{equation}
\label{Gtreemm} 
G_{\vec n,\vec m}\big|_{\text{tree}} = \big\langle O_{\vec n}(\vphi) 
\,\widebar O_{\vec m}(\widebar\vphi)\big\rangle~.
\end{equation}
Also the one-loop correlator (\ref{onelres}) can be written in a simple way using this formalism. 
Indeed, we have 
\begin{equation}
\label{onelresmm}
G_{\vec n,\vec m}\big|_{\text{1-loop}} 
= \,v_{2,1}\,\big\langle V_{2}(\vphi,\widebar\vphi)\,
O_{\vec n}(\vphi)\, \widebar O_{\vec m}(\widebar\vphi)
\big\rangle~,
\end{equation}
where
\begin{equation}
\label{defV21vphi}
V_{2}(\vphi,\widebar\vphi) = \,\,\delta^{ab}\nord{\vphi^a\,\widebar{\vphi}^b}
\,\,=2\nord{\tr \vphi\,\widebar{\vphi}} ~. 
\end{equation}
As usual, the notation $:\,\,:$ stands for normal ordering.
Using (\ref{treemc}), it is easy to check that inside a vacuum expectation
value (\ref{onelresmm}) we can use the relation
\begin{equation}
V_{2}(\vphi,\widebar\vphi)\,O_{\vec n}(\vphi) = n\,O_{\vec n}(\vphi)
\label{V2On}
\end{equation}
that follows from the SU$(N)$ fusion/fission rules
\begin{equation}
\begin{aligned}
\tr\, T^a B_1 T^a B_2 &=\frac{1}{2}\, \tr\,  B_1\,\tr\, B_2 -\frac{1}{2N}\,\tr\,B_1 B_2~,\\
\tr\,T^a B_1~\tr\, T^a B_2&=\frac{1}{2}\,\tr\,B_1 B_2 -\frac{1}{2N}\,\tr\,B_1~\tr\,B_2~,
\end{aligned}
\label{identities}
\end{equation}
valid for two arbitrary  $N\times N$  matrices $B_1$ and $B_2$.

Let us now compute the two-loop contribution to the two-point correlation function from diagrams shown in Fig.~\ref{fig:diagrams}. Using the reducible term (\ref{2p1lres}) we find that 
the contribution of the diagram in Fig.~\ref{fig:diagrams}$(c)$ is
\begin{equation}
G_{\vec n,\vec m}^{(c)}=\frac 12 n(n-1)\, v_{2,1}^2\, G_{\vec n,\vec m}\big|_{\text{tree}}~.
\label{G2lc}
\end{equation}
In terms of the effective vertex (\ref{defV21vphi}), this result can be rewritten as follows:
\begin{equation}
\begin{aligned}
G_{\vec n,\vec m}^{(c)}&=\frac 12\, v_{2,1}^2\,  \big\langle\! :\!
\left[{V}_{2}(\vphi,\widebar\vphi)\right]^2 \!: 
\,O_{\vec n}(\vphi)\,\widebar{O}_{\vec{m}}(\widebar{\vphi})
\big\rangle\\
&=\frac{1}{2}\, v_{2,1}^2\Big[\big\langle
\left[{V}_{2}(\vphi,\widebar\vphi)\right]^2
\,O_{\vec n}(\vphi)\,\widebar{O}_{\vec{m}}(\widebar{\vphi})\big\rangle
-2\,\big\langle {V}_{2}(\vphi,\widebar\vphi)
\,O_{\vec n}(\vphi)\,\widebar{O}_{\vec{m}}(\widebar{\vphi})\big\rangle\\
&\qquad\qquad-\big\langle
\left[{V}_{2}(\vphi,\widebar\vphi)\right]^2\big\rangle\,\big\langle
O_{\vec n}(\vphi)\,\widebar{O}_{\vec{m}}(\widebar{\vphi})\big\rangle
\Big]~,
\end{aligned}
\label{G2lc1}
\end{equation}
where in the second line we have used the identity
\begin{equation}
:\! {V}_{2}(\vphi,\widebar\vphi) {V}_{2}(\vphi,\widebar\vphi) \!: \,\,=\left[{V}_{2}(\vphi,\widebar\vphi)\right]^2-
2\,{V}_{2}(\vphi,\widebar\vphi)-\big\langle
\left[{V}_{2}(\vphi,\widebar\vphi)\right]^2\big\rangle 
\end{equation}
that follows from Wick's theorem.

The two-loop reducible correction (\ref{2loopv212}) to the scalar propagator can be inserted in any
of the $n$ internal lines. Thus, the contribution of the diagram 
in Fig.~\ref{fig:diagrams}$(d)$ is
\begin{equation}
G_{\vec n,\vec m}^{(d)} \approx  n\, v_{2,1}^2\, G_{\vec n,\vec m}\big|_{\text{tree}} =     v_{2,1}^2 \, \big\langle {V}_{2}(\vphi,\widebar\vphi) 
\,O_{\vec n}(\vphi)\,\widebar{O}_{\vec{m}}(\widebar{\vphi})
\big\rangle ~,
\label{G2ld1}
\end{equation}
where we used (\ref{V2On}).  
In a similar way, the irreducible two-loop correction (\ref{2loop2}) produces the following
contribution to the diagram in Fig.~\ref{fig:diagrams}$(e)$:
\begin{equation}
G_{\vec n,\vec m}^{(e)}=n \, v_{2,2} \,G_{\vec n,\vec m}\big|_{\text{tree}} =v_{2,2}\,\big\langle V_{2}(\vphi,\widebar\vphi)
\,O_{\vec n}(\vphi)\,\widebar{O}_{\vec{m}}(\widebar{\vphi})
\big\rangle~.
\label{V22Onb1}
\end{equation}

Let us now consider the two-loop contributions proportional to $v_{4,2}^{(A)}$ and $v_{4,2}^{(B)}$
given, respectively, in (\ref{v42Aform}) and  (\ref{2loop4}). 
To write the results in a compact form,
 it is convenient to 
introduce the quartic vertices
\begin{align}
V_{4}^{(A)}(\vphi,\widebar\vphi) & =  C_{4}^{(A)\,a_1 a_2 b_1 b_2} \,:\vphi^{a_1}\,\vphi^{a_2}\, \widebar\vphi^{b_1}\, \widebar\vphi^{b_2}: \phantom{\Big|}\notag\\
& = \frac{2}{N}  \,:\tr\comm{\,\vphi}{\widebar\vphi\,}^2\!: \,\, 
= \frac{4}{N}\,\big(\!:\tr \vphi\,\widebar\vphi\,\vphi\,\widebar\vphi : - :\tr\vphi^2\,\widebar\vphi^2:\big)~,\phantom{\Big|}\ \label{V4Ais}\\[3mm]
V_{4}^{(B)}(\vphi,\widebar\vphi)
&= {C}_{4}^{(B)\,a_1a_2b_1b_2}\,:\vphi^{a_1}\,\vphi^{a_2}\,\widebar{\vphi}^{b_1}\,
\widebar{\vphi}^{b_2}:\phantom{\Big|}\notag\\
&=-(2N-N_f):\tr\vphi\,\widebar{\vphi}\,\vphi\,\widebar{\vphi}:
-\,4:\big(\tr \vphi\,\widebar{\vphi}\big)^2\!:
-\,2:\tr\vphi\,\vphi\,\,\tr\widebar{\vphi}\,\widebar{\vphi}:~,\phantom{\Big|}\label{V4}
\end{align}
Then, the contribution of the effective vertex (\ref{v42Aform}) to the correlator can be written as
\begin{equation}
\label{v42V4A}
G^{(A)}_{\vec n,\vec m} =
v_{4,2}^{(A)} \,\big\langle V_{4}^{(A)}(\vphi,\widebar\vphi) 
\,O_{\vec n}(\vphi) \,\widebar O_{\vec m}(\widebar\vphi)\big\rangle~.
\end{equation}
Repeatedly using the fission/fusion identities (\ref{identities}), 
it is possible to show that, inside the vacuum expectation value \re{v42V4A}, 
the following relation holds:
\begin{equation}
V_{4}^{(A)}(\vphi,\widebar\vphi)\,\,\tr \vphi^n
=\frac{n}{N}\,\sum_{\ell=0}^{n-2}\big(\tr \vphi^{\ell+1}~\tr \vphi^{n-\ell-1}-
\tr \vphi^\ell~\tr \vphi^{n-\ell}\,\big) =-n\,\tr \vphi^n~.
\label{identity0V4hat}
\end{equation}
More generally, one can prove that
\begin{equation}
V_{4}^{(A)}(\vphi,\widebar\vphi)\,\,O_{\vec n}(\vphi)=-n\,O_{\vec n}(\vphi)~.
\label{identityV4hat}
\end{equation}
By comparing with (\ref{V2On}) we conclude that  the 
quartic vertex $V_{4}^{(A)}$ can be effectively replaced by $(-V_2)$  inside a vacuum expectation value, so that (\ref{v42V4A}) 
becomes
\begin{equation}
\label{G42la}
G^{(A)}_{\vec n,\vec m}= -v_{4,2}^{(A)}\,\big\langle V_{2}(\vphi,\widebar\vphi)
\,O_{\vec n}(\vphi)\,\widebar{O}_{\vec{m}}(\widebar{\vphi})
\big\rangle=
-n \, v_{4,2}^{(A)}\, G_{\vec n,\vec m}\big|_{\text{tree}}~.
\end{equation}

The contribution of the effective vertex (\ref{2loop4})  to the correlator  can be treated in an
analogous way, and it reads
\begin{equation}
G^{(B)}_{\vec n,\vec m}={v}_{4,2}^{(B)}\,\big\langle {V}_{4}^{(B)}(\vphi,\widebar\vphi)
\,O_{\vec n}(\vphi)\,\widebar{O}_{\vec{m}}(\widebar{\vphi})
\big\rangle~.
\label{V4On}
\end{equation}
Notice that, in distinction to the other two-loop contributions,  the expectation value (\ref{V4On}) is 
not proportional, in general, to the tree-level correlator 
$G_{\vec n,\vec m}\big|_{\text{tree}}$ due to the structure of the vertex 
$V_4^{(B)}$, and it has to be computed case by case. A few explicit examples with 
operators of even dimensions are:
\begin{equation}
\begin{aligned}
\!\!&\big\langle V_4^{(B)}(\vphi,\widebar{\vphi}) \,
O_{(2)}(\vphi)\,\widebar{O}_{(2)}(\widebar{\vphi}) \big\rangle=
-\frac{N^2-1}{2}\,\Big(N^2+\frac{N_f}{2N}\Big)~,\\[2mm]
\!\!\!\!&\big\langle V_4^{(B)}(\vphi,\widebar{\vphi}) \,
O_{(2,2)}(\vphi)\,\widebar{O}_{(2,2)}(\widebar{\vphi}) \big\rangle=-
\frac{N^2-1}{2}\,\Big(2 N^4+22 N^2-3 N N_f+\frac{7N_f}{N}\Big)
~,\\[2mm]
\!\!&\big\langle V_4^{(B)}(\vphi,\widebar{\vphi}) \,
O_{(4)}(\vphi)\,\widebar{O}_{(2,2)}(\widebar{\vphi}) \big\rangle=-\frac{N^2-1}{2}\,\Big(6 N^3- N^2 N_f + 6 N + 8 N_f - \frac{21 N_f}{N^2} \Big)~,\\[2mm]
\!\!&\big\langle V_4^{(B)}(\vphi,\widebar{\vphi}) \,
O_{(4)}(\vphi)\,\widebar{O}_{(4)}(\widebar{\vphi}) \big\rangle=
-\frac{N^2-1}{2}\,\Big(12 N^2+ N N_f-18-\frac{21N_f}{N}+\frac{63N_f}{N^3}\Big)~.
\end{aligned}
\label{V4onO}
\end{equation}

The contribution to the two-point correlator from the two-loop
diagram of Fig.~{\ref{fig:diagrams}$(f)$ is given by the sum of (\ref{G42la}) and (\ref{V4On}), 
namely
\begin{equation}
G^{(f)}_{\vec n,\vec m}=-n \, v_{4,2}^{(A)}\, G_{\vec n,\vec m}\big|_{\text{tree}}+
{v}_{4,2}^{(B)}\,\big\langle {V}_{4}^{(B)}(\vphi,\widebar\vphi)
\,O_{\vec n}(\vphi)\,\widebar{O}_{\vec{m}}(\widebar{\vphi})\big\rangle\,,
\label{Gfmn}
\end{equation}
where $ v_{4,2}^{(A)}$ and $ v_{4,2}^{(B)}$ are defined in \re{v42Ares} and \re{v42is}, respectively,	
	
\subsection{Summary of results}
	
Collecting our findings, up to two loops the bare correlator is given by
\begin{equation}
\begin{aligned}
G_{\vec n,\vec m}
& \approx \,\Big[1+n\,v_{2,1}+
\frac{n(n+1)}{2} \, v_{2,1}^2\Big]\,G_{\vec n,\vec m}\big|_{\text{tree}}\phantom{\Big|}\\
&\quad+ n\,\big(v_{2,2}-{v}_{4,2}^{(A)} \big)
\,G_{\vec n,\vec m}\big|_{\text{tree}} 
+{v}_{4,2}^{(B)}\,\big\langle {V}_{4}^{(B)}(\vphi,\widebar\vphi)
\,O_{\vec n}(\vphi)\,\widebar{O}_{\vec{m}}(\widebar{\vphi})\big\rangle \phantom{\Big|}~.
\end{aligned}
\label{Gnmtot}
\end{equation}
The first line contains the tree-level term, the one-loop correction and the reducible two-loop part,
while the irreducible two-loop terms are written in the second line. 

Eq.~(\ref{Gnmtot}) is the main result of this section. It expresses the  bare two-point correlator 
between chiral and anti-chiral operators up to order $g_0^4$, in terms of the tree-level correlator 
and the {matrix model} correlator with the insertion of the quartic effective vertex $V_4^{(B)}$.
Moreover, it {exhibits} a particularly simple structure of the UV divergences 
of $G_{\vec n,\vec m}$. First, we notice that all divergent terms 
in the difference $\big(v_{2,2}-{v}_{4,2}^{(A)}\big)$ exactly cancel.
Indeed, using (\ref{v2}) and (\ref{v42Ares}), we have
\begin{equation}
v_{2,2}-{v}_{4,2}^{(A)} \approx  -3\zeta(3) \Big(\frac{g_0^2}{8\pi^2}\Big)^2\,
\Big(8N^2-\frac{7N N_f}{2}+\frac{N_f}{2N}\Big)~.
\label{v42etc}
\end{equation}
This, together with the fact that
\begin{equation}
v_{4,2}^{(B)} \approx  3\,\zeta(3) \,\Big(\frac{g_0^2}{8\pi^2}\Big)^2 ~,
\label{v42Bexp}
\end{equation}
implies that the total two-loop irreducible contribution in the second line of (\ref{Gnmtot}) is finite
for $\epsilon\to 0$.
The only divergences remaining at two loops come from the square
of those present at one-loop. At the given loop order, they can be nicely combined into an overall factor (see the first line of (\ref{Gnmtot}))
\begin{equation}
1+n\,v_{2,1}+\frac{n(n+1)}{2} \, v_{2,1}^2+ \cdots =\frac{1}{(1-v_{2,1})^n}~,
\end{equation}
which depends only on the {bare} dimension $n$ of the operators but not on their detailed structure.
		
We conclude this section by showing that the two-loop result (\ref{Gnmtot}) can be rewritten 
in an alternative and elegant form as a correlator in the matrix model. 
 {Combining} the tree-level, one-loop and two-loop contributions given in
(\ref{Gtreemm}), (\ref{onelresmm}), (\ref{G2lc1}), (\ref{G2ld1}), (\ref{V22Onb1}) and
(\ref{Gfmn}), we obtain
\begin{equation}
\begin{aligned}
G_{\vec n,\vec m} &\approx \big\langle O_{\vec n}(\vphi)\,\widebar{O}_{\vec{m}}(\widebar{\vphi})
\big\rangle+\big(v_{2,1}+v_{2,2}-{v}_{4,2}^{(A)}\big)\,\big\langle
V_2(\vphi,\widebar{\vphi})\,O_{\vec n}(\vphi)\,\widebar{O}_{\vec{m}}(\widebar{\vphi})\big\rangle
\phantom{\Big|}\\
&~~~+\frac{1}{2}\, v_{2,1}^2\Big[\big\langle
[{V}_{2}(\vphi,\widebar\vphi)]^2
\,O_{\vec n}(\vphi)\,\widebar{O}_{\vec{m}}(\widebar{\vphi})\big\rangle
-\big\langle
[{V}_{2}(\vphi,\widebar\vphi)]^2\big\rangle\,\big\langle
O_{\vec n}(\vphi)\,\widebar{O}_{\vec{m}}(\widebar{\vphi})\big\rangle\Big]\\
&~~~+{v}_{4,2}^{(B)}\,\big\langle {V}_{4}^{(B)}(\vphi,\widebar\vphi)
\,O_{\vec n}(\vphi)\,\widebar{O}_{\vec{m}}(\widebar{\vphi})\big\rangle~.\phantom{\Big|}
\end{aligned}
\label{Gnmfin1}
\end{equation}
Then, defining the effective interaction vertex
\begin{equation}
V_{\text{eff}}(\vphi,\widebar{\vphi})=-\big(v_{2,1}+v_{2,2}-{v}_{4,2}^{(A)}\big)\,
V_2(\vphi,\widebar{\vphi})-v_{4,2}^{(B)}\,V_{4}^{(B)}(\vphi,\widebar{\vphi})~,
\label{Veff}
\end{equation}
we can recast (\ref{Gnmfin1}) in a very compact way as follows:
\begin{equation}
G_{\vec n,\vec m} \approx  \frac{\displaystyle{\big\langle \rme^{-V_{\text{eff}}(\vphi,\widebar{\vphi})}~O_{\vec{n}}(\vphi)\,\widebar{O}_{\vec{m}}(\widebar{\vphi})\big\rangle\phantom{\Big|}}}
{\displaystyle{\big\langle 
\rme^{-V_{\text{eff}}(\vphi,\widebar{\vphi})}\big\rangle}\phantom{\Big|}}~.
\label{Gnmfin}
\end{equation}
Indeed, expanding the exponentials up to order $g_0^4$, we precisely recover all terms of (\ref{Gnmfin1}).
In particular, from the insertion of a single $V_{\rm eff}$ in the correlator we obtain
the linear terms in the $v_{k\,\ell}$'s appearing in the first and third line of (\ref{Gnmfin1}), 
while the quadratic terms in the second line arise from two insertions of $V_{\rm eff}$.
Notice that since this effective vertex is normal-ordered, the denominator in (\ref{Gnmfin})
contributes up to order $g_0^4$ only with the term proportional 
to $\big\langle [V_2(\vphi,\widebar{\vphi})]^2\big\rangle$ appearing in the second line
of (\ref{Gnmfin1}).
	
In Section~\ref{secn:loc} we show that localization on a four sphere produces an 
expression similar to (\ref{Gnmfin}).
However, in order to compare the two expressions, we should first get rid of the UV divergences and scheme ambiguities that are present in the bare correlator $G_{\vec n,\vec m} $. 
This is the content of the next section.

\section{Renormalization}
\label{secn:rencorr}

The dimensionally regularized bare correlators $G_{\vec n, \vec m}$ given in 
(\ref{Gnmtot}) are divergent for $\epsilon \to 0$, since the one-loop coefficient $v_{2,1}$
defined in (\ref{v1loop}) behaves for small $\epsilon$ as
\begin{equation}
v_{2,1}\approx\frac{g_0^2}{16 \pi^2}\,(\pi x^2)^\epsilon\, (2N - N_f)\,\Big(\frac{1}{\epsilon}+2+\gamma_{\text{E}}\Big)\,.
\label{v21exp}
\end{equation}
As we have remarked before, the UV divergence due to $v_{2,1}$ is the only one that plagues the correlators, since all other terms in $G_{\vec n, \vec m}$ are finite for $\epsilon\to 0$.
To get rid of this divergence, we have to apply the standard renormalization procedure. First, we 
introduce the dimensionless renormalized gauge coupling constant $g$ through the relation
\begin{equation}
\label{gtog0}
g_0^2  =  \mu^{2\epsilon}g^2 \, Z( g^2, \epsilon) \,,
\end{equation}
where $\mu$ is an arbitrary scale, and $Z$ is a suitable function to be determined.
Then, we define the renormalized operators $O^R_{\vec n}(x)$ according to
\begin{equation}
O^R_{\vec n}(x) = \sum_{\vec{m}} Z_{\vec n}^{~\vec m}(g^2,\epsilon)\,O_{\vec m}(x) \,,
\label{Onren0}
\end{equation}
where $Z_{\vec n}^{~\vec m}$ is a matrix-valued function. However, in the previous section we have shown that the divergences of the two-point functions depend only on the scaling dimensions of the 
operators and not on the operator details; therefore a block-diagonal matrix 
can do the job, and we can simplify (\ref{Onren0}) by setting
\begin{equation}
\label{Onren}
O^R_{\vec{n}}(x)
= Z_n(g^2,\epsilon) \, O_{\vec{n}}(x)~.
\end{equation} 
A similar formula holds for the anti-chiral renormalized operators $\widebar{O}^R_{\vec{n}}(x)$.

The singular terms of the functions $Z(g^2, \epsilon)$ and $Z_n(g^2,\epsilon)$ are determined 
by requiring that the correlator 
$\big\langle O^R_{\vec{n}}(x)\,\widebar{O}^R_{\vec{m}}(0)\big\rangle$ should be finite 
when expressed in terms of the renormalized coupling $g$. This means that the renormalized 
correlator
\begin{equation}
\label{renG}
G^R_{\vec{n},\vec{m}} = Z_n^2( g^2, \epsilon)\, G_{\vec n,\vec m}
\Big|_{g_0^2=\mu^{2\epsilon}g^2\,Z(g^2, \epsilon)}
\end{equation}  
is well-defined and free of divergences in the limit $\epsilon\to 0$.

\subsection{The $\beta$-function and anomalous dimensions}

The dependence of the renormalized coupling $g^2$ and 
of the renormalization constant $Z_n(g^2,\epsilon)$ on the energy scale $\mu$ is described, respectively,
by the $\beta$-function and by the anomalous dimensions $\gamma_n(g^2)$
of the operators $O_{\vec n}$. They are defined as follows:
\begin{equation}
\label{defbeta}
\beta(g^2)  \equiv  \mu\,\frac{d g^2}{d\mu}=-2\epsilon g^2
- g^2 \,\mu\,\frac{d \ln Z(g^2,\epsilon)}{d\mu}\,,
\end{equation}
where the last equality stems from the $\mu$-independence of $g_0$, and
\begin{equation}
\label{defgamma}
\gamma_n(g^2)  \equiv -  \mu\,\frac{d \ln Z_n (g^2,\epsilon)}{d\mu}  
= - \beta(g^2)\,\frac{d \ln Z_n(g^2,\epsilon)}{d g^2}\,,
\end{equation} 
where in the second step we used (\ref{defbeta}). Using the perturbative expansions 
\begin{equation}
\begin{aligned}
\beta(g^2) &= -2\epsilon g^2+\beta_0\, g^4 + \beta_1\, g^6 + \ldots~,\\
\gamma_n(g^2)&=\gamma_{n,0}\, g^2 + \gamma_{n,1}\, g^4+\ldots~,
\end{aligned}
\label{betagammaexp}
\end{equation}
we can explicitly integrate (\ref{defbeta}) and (\ref{defgamma}) and get in the minimal subtraction (MS) scheme
\begin{equation}
\begin{aligned}
Z(g^2,\epsilon)&= \exp\left[-\int_0^{g^2 } \!\frac{d t}{t}
\, \frac{\beta(t)+2\epsilon t}{\beta(t)}\right] =
1+ g^2 \,\frac{\beta_0}{2\epsilon} + g^4 \,\Big( \frac{\beta_0^2}{4\epsilon^2} 
+ \frac{\beta_1}{4\epsilon}\Big)+
\ldots~,\\[3mm]
Z_n(g^2,\epsilon)
& =\exp\left[-\int_0^{g^2} \!d t \,\frac{\gamma_n(t)}{\beta(t)}\right]
= 1+ g^2 \,\frac{\gamma_{n,0}}{2\epsilon} + g^4 \,\Big(\frac{\beta_0 \,\gamma_{n,0}+\gamma_{n,0}^2}{8\epsilon^2}+\frac{\gamma_{n,1}}{4\epsilon}\Big) +\ldots~.
\end{aligned}
\label{ZZnexp}
\end{equation}
This shows that the expansion coefficients of the $\beta$-function and of the anomalous dimensions 
$\gamma_n$ are directly related to the coefficients of the $1/\epsilon\,$-terms in $Z$ and $Z_n$ respectively.

Differentiating (\ref{renG}) and using $d G_{\vec n, \vec m}/d\mu=0$, one obtains the Callan-Symanzik  equation 
\begin{equation}
\label{CS}
\left( \mu\,\frac{\partial}{\partial \mu} + \beta(g^2)\,\frac{\partial}{\partial g^2} 
+2 \gamma_n(g^2)  \right)\,G^R_{\vec n,\vec m}=0
\end{equation}
on which we can safely take the limit $\epsilon\to 0$.

We now determine the coefficients of the $\beta$-function and the
anomalous dimensions, using the explicit results of the previous section.
To do so, we expand the bare correlator (\ref{Gnmtot})
for small $\epsilon$ using (\ref{v21exp}), and write
\begin{equation}
G_{\vec n,\vec m}=\alpha_0+g_0^2\Big(\frac{\alpha_{1,1}}{\epsilon}+\alpha_{1,0}+\ldots\Big)+
g_0^4\Big(\frac{\alpha_{2,2}}{\epsilon^2}+\frac{\alpha_{2,1}}{\epsilon}+\ldots\Big)+O(g_0^6)\,,
\label{Gbareexp}
\end{equation}
where $\alpha_0= G_{\vec n,\vec m}\big|_{\mathrm{tree}}$ and
\begin{align}
\alpha_{1,1}&=n\,\frac{2N-N_f}{16\pi^2}\,\alpha_0
~, 
&& \alpha_{1,0}=n\,\frac{(2N-N_f)(2+\gamma_{\text{E}}+\ln \pi x^2)}{16\pi^2}\,\alpha_0
~,\label{alphas}\\[2mm]
\alpha_{2,2}&=n(n+1)\,\frac{(2N-N_f)^2}{512\pi^4}\,\alpha_0~, 
&&\alpha_{2,1}=n(n+1)\,\frac{(2N-N_f)^2(2+\gamma_{\text{E}}+\ln \pi x^2)}{256\pi^4}\,\alpha_0~.
\notag
\end{align}
Plugging the expansion (\ref{Gbareexp}) into the renormalized correlator 
(\ref{renG}), using (\ref{ZZnexp}) and requiring that all divergent terms cancel, one finds 
\begin{align}
\label{betagammares0}
\beta_0= \frac{2\,\big(\alpha_{1,1}^2-2\,\alpha_0 \,\alpha_{2,2}\big)}{\alpha_0\, \alpha_{1,1}}
,\qquad  
\gamma_{n,0} = -\frac{\alpha_{1,1}}{\alpha_0}
,\qquad
\gamma_{n,1} =\frac{2\,\big(2\,\alpha_{1,0} \,\alpha_{2,2}-\alpha_{1,1} \,\alpha_{2,1} \big)}
{\alpha_0\,\alpha_{1,1}},
\end{align}
leading to 
\begin{equation}
\begin{aligned}
\beta_0 =-\frac{2N-N_f}{8\pi^2}~,\qquad
\gamma_{n,0} & = -n \frac{2N-N_f}{16\pi^2}~,\qquad
\gamma_{n,1} = 0~.
\end{aligned}
\label{betagammares}
\end{equation}
This value of $\beta_0$ is in agreement with the well-known result for the one-loop
coefficient of the $\beta$-function in $\cN=2$ SQCD. We also notice that 
$\gamma_{2,0}=\beta_0$. This is consistent with $\mathcal{N}=2$ 
supersymmetry, since the chiral operator $O_{(2)} = \tr \varphi^2$
and the Yang-Mills Lagrangian $-\tr(F^2/4)+\dots $  belong to the same supermultiplet, and thus should renormalize in the same way, that is
\begin{equation}
\label{Z2isZ}
Z_2(g^2,\epsilon) = Z(g^2,\epsilon)~.
\end{equation}
Moreover, using the fact that in $\mathcal{N}=2$ SYM theories
the $\beta$-function receives only one-loop correction%
\footnote{This fact has been tested by explicit computations at two loops,
see for example \cite{Jones:1974mm,Capper:1979ns,Howe:1984xq},
and then extended to all loops using non-renormalization and
anomaly arguments, and further strengthened at the non-perturbative level 
\cite{Novikov:1983uc,Novikov:1985rd,Seiberg:1988ur}.}, {\it{i.e.}}
$\beta_\ell=0$ for all $\ell\geq 1$, we conclude that also 
the anomalous dimensions of  $\tr \varphi^2$ are corrected only at one loop, 
{\it{i.e.}} $\gamma_{2,\ell}=0$ for all $\ell\geq 1$.
This implies that
\begin{equation}
\label{wavefunctions}
Z_2(g^2,\epsilon) = Z(g^2,\epsilon) =1-\frac{g^2(2N-N_f) } {16 \pi^2\, \epsilon}+
\frac{g^4(2N-N_f)^2} {256 \pi^4\, \epsilon^2}+\ldots
=\frac{1}{1+\frac{g^2(2N-N_f)} {16 \pi^2\, \epsilon}}~.
\end{equation}
Furthermore, we deduce from (\ref{betagammares}) that the following 
relation
\begin{equation}
\label{ZnisZ2n2}
Z_n(g^2,\epsilon)  = \left[Z(g^2,\epsilon)\right] ^{\frac n2}
\end{equation}
holds up to two loops. It would be very interesting to investigate whether this relation holds also at higher loops. While this issue is not relevant for the two-loop analysis of the present paper, it is
tempting to speculate that (\ref{ZnisZ2n2}) might actually be true in general. Indeed, in our set-up the anomalous dimensions of the chiral and anti-chiral operators arise because of the breaking of conformal invariance at the quantum level due to dimensional transmutation. The fact that the coefficients $\gamma_{n,0}$ and $\beta_0$ are proportional to each other and that the proportionality factor is $n/2$ (see (\ref{betagammares})), together with $\mathcal{N}=2$ supersymmetry, naturally leads one
to propose the relation (\ref{ZnisZ2n2}). Notice that in the conformal case $N_f=2N$, the renormalization functions simply reduce to 1, due to the absence of divergences,
so that (\ref{ZnisZ2n2}) is trivially satisfied in this case.

\subsection{Renormalized correlators}

Using the previous results, it is easy to see that up to two loops and in the limit $\epsilon\to 0$,
the renormalized correlators (\ref{renG}) take a form completely analogous to the
bare correlators (\ref{Gnmtot}), namely
\begin{equation}
\begin{aligned}
G^R_{\vec n,\vec m}& =\,\Big[1+n\,c_1+
\frac{n(n+1)}{2} \, c_1^2+ n\,c_2\,\Big]\,G_{\vec n,\vec m}\big|_{\text{tree}}\phantom{\Big|} 
+c_3\,\big\langle {V}_{4}^{(B)}(\vphi,\widebar\vphi)
\,O_{\vec n}(\vphi)\,\widebar{O}_{\vec{m}}(\widebar{\vphi})\big\rangle\phantom{\Big|} ~,
\end{aligned}
\label{GnmRtot}
\end{equation}
where
\begin{equation}
\begin{aligned}
c_1&=\frac{g^2}{16\pi^2}\,\nu\,(2N-N_f)~,\\
c_2&= -3\,\zeta(3) \Big(\frac{g^2}{8\pi^2}\Big)^2\,
\Big(8N^2-\frac{7N N_f}{2}+\frac{N_f}{2N}\Big)~,\\
c_3&=3\,\zeta(3) \,\Big(\frac{g^2}{8\pi^2}\Big)^2~,
\end{aligned}
\label{c123}
\end{equation}
and
\begin{equation}
\label{nudef}
\nu= 2 + \gamma_{\text{E}} + \ln \pi \mu^2 x^2~.
\end{equation}
The coefficients $c_i$ are obtained from (\ref{v21exp}), (\ref{v42etc}) and (\ref{v42Bexp}). 
In the conformal case $c_1$ vanishes and the first perturbative correction to the correlator 
appears at order $g^4$.

Performing the same manipulations as described in Section~\ref{secn:pert} 
for the bare correlators, we can rewrite (\ref{GnmRtot}) in the following form:
\begin{align}
{}& G^R_{\vec n,\vec m} = \frac{\displaystyle{\big\langle \rme^{-V^R_{\text{eff}}(\vphi,\widebar{\vphi})}
		~O_{\vec{n}}(\vphi)\,\widebar{O}_{\vec{m}}(\widebar{\vphi})\big\rangle\phantom{\Big|}}}
{\displaystyle{\big\langle \rme^{-V^R_{\text{eff}}(\vphi,\widebar{\vphi})}\big\rangle}\phantom{\bigg|}}
+O(g^6)\,,
\label{GRnmfin}
\\[2mm]
{}& V^R_{\text{eff}}(\vphi,\widebar{\vphi})=-\big(c_1+c_2\big)\,V_2(\vphi,\widebar{\vphi})
-c_3\,V_{4}^{(B)}(\vphi,\widebar{\vphi})\,,
\label{VReff}
\end{align}
where the two- and four-point vertices $V_2$ and $V_4^{(B)}$ are defined, respectively, 
in (\ref{defV21vphi}) and (\ref{V4}).

We remark that for $N_f\neq 2N$, the renormalized correlator $G^R_{\vec{n},\vec{m}}$ 
is \emph{not} a constant, but it depends on $x$ through the $\ln \pi \mu^2 x^2$ term contained 
in $\nu$. At first sight, this fact makes it unlikely that the correlator can be encoded 
in a matrix model. However, the dependence of $G^R_{\vec{n},\vec{m}}$ on $\nu$ is determined by the Callan-Symanzik equation (\ref{CS}) %
\footnote{This equation requires that the correlator must have the form
\begin{equation*}
\label{GRfinite}
G^R_{\vec{n},\vec{m}} = d_0 + g^2 (d_1-d_0 \gamma_{n,0}\,\nu) + g^4 
\left[d_{2}-\nu\,d_0 \gamma_{n,1}+\frac{\nu}{4}(\nu\,d_0 \gamma_{n,0} -2 d_1)  (\beta_0+2\gamma_{n,0}) \right] + O(g^6)~.
\end{equation*}
It is easy to check that (\ref{GnmRtot}) satisfies this requirement. Notice that
the whole $\nu$-dependence of $G^R_{\vec{n},\vec{m}}$ can be reconstructed order by order 
in $g^2$ from the correlator at a given value $\widehat{\nu}$, for instance $\widehat\nu=0$, and 
the coefficients $\beta_0$, $\gamma_{n,0}$ and $\gamma_{n,1}$.}
and by its value at a reference point $\widehat\nu$. 
For instance, setting
\begin{equation}
\label{msbar}
\mu^2 x^2= \frac{\rme^{\gamma_{\text{E}}}}{\pi}~,
\end{equation} 
we get {from (\ref{nudef})}
\begin{equation}
\label{Lmsbar}
\widehat{\nu} = 2(1+\gamma_{\text{E}})~,
\end{equation}
which, as we will see in Section~\ref{secn:loc}, is the combination that matches the one-loop matrix model  results from localization.
In particular, with the choice \re{msbar} the coefficient $c_1$ becomes
\begin{equation}
c_1=\frac{g^2}{8\pi^2}\,(1+\gamma_{\text{E}})\,(2N-N_f) ~.
\label{c1msbar}
\end{equation}
Using \re{Trees} and \re{V4onO} we find from (\ref{GnmRtot}) 
\begin{equation}
\begin{aligned}
G^R_{(2), (2)} &=
\frac{N^2-1}{2}
\bigg[1+ 2\, c_1+3\, c_1^2+2 \,c_2-c_3\Big(N^2+\frac{N_f}{2N}\Big) \bigg]+
O(g^6)~.
\end{aligned}
\label{GR22}
\end{equation}
Similar explicit formulae can be worked out for correlators involving higher dimensional operators. For example, at dimension 4 we find
\begin{subequations}
\begin{align}
G^R_{(2,2),(2,2)}&=\frac{N^4-1}{2}
\Big(1+4\, c_1+10\, c_1^2 +4 \,c_2 \Big)\notag\\
&\qquad
-c_3\,\frac{(N^2-1)}{2}\,\Big(2 N^4+22 N^2-3 N N_f+\frac{7N_f}{N}\Big)+O(g^6)~,\\[3mm]
G^R_{(4),(2,2)}&=\frac{2N^4-5N^2+3}{2N}
\Big(1+4\, c_1+10\, c_1^2 +4 \,c_2 \Big)\notag\\
&\qquad
-c_3\,\frac{(N^2-1)}{2}\,\Big(6 N^3- N^2 N_f + 6 N + 8 N_f - \frac{21 N_f}{N^2} \Big)+O(g^6)
~,\\[3mm]
G^R_{(4),(4)}&=\frac{N^6- 7 N^4+ 24 N^2-18}{4 N^2}
\Big(1+4\, c_1+10\, c_1^2 +4 \,c_2 \Big)\notag\\
&\qquad
-c_3\,\frac{(N^2-1)}{2}\,\Big(12 N^2+ N N_f-18-\frac{21N_f}{N}+\frac{63N_f}{N^3}\Big)+O(g^6) ~,
\end{align}
\label{GR44}
\end{subequations}
where the coefficients $c_1$, $c_2$ and $c_3$ are defined in \re{c123}.
Notice that $c_1$ and $c_2$ enter into these expressions in the same combination.

\subsection{Normalized correlators}
\label{subsecn:normcorr}

The renormalized correlators $G^R_{\vec{n},\vec{m}}$ are finite but they cannot be 
considered as physical observables since they depend on the choice of 
the renormalization scheme. In particular, given a renormalized correlator 
at a certain normalization scale $\mu$, one can always perform a finite 
renormalization of the operators by multiplying the renormalization factors 
$Z$ and $Z_n$ defined in (\ref{gtog0}) and (\ref{Onren}), respectively, by an arbitrary 
finite function of the coupling. This transformation preserves the UV finiteness of the correlator 
and corresponds to a change of the renormalization scheme 
(see, for example, \cite{Sint:1998iq} for a discussion of this point in a different context). 

Since, up to two loops we have $Z_n=(Z_2)^{\frac{n}{2}}$, we can eliminate the scheme 
dependence by considering dimensionless ratios of correlators.
In fact, the renormalized correlators can be written as 
\begin{equation}
\label{GRexp3}
G^R_{\vec{n},\vec{m}} = \frac{1}{(1-c_1-c_2)^n}
\bigg[G_{\vec{n},\vec{m}}\big|_{\text{tree}}+
c_3\,\big\langle {V}_{4}^{(B)}(\vphi,\widebar\vphi)
\,O_{\vec{n}}(\vphi)\,\widebar{O}_{\vec{m}}(\widebar{\vphi})\big\rangle \bigg]
+ O(g^6)~.
\end{equation}
This relation shows that the dependence on $x^2$ and on the renormalization scale 
$\mu$ coming from the coefficient $c_1$ is entirely encoded in a prefactor which only 
depends on the bare scaling dimension $n$ of the operators but not on their specific form. 
Therefore, this prefactor cancels in the ratio of correlators of operators of the same 
dimension. Choosing, for example, as a reference the correlator between two 
operators $O_{(2)}$, which are the only ones with dimension 2, we are led to 
introduce the normalized correlators
\begin{equation}
{A}^R_{\vec{n},\vec{m}} = 
\frac{G^R_{\vec{n},\vec{m}}}{ \big[ G^R_{(2),(2)} \big]^{\frac{n}{2}}\phantom{\Big|}}~.
\label{defARratio}
\end{equation}
These ratios are independent of the choice of the renormalization scale $\mu$ and scheme, 
and as such they represent physical quantities.  

It is interesting to observe that the two-loop contribution to the ratio ${A}^R_{\vec{n},\vec{m}}$ 
only comes from the irreducible diagram represented in Fig.~\ref{fig:v42diag}, which is finite in 
the limit $\epsilon\to 0$ (see (\ref{2loop4}) and (\ref{v42is})). 
This shows that also the bare ratios
\begin{equation}
{A}_{\vec{n},\vec{m}} = 
\frac{G^{\phantom{R}}_{\vec{n},\vec{m}}}{ \big[ G_{(2),(2)} \big]^{\frac{n}{2}} 
	\phantom{\Big|}}={A}^R_{\vec{n},\vec{m}} 
\label{defAratio}
\end{equation}
are finite. The equality between ${A}_{\vec{n},\vec{m}}$ and ${A}^R_{\vec{n},\vec{m}} $ 
comes from the fact that the $Z_n$-factors cancel between the numerator and denominators, 
and the $Z$-renormalization of the gauge coupling starts to contribute 
at the next order. One can check this explicitly, by writing the bare correlators (\ref{Gnmtot}) as
\begin{equation}
\label{GRexp2}
G_{\vec{n},\vec{m}} = \frac{1}{\big(1-v_{2,1}-v_{2,2}+v_{4,2}^{(A)}\big)^n}
\bigg[G_{\vec{n},\vec{m}}\big|_{\text{tree}}
+{v}_{4,2}^{(B)}\,\big\langle {V}_{4}^{(B)}(\vphi,\widebar\vphi)
\,O_{\vec{n}}(\vphi)\,\widebar{O}_{\vec{m}}(\widebar{\vphi})\big\rangle \bigg]
+ O(g_0^6)~,
\end{equation}
which shows that the divergence encoded in the one-loop coefficient $v_{2,1}$ 
cancels in the ratios $A_{\vec{n},\vec{m}}$. Furthermore, by 
comparing (\ref{GRexp2}) and (\ref{GRexp3}), we can realize that the bare and the 
renormalized ratios match at two loops, apart from the obvious replacement of $g_0$ 
with $g$.

The explicit expressions of the normalized correlators for operators of 
dimension 4 are:
\begin{subequations}
\begin{align}
{A}^R_{(22),(22)}   &=   \frac{2 \left(N^2+1\right)}{N^2-1}-
\frac{3g^4\zeta(3)}{16 \pi^4} \frac{  \left(10 N^3 -2\, N^2 N_f  +3 N_f\right)}
{N \left(N^2-1\right)}~,\\[3mm]
{A}^R_{(4),(22)}   &=\frac{ 2(2 N^2-3) }{N(N^2-1) }- \frac{3g^4\zeta(3)}{32 \pi^4 } 
\frac{ \left(2 N^5 +12	N^3-N^4 N_f+ 
	6 N^2 N_f -18  N_f \right)}{N^2 \left(N^2-1\right)} ~,\\[3mm]
{A}^R_{(4),(4)}  &= \frac{N^4-6 N^2+18}{N^2 \left(N^2-1\right)}\notag\\
&\qquad+ \frac{3 g^4 \zeta (3)}{64 \pi^4 }
\frac{\left(2 N^7-36 N^5 +72 N^3-N^4N_f +  
	36 N^2N_f  -108N_f  \right)}{N^3\left(N^2-1\right)}~.
\label{aqft}
\end{align}
\end{subequations}
Similar formulae can be easily found also for operators of other dimensions.

In the next section we will recover this same result from the matrix model obtained 
by applying localization on the sphere $S^4$.

\section{Matrix model approach}
\label{secn:loc}

In \cite{Pestun:2007rz} it was shown, using localization techniques, that the partition function of a 
$\mathcal{N}=2$ SYM theory with gauge group SU$(N)$ defined on a four-sphere 
$S^4$ can be written in terms of a traceless  Hermitian  $N\times N$
matrix $a$ in the following way:
\begin{equation}
\cZ_{S^4}=\int \prod_{u=1}^{N}\! da_u~\Delta(a)\,  \big| Z(\ii a,\tau)\big|^2
\,\delta\Big(\sum_{v=1}^Na_v\Big)~.
\label{ws444}
\end{equation} 
Here we have denoted by $a_u$ the (real) eigenvalues of $a$, by $\Delta(a)$ 
the Vandermonde determinant
\begin{equation}
\Delta(a) =   \prod _{u<v=1}^{N}a_{uv}^2~,
\label{vandermon}
\end{equation}
where $a_{uv}=a_u-a_v$, and for simplicity have set to 1 the radius $R$ of the four-sphere\,\footnote{The dependence on $R$ can be easily restored by replacing $a_u$ with $a_u\,R$.}. Furthermore,
$Z(\ii a,\tau)$ is the gauge theory partition function with $\tau$ being the complexified gauge coupling:
\begin{equation}
\tau=\frac{\theta}{2\pi }+\ii \frac{4\pi}{g^2}\,.
\label{tau}
\end{equation}
In this paper we actually set the $\theta$-angle to zero. We remark that in the non-conformal theories the coupling $g$ appearing in the matrix model has to be interpreted as the renormalized 
gauge coupling at a scale proportional to the inverse radius of the four-sphere \cite{Pestun:2007rz}.

The gauge theory partition function $Z(\ii a,\tau)$ is computed using the localization techniques of 
\cite{Nekrasov:2002qd,Nekrasov:2003rj} with a purely imaginary vacuum
expectation value $\vev{\varphi}=\ii\,a$ for the adjoint scalar, and an 
$\Omega$-background with parameters $\epsilon_1=\epsilon_2=1/R$. 
This partition function can be written as a product of the classical, one-loop and instanton contributions, namely
\begin{equation}
Z(\ii a,\tau)  =  Z_{\mathrm{class}}(\ii a,\tau)\,  Z_{\mathrm{one-loop}} (\ii a)\,  
Z_{\mathrm{inst}} (\ii a,\tau) \,.
\end{equation}
Since we work at weak coupling $g^2 \ll 1$, where instantons are exponentially suppressed, 
we can set $Z_{\mathrm{inst}}(\ii a)=1$.
The classical part produces a simple Gaussian term in the matrix model:
\begin{equation}
\left|Z_{\mathrm{class}}(\ii a,\tau)\right|^2
=  \rme^{-\frac{8\pi^2}{g^2}  \sum_u a_u^2  }
\, = \, \rme^{-\frac{8\pi^2}{g^2} \,\tr a^2  }~.
\end{equation}
The one-loop contribution arising from the gauge multiplet and $N_f$ matter multiplets can be written as
\cite{Pestun:2007rz} (see also \cite{Billo:2017glv} for details)
\begin{equation}
\left|Z_{\mathrm{1-loop}}(\ii a)\right|^2 \,=\,
\rme^{-S_2(a)  -S_4(a) +\,\cdots}
\label{zloops4}
\end{equation}
where $S_n(a)$ are homogeneous polynomials in $a$ of order $n$.
The first few of them are:
\begin{subequations}
	\begin{align}
	S_2(a) &= -(1+\gamma_{\text{E}}) 
	\bigg(\sum_{u,v=1}^N  a_{uv}^2-N_f  \sum_{u=1}^N  a_{u}^2 \bigg)
	=\, -(1+\gamma_{\text{E}}) \,(2N-N_f)\, \mathrm{tr}\,a^2 ~, \label{S2intexp}\\[1mm]
	S_4(a) &=  \frac{\zeta(3)}{2} \, \bigg(\sum_{u,v=1}^N
	a_{uv}^4-N_f  \sum_{u=1}^N  a_{u}^4 \bigg) =\,\frac{\zeta(3)}{2} \, \Big[ (2 N-N_f) \,
	{\mathrm{tr}}\,a^4 + 6 \left(\mathrm{tr}\,a^2\right)^2  \Big] ~,  \label{S4intexp}\\[1mm]
	S_6(a) &=  -\frac{\zeta(5)}{3}\, \bigg(\sum_{u,v=1}^N  a_{uv}^6-N_f  \sum_{u=1}^N  a_{u}^6 \bigg)
	=\,- \frac{\zeta(5)}{3} \,  \Big[ (2 N-N_f) \, \mathrm{tr} \,a^6\label{S6intexp}\\
	&\qquad\qquad\qquad\qquad\qquad\qquad\qquad\qquad\qquad~
	+30\, \mathrm{tr} \,a^4 \, \mathrm{tr} \,a^2 -20 \,\left(\mathrm{tr} \,a^3\right)^2
	\Big]~. \notag
	\end{align}
\end{subequations}
Performing the rescaling
\begin{equation}
\label{resca}
a \to \Big(\frac{g^2}{8\pi^2}\Big)^{\frac 12}\, a~,
\end{equation}
the matrix model gets a canonically normalized Gaussian factor and the sphere partition function 
(\ref{ws444}) becomes
\begin{equation}
\label{rescaledmm}
\cZ_{S^4} = \Big(\frac{g^2}{8\pi^2}\Big)^{\frac{N^2-1}{2}} \, \int \prod_{u=1}^{N}
da_u~\Delta(a)~\rme^{-\tr a^2 - S_{\mathrm{int}}(a)}\,\delta\Big(\sum_{v=1}^Na_v\Big)
\end{equation}
with
\begin{equation}
\label{Sintresc}
S_{\mathrm{int}}(a) = \frac{g^2}{8\pi^2}~S_2(a) +  \Big(\frac{g^2}{8\pi^2}\Big)^2 \, S_4(a) +
\Big(\frac{g^2}{8\pi^2}\Big)^3\, S_6(a) + \cdots ~.
\end{equation}
The term of order $g^{2\ell}$ in $S_{\mathrm{int}}(a)$ accounts
for effects that take place at $\ell$ loops in the corresponding field theory computation. 
Therefore we will refer to the $g^2$-expansion of $S_{\mathrm{int}}$ as a loop expansion. 

Exploiting the Vandermonde determinant $\Delta(a)$ and writing $a=a^b\,T^b$, we can alternatively
express the integral (\ref{rescaledmm}) using a flat integration measure $da$ over all matrix components
$a^b$ as follows
\begin{equation}
\cZ_{S^4}= c_N\,\Big(\frac{g^2}{8\pi^2}\Big)^{\frac{N^2-1}{2}} \, 
\int da~\rme^{-\tr a^2 - S_{\mathrm{int}}(a)}
\label{ZS41}
\end{equation}
where $c_N$ is a $g$-independent constant and $da\propto \prod_b da^b$. The overall prefactors 
in (\ref{ZS41}) are irrelevant when computing correlators and thus can be neglected.

Given any function $f(a)$, its vacuum expectation value in the matrix model described above 
is defined as follows
\begin{equation}
\label{vevmat}
\begin{aligned}
\big\langle\,f(a)\,\big\rangle  &= 
\frac{1}{\cZ_{S^4}}
\int \prod_{u=1}^{N}da_u~\Delta(a)\,  \big| Z(\ii a,\tau)\big|^2
\,\delta\Big(\sum_{v=1}^N a_v\Big) \, f(a) \\[3mm]
&=\,\frac{\displaystyle{ \int da ~\rme^{-\tr a^2-S_{\mathrm{int}}(a)}\,
		f(a)}}{ \displaystyle{\int da~\rme^{-\tr a^2-S_{\mathrm{int}}(a)}} }
\,=\,
\frac{\displaystyle{\big\langle\,\rme^{-S_{\mathrm{int}}(a)}\,f(a)\,\big\rangle_0}}
{\displaystyle{\big\langle\,\rme^{-S_{\mathrm{int}}(a)}\,\big\rangle_0}}~,
\end{aligned}
\end{equation}
where in the second line we have used (\ref{ZS41}). The subscript ``0'' 
denotes the vacuum expectation values taken with respect to the Gaussian measure, which can be 
computed by repeatedly using Wick's theorem to reduce them to the basic contraction
\begin{equation}
\label{2pfa}
\big\langle\,a^b\, a^c\,\big\rangle_{0} = \delta^{bc}~.
\end{equation}

\subsection{Chiral and anti-chiral operators in the matrix model}

We are interested in extracting from the matrix model (\ref{vevmat}) 
the two-point functions (\ref{OnOm}).
To this aim we have first to find counter-partners of the chiral and anti-chiral operators 
in the matrix model. It would seem natural to associate to
the multi-trace operator $O_{\vec{n}}(x)$ defined in (\ref{On}), an analogous function 
$O_{\vec{n}}(a)$ in the matrix model, given by the same expression (\ref{On}) but with 
$\varphi(x)$ replaced by $a$, namely
\begin{equation}
O_{\vec{n}}(a) = \tr a^{n_1}\, \tr a^{n_2} \, \ldots \,\tr  a^{n_\ell}~.
\end{equation}
However, the operator $O_{\vec{n}}(x)$ has vanishing vacuum expectation value in the field theory, while in the matrix model $\vev{O_{\vec{n}}(a)}\neq 0$ due to the self-contractions of $a$. 
This means that we have to refine the dictionary and make $O_{\vec{n}}(a)$
normal-ordered. This can be done by subtracting from $O_{\vec{n}}(a)$ all possible self-contractions
and making it orthogonal to all operators with lower dimensions. 

As discussed in \cite{Billo:2017glv,Billo:2018oog,Sysoeva:2017fhr}, the
prescription to define the normal ordering of any operator $O(a)$ in the matrix model
is the following. Let be $\Delta$ the dimension of
$O(a)$ and $\big\{O_p(a)\big\}$ a basis in the finite-dimensional space of 
matrix operators with dimension smaller than $\Delta$. Denoting by $C_\Delta$ the (finite-dimensional) matrix of correlators
\begin{equation}
\big(C_\Delta\big)_{pq} = \big\langle\,O_p(a)\,O_q(a)\,\big\rangle\,,
\label{Cpq}
\end{equation}
which are computed according to (\ref{vevmat}), we define the normal-ordered operator 
as
\begin{equation}
\nordg{O(a)}\,\,
=\, O(a) - \sum_{p,q} \big\langle\,O(a)\, O_{p}(a)\,\big\rangle\, (C_\Delta^{-1})^{pq}\, O_q(a)~.
\label{normalo}
\end{equation}
Our notation stresses the fact that this normal-ordering is $g$-dependent, 
since the correlators on the right hand side of (\ref{normalo}) are computed in the interacting matrix model. The proposal is then to associate to the
field theory operators the corresponding normal-ordered matrix operators, namely 
\begin{equation}
\label{corrOcO}
O_{\vec{n}}(x) \,\quad \to\,\quad \cO_{\vec{n}}(a) =\,\, \nordg{O_{\vec{n}}(a)}~.
\end{equation} 
A similar replacement holds for the anti-chiral operators. 

For example, using the definition \re{normalo} we find
\begin{eqnarray}
\cO_{(2)}(a)&=& \tr a^2 -\frac{N^2-1}{2}\bigg[1+\frac{g^2}{8\pi^2}\,
(1+\gamma_{\text{E}})(2N-N_f)+
\Big(\frac{g^2}{8\pi^2}\Big)^2
(1+\gamma_{\text{E}})^2(2N-N_f)^2\nonumber \\
&&\qquad\qquad\qquad
-\Big(\frac{g^2}{8\pi^2}\Big)^2\zeta(3)\Big(5N^2-N\,N_f
+\frac{3N_f}{2N}\Big)\bigg]+O(g^6)~.
\label{O2g}
\end{eqnarray}
The term of order $g^0$ inside the square brackets represents the self-contraction of $\tr a^2$, 
while the terms of higher order in $g^2$ represent the self-contractions of the operator through the interaction vertices coming from the matrix model action. 
Analogous expressions can be worked out for operators of higher dimension.

\subsection{Correlators in the matrix model}

Once the operators have been identified, their correlators can be computed in a straightforward way
using the definition (\ref{vevmat}). 
In particular the two-point correlators are defined as
\begin{equation}
\label{Gmmis}
\cG_{\vec{n}, \vec{m}} = \big\langle\,\cO_{\vec{n}}(a)\,\cO_{\vec{m}}(a)\,\big\rangle
= \frac{\displaystyle{\big\langle\,\rme^{-S_{\mathrm{int}}}
		\,\cO_{\vec{n}}(a)\,\cO_{\vec{m}}(a)\,\big\rangle_0}}
{\displaystyle{\big\langle\,\rme^{-S_{\mathrm{int}}}\,\big\rangle_0}}~.
\end{equation}
Since normal-ordered operators {with different dimensions} are orthogonal to each other, $\cG_{\vec{n}, \vec{m}}$ vanishes for $n\neq m$.

For instance, for the simplest operator $\cO_{(2)}(a)$ defined in (\ref{O2g}) we find
\begin{equation}
\begin{aligned}
\cG_{(2), (2)} &=
\frac{N^2-1}{2}
\bigg[1+\frac{g^2}{8\pi^2}\,2\,(1+\gamma_{\text{E}})(2N-N_f)+
\Big(\frac{g^2}{8\pi^2}\Big)^2 3\,(1+\gamma_{\text{E}})^2(2N-N_f)^2\\
&\quad-\Big(\frac{g^2}{8\pi^2}\Big)^2\zeta(3)\Big(15N^2-3N N_f+\frac{9N_f}{2N}\Big)\bigg]+
O(g^6)~.
\end{aligned}
\label{Gmm22}
\end{equation}
The explicit expressions of correlators with higher dimensional operators can be computed 
in a similar way. At dimension 4 we find
\begin{subequations}
	\begin{align}
	\cG_{(2,2),(2,2)}&=\frac{N^4-1}{2}
	\bigg[1+\Big(\frac{g^2}{8\pi^2}\Big)4\,(1+\gamma_{\text{E}})(2N-N_f)+
	\Big(\frac{g^2}{8\pi^2}\Big)^2 10\,(1+\gamma_{\text{E}})^2(2N-N_f)^2\notag\\
	&\quad-\Big(\frac{g^2}{8\pi^2}\Big)^2\,12\,
	\zeta(3)\Big(2N^2-\frac{N N_f}{2}+\frac{N_f}{2N}\Big)\bigg]\notag\\
	&\quad
	-\Big(\frac{g^2}{8\pi^2}\Big)^2\,\frac{3}{2}\,\zeta(3)\,(N^2-1)\,\Big(2 N^4+22 N^2-3 N N_f+\frac{7N_f}{N}\Big)+O(g^6)~,\\[4mm]
	\cG_{(4),(2,2)}&=\frac{2N^4-5N^2+3}{2N}
	\bigg[1+\Big(\frac{g^2}{8\pi^2}\Big)4\,(1+\gamma_{\text{E}})(2N-N_f)\notag\\
	&\quad+
	\Big(\frac{g^2}{8\pi^2}\Big)^2 10\,(1+\gamma_{\text{E}})^2(2N-N_f)^2
	-\Big(\frac{g^2}{8\pi^2}\Big)^2\,12\,
	\zeta(3)\Big(2N^2-\frac{N N_f}{2}+\frac{N_f}{2N}\Big)\bigg]\notag\\
	&\quad
	-\Big(\frac{g^2}{8\pi^2}\Big)^2\,\frac{3}{2}\,\zeta(3)(N^2-1)\,\Big(6 N^3- N^2 N_f + 6 N + 8 N_f - \frac{21 N_f}{N^2} \Big)+O(g^6)~,\\
	\cG_{(4),(4)}&=\frac{N^6- 7 N^4+ 24 N^2-18}{4 N^2}
	\bigg[1+\Big(\frac{g^2}{8\pi^2}\Big)4\,(1+\gamma_{\text{E}})(2N-N_f)\notag\\
	&\quad+
	\Big(\frac{g^2}{8\pi^2}\Big)^2 10\,(1+\gamma_{\text{E}})^2(2N-N_f)^2
	-\Big(\frac{g^2}{8\pi^2}\Big)^2\,12\,
	\zeta(3)\Big(2N^2-\frac{N N_f}{2}+\frac{N_f}{2N}\Big)\bigg]\notag\\
	&\quad
	-\Big(\frac{g^2}{8\pi^2}\Big)^2\,\frac{3}{2}\,\zeta(3)(N^2-1)\,\Big(12 N^2+ N N_f-18-\frac{21N_f}{N}+\frac{63N_f}{N^3}\Big)+O(g^6) ~.
	\end{align}
\label{Gmm44}
\end{subequations}
Here we have split the $O(g^4)$ contribution into the sum of a few terms in order to facilitate a comparison with the field theory calculation.

It is easy to check that for $N_f = 2N$ the matrix model correlators 
$\cG_{\vec{n}, \vec{m}}$ exactly match, up to two loops, the correlators $G^R_{\vec{n}, \vec{m}}$
computed in perturbation theory (see (\ref{GR22}) and (\ref{GR44})), thus 
confirming the general results obtained in \cite{Baggio:2014sna,Baggio:2015vxa,Gerchkovitz:2016gxx,Baggio:2016skg,Rodriguez-Gomez:2016ijh,Rodriguez-Gomez:2016cem,Baggio:2014ioa,Billo:2017glv} %
\footnote{In the recent paper \cite{Bourget:2018fhe} a discrepancy at six loops, 
	proportional to $\zeta^2(5) g^{12}$,
	has been pointed out in the comparison between the matrix model results and the correlators 
	obtained by solving Toda equations.}.
The fact that the partition function on the sphere $S^4$ and its associated matrix model
contain information on the correlators in the flat space $\mathbb{R}^4$ is not too surprising in 
the conformal case. We now want to investigate to what extent this relation 
holds in the non-conformal theories with $N_f\neq 2N$.

\subsection{Comparison between matrix model and field theory correlators}

Comparing (\ref{GR22}) with (\ref{Gmm22}), and (\ref{GR44}) with (\ref{Gmm44}), we see that 
they have the same structure and that many terms exactly match. 
However, for $N_f\neq 2N$ there are some differences in the terms proportional
to $\zeta(3)$.  To make the comparison simpler, it is convenient to write
$\cG_{\vec{n},\vec{m}}$ in terms of the complex matrices $\vphi$ and $\widebar{\vphi}$ 
using the formalism introduced in Section~\ref{secn:pert}.
Indeed, it is possible to explicitly check that, up to two loops,
the matrix model correlators (\ref{Gmmis}) can be expressed as follows
\begin{equation}
\cG_{\vec n,\vec m} = \frac{\displaystyle{\big\langle \rme^{-\widehat{V}_{\text{eff}}(\vphi,\widebar{\vphi})}
		~O_{\vec{n}}(\vphi)\,\widebar{O}_{\vec{m}}(\widebar{\vphi})\big\rangle\phantom{\Big|}}}
{\displaystyle{\big\langle \rme^{-\widehat{V}_{\text{eff}}(\vphi,\widebar{\vphi})}\big\rangle}\phantom{\bigg|}}
+O(g^6)\,,
\label{Gmmfin}
\end{equation}
where
\begin{equation}
\widehat{V}_{\text{eff}}(\vphi,\widebar{\vphi})=-\big(\widehat{c}_1+
\widehat{c}_2\big)\,V_2(\vphi,\widebar{\vphi})
-\widehat{c}_3\,V_{4}^{(B)}(\vphi,\widebar{\vphi})
\label{Vmmeff}
\end{equation}
with $V_2$ and $V_4^{(B)}$ defined, respectively, 
in (\ref{defV21vphi}) and (\ref{V4}), and 
\begin{equation}
\begin{aligned}
\widehat{c}_1& =\frac{g^2}{8\pi^2}\,(1+\gamma_{\text{E}})(2N-N_f)~,\\
\widehat{c}_2&= -3\,\zeta(3) \Big(\frac{g^2}{8\pi^2}\Big)^2\,
\Big(2N^2-\frac{N N_f}{2}+\frac{N_f}{2N}\Big)~,\\
\widehat{c}_3&= 3\,\zeta(3) \,\Big(\frac{g^2}{8\pi^2}\Big)^2~.
\end{aligned}
\label{chat123}
\end{equation}
Notice that the effective vertex (\ref{Vmmeff}) has the same form as the renormalized
vertex (\ref{VReff}) obtained from perturbation theory.
Comparing (\ref{chat123}) with (\ref{c123}) and (\ref{c1msbar}), we find
\begin{equation}
\widehat{c}_1=c_1~,~~\widehat{c}_2=c_2+9\,\zeta(3)
\Big(\frac{g^2}{8\pi^2}\Big)^2 \,N(2N-N_f)~,~~\widehat{c}_3
=c_3~.
\end{equation}
Therefore, the difference between the effective vertex 
$\widehat{V}_{\text{eff}}$ of the matrix model and
the renormalized effective vertex ${V}^R_{\text{eff}}$ is
\begin{equation}
\delta=\widehat{V}_{\text{eff}}-{V}^R_{\text{eff}}=9\,\zeta(3)
\Big(\frac{g^2}{8\pi^2}\Big)^2 \,N(2N-N_f)\,V_2(\vphi,\widebar{\vphi})~.
\label{delta}
\end{equation}
It is interesting to observe that $\delta$ vanishes in the conformal case. Moreover, it is proportional to $V_2$ which, as follows from (\ref{V2On}), computes the scaling
dimension of the operators %
\footnote{Notice that a non-zero value of $\delta$ can be compensated by performing a finite renormalization of the scalar operators.}. This fact suggests that it might be interpreted as due to a 
conformal anomaly which, in non-conformal theories, affects the correlation functions in going from
the four-sphere $S^4$ to the flat space $\mathbb{R}^4$, or vice versa.

In the two-loop approximation, we can rewrite (\ref{Gmmfin}) as follows:
\begin{equation}
\label{Gmmexp}
\begin{aligned}
\cG_{\vec{n},\vec{m}} &= \frac{1}{(1-\widehat{c}_1-\widehat{c}_2)^n}
\bigg[G_{\vec{n},\vec{m}}\big|_{\text{tree}}+
\widehat{c}_3\,\big\langle {V}_{4}^{(B)}(\vphi,\widebar\vphi)
\,O_{\vec{n}}(\vphi)\,\widebar{O}_{\vec{m}}(\widebar{\vphi})\big\rangle \bigg]   + O(g^6)~.
\end{aligned}
\end{equation}
This formula clearly shows that the dependence on $\widehat{c}_1$ 
and $\widehat{c}_2$ drops out in the ratio between correlators of operators
with the same scaling dimensions. Thus, in analogy with (\ref{defARratio}),
we are led to define the ratio of correlators in the matrix model
\begin{equation}
\cA_{\vec{n},\vec{m}} =
\frac{\cG_{\vec{n},\vec{m}}}{ \big[ \cG_{(2),(2)} \big]^{\frac{n}{2}}\phantom{\Big|}}~.
\label{defAmmratio}
\end{equation}
Since $\widehat{c}_3=c_3$, it exactly matches the normalized correlator
$A^R_{\vec{n},\vec{m}}$, namely
\begin{equation}
\cA_{\vec{n},\vec{m}}  = A^R_{\vec{n},\vec{m}} ~.
\label{Amatch}
\end{equation}
We have checked this relation in many explicit examples, with operators of dimensions up to 6.

\section{Two-point correlators on the four-sphere}
\label{secn:sphere}

In this section we study in more detail the relation between the correlators
in flat space, discussed in Sections~\ref{secn:pert} and \ref{secn:rencorr}, and those on 
the four-sphere $S^4$. The latter are closely related to the correlators derived
from matrix model presented in Section~\ref{secn:loc}.  In particular we consider
the one-loop correction to the scalar propagator on $S^4$ and compare it 
with the one-loop propagator in flat space defined in (\ref{oneloopa2}).

To this aim, it is convenient to describe a sphere in $D-$dimensions by using 
flat embedding coordinates $\{\eta_0,\eta_\mu\}$ satisfying the quadratic constraint
\begin{align}
\label{sp}
\eta_0^2 + \sum_{\mu=1}^D \eta_\mu^2 = R^2~,
\end{align}
where $R$ is the radius of the $D$-sphere. Following \cite{Adler:1972qq,Adler:1973ty,Drummond:1975yc,Drummond:1977dn,Drummond:1977uy}, we use the stereographic projection
\begin{equation}
\label{map}
\eta_0= R\, \frac{x^2-R^2}{x^2+R^2}~,\qquad \eta_\mu = R^2\, \frac{2x_\mu}{x^2+R^2}
\qquad\mbox{with}~~x^2=\sum_{\mu=1}^Dx_\mu^2~,
\end{equation}
to relate a theory defined on a $D$-sphere to a theory in 
$\mathbb{R}^D$, parametrized by the flat coordinates $x_\mu$. 
One of the advantages of this formalism is that the scalar propagator
on the sphere, denoted by a subscript $\cS$, takes a very simple form given by
\begin{equation}
\label{propsphere}
\big\langle\,\varphi^a(\eta_1)\,\widebar{\varphi}^{\,b}(\eta_2)\,\big\rangle_{S}=
\Delta_{\cS}(\eta_{12})\,\delta^{ab}~,
\end{equation}
where $\eta_{12}=\eta_1-\eta_2$ and
\begin{equation}
\label{DeltaS}
\Delta_{\cS}(\eta_{12}) = \frac{\Gamma(1-\epsilon)}{
	4\pi\,(\pi \eta_{12}^2)^{1-\epsilon}}~.
\end{equation}
Here we have used $D=4-2\epsilon$ and defined
\begin{equation}
\eta_{12}^2= \frac{(x_1-x_2)^2}{\kappa(x_1)\kappa(x_2)}~,\qquad
\kappa(x)= \frac{x^2+R^2}{2R^2}~.
\label{kappa}
\end{equation}
Inserting this into (\ref{DeltaS}) and comparing with (\ref{Delta}), we get
\begin{equation}
\Delta_{\cS}(\eta_{12})  = \big[\kappa(x_1)\,\kappa(x_2)\big]^{1-\epsilon}\,\Delta(x_{12})~.
\label{DeltaS1}
\end{equation}
The scalar propagator on the sphere is thus proportional to the one in flat space, with a scaling factor
raised to the engineering dimensions of the scalar fields.
Notice that this is the same scaling factor that defines the induced metric 
on the sphere through the conformal map (\ref{map}); indeed
\begin{equation}
ds^2 = d\eta_0^2 +\sum_{\mu=1}^D d\eta_\mu^2 = \frac{1}{\kappa^2(x)}
\sum_{\mu=1}^D dx_\mu^2~.
\end{equation}

Let us now consider the correlators between two operators
on the sphere. They have a structure similar to the ones in flat space given in (\ref{formOO}), 
namely 
\begin{equation}
\label{formOOsphere}
\big \langle \,O_{\vec n}(\eta_1) \, \widebar{O}_{\vec m}(\eta_2)\,\big\rangle_{\cS}
= \Delta^n_{\cS}(\eta_{12}) \, G^{(\cS)}_{\vec n,\vec m}(g_0,\epsilon,\eta_{12}) 
\, \delta_{nm} ~.
\end{equation}
The correlators $G^{(\cS)}_{\vec n,\vec m}(g_0,\epsilon,\eta_{12})$, which we will simply
denote as $G^{(\cS)}_{\vec n,\vec m}$, can be computed order by order in perturbation theory. 
At tree level, we have just to contract the color indices of the constituent fields, so that
\begin{equation}
G^{(\cS)}_{\vec n,\vec m}\Big|_{\text{tree}} =G_{\vec n,\vec m}\Big|_{\text{tree}}~.
\label{GStree}
\end{equation}
Inserting this into (\ref{formOOsphere}), using the propagator (\ref{DeltaS}) and taking the 
limit $\epsilon\to 0$, we can easily obtain
\begin{equation}
\label{OOspheretree}
\big \langle \,O_{\vec n}(\eta_1) \, \widebar{O}_{\vec m}(\eta_2)\,\big\rangle_{\cS}
\,\Big|_{\text{tree}}
= \kappa^n(x_1)\,\kappa^m(x_2)\,\big \langle \,O_{\vec n}(x_1) \, \widebar{O}_{\vec m}(x_2)\,\big\rangle\,\Big|_{\text{tree}}~.
\end{equation}
This is the expected relation between correlators on the sphere and correlators 
in flat space that follows from the conformal map (\ref{map}).

Let us now consider the one-loop correction. Before analyzing the correlators on the sphere, 
it is convenient to revisit the calculation of one-loop correction to the scalar propagator in 
flat space, given in (\ref{oneloopa2}) in coordinate space.
The one-loop correction to $\big\langle \varphi^a(x_1)\,\widebar{\varphi}^{\,b}(x_2) \big\rangle$ 
can be written as
\begin{equation}
\label{WR1}
W_1^{ab}(x_{12}) = -g_0^2\, (2N-N_f)\,W_1(x_{12})\,\delta^{ab} ~,
\end{equation}
where
\begin{equation}
W_1(x_{12})=\!\int \!
d^D x_3\,  d^2\bar\theta_3 \,d^D x_4\, d^2\theta_4 \,\Delta(x_{13})
\big( \rme^{-2\ii \theta_4 \partial_{x_{43}}\bar\theta_3}\Delta(x_{43})\big)^2
\Delta(x_{42})~.
\label{Wintis}
\end{equation}
Its Fourier transform is the function
$\cW_1(p)$ discussed in Appendix~\ref{app:Feynman diagrams} (see in particular (\ref{cW1is})
and (\ref{cW1nc})). Computing the integrals, we find 
\begin{equation}
W_1(x_{12})=-\frac{(\pi x_{12}^2)^\epsilon\,\Gamma(1-\epsilon)}{(4\pi)^2\,\epsilon(1-2\epsilon)}
\,\Delta(x_{12})~.
\label{Wintexpl}
\end{equation}
Using this in (\ref{WR1}), one recovers the result presented in (\ref{oneloopa2})
and (\ref{v1loop}). 

Going to the sphere, we find that the one-loop correction to the scalar propagator has a 
form similar to (\ref{WR1}), that is
\begin{equation}
\label{WS1}
W_{1\,\cS}^{ab}(\eta_{12}) = -g_0^2\, (2N-N_f)\,W_{1\,\cS}(\eta_{12})\,\delta^{ab} ~,
\end{equation}
where the function $W_{1\,\cS}$ is the sphere generalization of $W_1$.
Applying the embedding formalism \cite{Adler:1972qq,Adler:1973ty,Drummond:1975yc,Drummond:1977dn,Drummond:1977uy}, the expression of $W_{1\,\cS}$ can be obtained by performing the conformal transformation (\ref{map}) to (\ref{Wintis}). 
Under this map, both the integration measure and the scalar propagators acquire scale factors
according to
\begin{equation}
\label{scaling}
\begin{aligned}
\int \! d^D x_i \,d^2\theta_i  {}& \to \int \! d^D x_i \,d^2\theta_i  \, \kappa^{-D+1}(x_i)~,\\
\Delta(x_{ij}) {}&  \to \Delta(x_{ij})\, \big[\kappa(x_i)\,\kappa(x_j)\big]^{\frac{D-2}{2}}~,
\end{aligned}
\end{equation}
so that $W_1(x_{12})$ becomes
\begin{equation}
\label{WS2}
W_{1\,\cS}(\eta_{12}) = \big[\kappa(x_1)\,\kappa(x_2)\big]^{1-\epsilon}\,I(x_1,x_2)~,
\end{equation}
where
\begin{equation}
I(x_1,x_2)=\!\int \!
d^D x_3\,  d^2\bar\theta_3 \,d^D x_4\, d^2\theta_4 \,\Delta(x_{13})
\big(\rme^{-2\ii \theta_4 \partial_{x_{43}}\bar\theta_3}\Delta(x_{43})\big)^2
\Delta(x_{42})\,\big[\kappa(x_3)\,\kappa(x_4)\big]^{-\epsilon}~.
\label{Iis}
\end{equation}
Comparing this integral with (\ref{Wintis}), we notice the presence of the additional scaling factor
$\big[ \kappa(x_3)\,\kappa(x_4)\big]^{-\epsilon}$, which clearly becomes 1 in four dimensions.

Therefore, if the integrals in (\ref{Wintis}) and (\ref{WS2}) were finite, $W_{1\,\cS}$ and $W_1$ 
would only differ by the overall scaling factor $ \kappa(x_1)\,\kappa(x_2)$. In other words,
if no UV divergences are present, one can safely perform the limit $\epsilon\to 0$
inside the integrals. However, the integral in (\ref{Iis}) is divergent, and thus the scaling factor 
$\big[\kappa(x_3)\,\kappa(x_4)\big]^{-\epsilon}$ in the integrand cannot be neglected. 
The evaluation of this integral is presented in Appendix~\ref{app:feynsphere}, and the result is
\begin{equation}
\label{WS3}
W_{1\,\cS}(\eta_{12}) \approx -\frac{(\pi \eta_{12}^2)^\epsilon\,\Gamma(1-\epsilon)}{(4\pi)^2\,\epsilon(1-2\epsilon)}
\,\Delta_{\cS}(\eta_{12})~.
\end{equation}
Comparing with (\ref{Wintexpl}), we see that, up to terms $O(\epsilon)$, 
the two expressions coincide upon replacing $\Delta_{\cS}$ with $\Delta$, and $\eta_{12}^2$ 
with $x_{12}^2$. The fact that the divergent parts of $W_{1\,\cS}$ and  $W_1$ coincide, 
is not surprising since the UV divergences come from integration at short distances 
where there is no distinction between the sphere and flat space.
What is non trivial, however, is that the finite parts coincide, modulo the obvious 
replacement of $x_{12}$ with $\eta_{12}$.

Putting everything together, we see that the one-loop correction to the scalar propagator on the 
sphere is
\begin{equation}
W^{ab}_{1\,\cS}(\eta_{12})=v^{(\cS)}_{2,1}\, \Delta_{\cS}(\eta_{12})\, \delta^{ab}~,
\label{WSfin}
\end{equation}
with
\begin{equation}
v^{(\cS)}_{2,1}
\approx \frac{g_0^2}{8 \pi^2}\, (2N - N_f)\, \frac{(\pi \eta_{12}^2)^\epsilon\,\Gamma(1-\epsilon)}{2\epsilon(1-2\epsilon)}~,
\label{v1loopS}
\end{equation}
in full analogy with (\ref{oneloopa2}) and  (\ref{v1loop}).
This implies that 
\begin{equation}
\label{onelresS}
G^{(\cS)}_{\vec n,\vec m}\,\big|_{\text{1-loop}} \,=\, n\,v^{(\cS)}_{2,1}
\, G_{\vec n,\vec m}\,\big|_{\text{tree}}~.
\end{equation}
Thus, the renormalization procedure can be done following the same steps we described
in Section~\ref{secn:rencorr}.
Choosing the renormalization scale $\mu^2$ as in (\ref{msbar}) with $x^2$ 
replaced by $\eta_{12}^2$ on the sphere and by $x_{12}^2$ in flat space, then
\begin{equation}
\label{OOsphere1loop}
\big \langle \,O^R_{\vec n}(\eta_1) \, \widebar{O}^R_{\vec m}(\eta_2)\,\big\rangle_{\cS}
\,\Big|_{\text{1-loop}}
= \kappa^n(x_1) \,\kappa^m(x_2) \,\big \langle \,O^R_{\vec n}(x_1) \, \widebar{O}^R_{\vec m}(x_2)\,\big\rangle\,\Big|_{\text{1-loop}}~.
\end{equation}
The relation (\ref{onelresS}) and the explicit expression of $v^{(\cS)}_{2,1}$ 
explain why the correlators $\cG_{\vec{n},\vec{m}}$ obtained from the matrix model perfectly agree with those computed in field theory at one loop.

The same analysis can be carried out at two loops, even though the resulting integrals on the sphere become way more complicated. Most of the two-loop diagrams develop UV divergences 
and need to be regularized. As in the one-loop case, the integrals on the sphere differ 
from those in flat space because of scaling factors $[\kappa(x_i)]^{-\epsilon}$ appearing in the integrands. Such factors do not modify the leading UV divergent contribution but they 
do affect the finite part. As a result, there is no reason {\it a priori} to expect that the finite 
part of the correlation functions on the sphere and in flat space should coincide. However, 
we stress the fact that the finite part of the two-point correlator is not a physical observable since it depends on the regularization scheme. The explicit one-loop calculation shows a perfect agreement between the matrix model and the field theory results for the two-point correlators for the special choice of the renormalization scale. It is natural to ask whether such identification holds also 
at higher loops. At two loops, the results of Sections~\ref{secn:rencorr} and \ref{secn:loc} reveal 
that the finite part of the correlation functions are different in flat space and in the matrix model. 
Still, a perfect match is found for physical observables that are independent of the renormalization
scheme, such as the ratios of correlators with operators of the same dimension. 
In such ratios, all divergent two-loop diagrams cancel and the whole contribution 
at order $g^4$ is due to a single and finite Feynman diagram, namely the irreducible diagram 
represented in Fig.~\ref{fig:v42diag}. The corresponding Feynman integral is finite in 
$\mathbb R^4$ and does not require a regularization. As a consequence, it
possesses the four-dimensional conformal symmetry and takes the same form 
in $\mathbb R^4$ and $S^4$. This explains why the ratios of the correlation functions
match the prediction from localization at two loops, as shown in (\ref{Amatch}).

\section{Summary and conclusions}
\label{secn:concl}

We have explicitly computed the two-point correlation functions between chiral and anti-chiral 
operators in the $\cN=2$ SYM theory with gauge group SU($N$) and $N_f$ fundamental flavors up to 
two loops, using standard (super) Feynman diagrams in dimensional regularization. 
Our results show that these correlators have a remarkably simple structure of UV divergences stemming from the fact that the anomalous dimensions of the operators are proportional to
the $\beta$-function.
We demonstrated that when the renormalization scale $\mu$ and the separation $x$ 
between the operators
are inversely proportional to each other, these correlators can be
obtained via a matrix model which is strikingly similar to the matrix model that
computes the partition function and the chiral/anti-chiral correlators on the four-sphere 
using localization.
Up to two loops, the difference between the two matrix models is just a term of order
$g^4$ proportional to $(2N-n_f)\,V_2$, which acting on the operators gives their
anomalous dimensions.
This suggests that this difference that vanishes in the conformal theories, might be interpreted 
as a conformal anomaly. In the non-conformal cases this could explain the difference between 
the correlators on the four-sphere and those in flat space.

We have also constructed normalized correlators, which are scheme independent and, as such, 
represent physical quantities. Up to two-loops, these normalized correlators are the same on the 
four-sphere and in flat space, and can be computed either using the field theory approach 
with Feynman diagrams, or using localization methods via a simple matrix model.

Our analysis clarifies the relation between the perturbative field calculations and the localization 
results in $\cN=2$ SYM theories. It would be interesting to generalize it in various directions, 
for example to compute the correlators at three or more loops, or to compute the one-point or higher-point correlation functions in presence of Wilson loops. 
In particular it would be interesting to explore in detail the 
two-loop calculations of the correlators using Feynman diagrams on the sphere, and/or 
obtain a ``first-principle" derivation of the difference between the two matrix models that 
yield the correlators on the four-sphere and in flat space. We hope to be able to return 
to some of these points in future works.

\vskip 1cm
\noindent {\large {\bf Acknowledgments}}
\vskip 0.2cm
We would like to thank A.~Belitsky, L.~Bianchi, M.~Frau, R.~Frezzotti, F.~Galvagno, P.~Gregori,
K.~Papadodimas, N.~Tantalo and T.~Vladikas for many useful discussions.

The work of M.B., A.L., F.F., J.F.M. is partially supported by the MIUR PRIN Contract 2015MP2CX4 ``Non-perturbative Aspects Of Gauge Theories And Strings''. The work of G.P.K. is supported by the 
French National Agency for Research grant ANR-17-CE31-0001-01. The work of A.L. is partially supported by the ``Fondi Ricerca Locale dell'Universit\`a del Piemonte Orientale''.
All authors would like thank the Galileo Galilei Institute for Theoretical Physics
for hospitality during the course of this work.  G.P.K. is grateful to INFN and the Simons Foundation 
for partial support.

\appendix

\section{Loop integrals}
\label{app:loop_integrals}
In this appendix we follow closely \cite{Chetyrkin:1981qh} (see also \cite{Grozin:2005yg} 
for a  review) and collect some useful formulae necessary to evaluate the Feynman integrals.
We work in $D=4-2\epsilon$ dimensions and use the propagator of a massless scalar field given 
in (\ref{Delta}), namely
\begin{equation}
\Delta(x) =  \int \frac{d^D k}{(2\pi)^D} \frac{\rme^{\ii k \cdot x}}{k^2}  = 
\frac{\Gamma(1-\epsilon)}{(4\pi)\,(\pi x^2)^{1-\epsilon}}~.
\label{DeltaApp}
\end{equation}
For later convenience, we introduce the graphical notation for Feynman integrals in the momentum representation
\vspace{0.2cm}
\begin{equation}
\parbox[c]{.08\textwidth}{
\begingroup%
  \makeatletter%
  \providecommand\color[2][]{%
    \errmessage{(Inkscape) Color is used for the text in Inkscape, but the package 'color.sty' is not loaded}%
    \renewcommand\color[2][]{}%
  }%
  \providecommand\transparent[1]{%
    \errmessage{(Inkscape) Transparency is used (non-zero) for the text in Inkscape, but the package 'transparent.sty' is not loaded}%
    \renewcommand\transparent[1]{}%
  }%
  \providecommand\rotatebox[2]{#2}%
  \newcommand*\fsize{\dimexpr\f@size pt\relax}%
  \newcommand*\lineheight[1]{\fontsize{\fsize}{#1\fsize}\selectfont}%
  \ifx\svgwidth\undefined%
    \setlength{\unitlength}{35bp}%
    \ifx\svgscale\undefined%
      \relax%
    \else%
      \setlength{\unitlength}{\unitlength * \real{\svgscale}}%
    \fi%
  \else%
    \setlength{\unitlength}{\svgwidth}%
  \fi%
  \global\let\svgwidth\undefined%
  \global\let\svgscale\undefined%
  \makeatother%
  \begin{picture}(1,0.53415317)%
    \lineheight{1}%
    \setlength\tabcolsep{0pt}%
    \put(0.4429747,0.73821547){\color[rgb]{0,0,0}\makebox(0,0)[lt]{\lineheight{1.25}\smash{\begin{tabular}[t]{l}\textbf{$\alpha$}\end{tabular}}}}%
    \put(0.4429747,-0.45821547){\color[rgb]{0,0,0}\makebox(0,0)[lt]{\lineheight{1.25}\smash{\begin{tabular}[t]{l}\textbf{$\beta$}\end{tabular}}}}%
    \put(0,-0.189){\includegraphics[width=\unitlength,page=2]{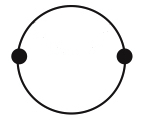}}%
  \end{picture}%
\endgroup%
} ~\equiv~
\int \frac{d^D k}{(2\pi)^D}\frac{1}{(k^2)^{\alpha}\,((p-k)^2)^{\beta}} 
\,=\, \frac{I_{\alpha,\beta}}{(p^2)^{\alpha+\beta-2+\epsilon}}~,
\label{int1}
\end{equation}
\vspace{0.3cm}
where
\begin{equation}
I_{\alpha,\beta}=
\frac{\Gamma(2-\epsilon-\alpha)\,\Gamma(2-\epsilon-\beta)\,\Gamma(\alpha
	+\beta-2+\epsilon)}{(4\pi)^{2-\epsilon}\,\Gamma(\alpha)\,\Gamma(\beta)\,\Gamma(4-2\epsilon-\alpha
	-\beta)}~.
\label{Ialphabeta}
\end{equation}
The black dots on the left and the right of the diagram in (\ref{int1}) denote, respectively, 
the incoming and outgoing momentum $p$. Furthermore, in each interaction vertex the momentum conservation is enforced. When $\alpha$ or $\beta$ is 1, for simplicity we 
do not write the labels. With these notations, we then have
\begin{equation}
{\parbox[c]{.08\textwidth}{ \includegraphics[width = .08\textwidth]{Iab.jpg}}}~=~
\frac{\Gamma^2(1-\epsilon) \,\Gamma(\epsilon)}{(4\pi)^{2-\epsilon}\,\Gamma(2-2\epsilon)}\,\frac{1}{(p^2)^{\epsilon}}~.
\label{int11}
\end{equation}
We will also make use of the Fourier transform integral
\begin{equation}
\begin{aligned}
\Pi_{\alpha}(x)=\int \frac{d^D k}{(2\pi)^D}\frac{\rme^{\ii k\cdot x}}{(k^2)^{\alpha}}&=
\frac{(x^2)^{\alpha+\epsilon-2}\,\Gamma(2-\epsilon-\alpha)}
{4^\alpha\pi^{2-\epsilon}\Gamma(\alpha)}
=\frac{(x^2)^{\alpha-1}\,\Gamma(2-\epsilon-\alpha)}{4^{\alpha-1}\Gamma(\alpha)\,
	\Gamma(1-\epsilon)}\,\Delta(x)~,
\end{aligned}
\label{Pialpha}
\end{equation}
which for $\alpha=1$ reduces to (\ref{DeltaApp}). In particular we will need
the following explicit formulae
\begin{subequations}
	\begin{align}
	\Pi_{1+\epsilon}(x)&=\frac{(x^2)^{\epsilon}\,\Gamma(1-2\epsilon)}{4^\epsilon\,\Gamma(1+\epsilon)
		\,\Gamma(1-\epsilon)}\,\Delta(x)~,\label{Pialphaa}\\[2mm]
	\Pi_{1+2\epsilon}(x)&=\frac{(x^2)^{2\epsilon}\,\Gamma(1-3\epsilon)}{4^{2\epsilon}\,\Gamma(1+2\epsilon)
		\,\Gamma(1-\epsilon)}\,\Delta(x)~,\label{Pialphab}\\[2mm]
	\Pi_{3\epsilon}(x)&=\frac{(x^2)^{2\epsilon}\,\Gamma(2-4\epsilon)}{4^{3\epsilon-2}\,
		\pi^{\epsilon-2}\,\Gamma(3\epsilon)
		\,\Gamma(1-\epsilon)^2}\,\Delta^2(x) ~.\label{Pialpha1c}
	\end{align}
	\label{Pialpha1}
\end{subequations}

\subsection{Triangle identity}
Let us consider the integral
\begin{equation}
J\big(\{\alpha_i\}\big)=\int\! \frac{d^D k}{(2\pi)^D}\,
\frac{1}{(k^2)^{\alpha_1}\,((k-q)^2)^{\alpha_2}\,((k-p)^2)^{\alpha_3}\,(q^2)^{\alpha_4}
	\,((q-p)^2)^{\alpha_5}\,(p^2)^{\alpha_6}}~,
\label{Jalpha1}
\end{equation}
which corresponds to the triangle diagram of Fig.~\ref{fig:triangle}.
\begin{figure}[H]
	\vspace{0.1cm}
	\begin{center}
		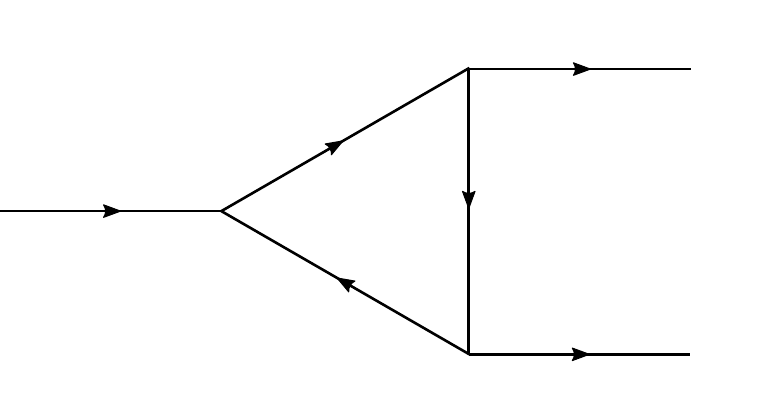
	\end{center}
	\vspace{-0.2cm}
	\caption{The one-loop diagram corresponding to the integral (\ref{Jalpha1}). 
		Here the labels $\alpha_i$ on the various lines denote the exponents of the propagators  
		appearing	in the integrand.}
	\label{fig:triangle}
\end{figure}
\noindent
Following \cite{Chetyrkin:1981qh,Grozin:2005yg}, we have
\begin{equation}
\begin{aligned}
0&=\int\! \frac{d^D k}{(2\pi)^D}\,
\frac{\partial}{\partial k^\mu}
\bigg[\frac{(k-q)^\mu}{(k^2)^{\alpha_1}\,((k-q)^2)^{\alpha_2}\,((k-p)^2)^{\alpha_3}\,(q^2)^{\alpha_4}\,((q-p)^2)^{\alpha_5}\,(p^2)^{\alpha_6}}\bigg]
\phantom{\Bigg|}\\[2mm]
&=\int\! \frac{d^D k}{(2\pi)^D}\,
\frac{D-\alpha_1\,\frac{2k\cdot(k-q)}{k^2}-2\alpha_2
	-\alpha_3\,\frac{2(k-p)\cdot(k-q)}{(k-p)^2}}{(k^2)^{\alpha_1}\,((k-q)^2)^{\alpha_2}\,((k-p)^2)^{\alpha_3}\,(q^2)^{\alpha_4}\,((q-p)^2)^{\alpha_5}\,
	(p^2)^{\alpha_6}}\phantom{\Bigg|}\\[2mm]
&=\int\! \frac{d^D k}{(2\pi)^D}\,
\frac{D-\alpha_1-\alpha_1\,\frac{(k-q)^2-q^2}{k^2}-2\alpha_2-\alpha_3
	-\alpha_3\,\frac{(k-q)^2-(q-p)^2}{(k-p)^2}}{(k^2)^{\alpha_1}\,((k-q)^2)^{\alpha_2}
	\,((k-p)^2)^{\alpha_3}\,(q^2)^{\alpha_4}\,((q-p)^2)^{\alpha_5}\,(p^2)^{\alpha_6}}~.\phantom{\Bigg|}
\end{aligned}
\end{equation}
{From} this, it is easy to obtain the so-called triangle identity:
\begin{equation}
(D-\alpha_1-2\alpha_2-\alpha_3)\,J\big(\{\alpha_i\}\big)
=\Big[\alpha_1 \,  \mathbf{1}^+ \big(\mathbf{2}^- \!-\mathbf{4}^-\big)
+\alpha_3 \,  \mathbf{3}^+ \big(\mathbf{2}^- \!- \mathbf{5}^-\big)\Big] J\big(\{\alpha_i\}\big)\,,
\label{triangle}
\end{equation}
where the notation $\mathbf{n}^\pm J\big(\{\alpha_i\}\big)$ means the
integral (\ref{Jalpha1}) with $\alpha_n$ replaced by $\alpha_n\pm 1$.
For example, we have
\begin{equation}
\mathbf{1}^+\mathbf{2}^- J\big(\{\alpha_i\}\big)=\!\int\! \frac{d^D k}{(2\pi)^D}\,
\frac{1}{(k^2)^{\alpha_1+1}\,((k-q)^2)^{\alpha_2-1}
	\,((k-p)^2)^{\alpha_3}\,(q^2)^{\alpha_4}\,((q-p)^2)^{\alpha_5}\,(p^2)^{\alpha_6}}~.
\end{equation}
Repeated applications of the triangle identity allow us to reduce the power of one of the 
propagators to zero and to express in the end the result in terms of the basic integrals (\ref{int1}).
A few examples are described in the next subsection.

\subsection{Scalar integrals}
Let us consider the two-loop integral
\begin{equation}
\int\! \frac{d^D k\,d^D q}{(2\pi)^{2D}}\,\frac{1}{k^2\,(k-q)^2\,(k-p)^2\,q^2\,(q-p)^2}
~\equiv~{\parbox[c]{.10\textwidth}{ \includegraphics[width = .10\textwidth]{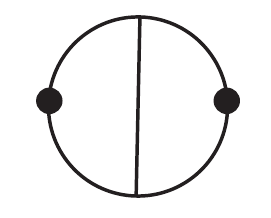}}}
\end{equation}
Here we have adopted the same graphical conventions as in (\ref{int11}).

Applying the triangle identity (\ref{triangle}), we obtain
\begin{equation}
{\parbox[c]{.10\textwidth}{ \includegraphics[width = .10\textwidth]{Ydot1}}}
~=~\frac{1}{\epsilon}~\Bigg[{\parbox[c]{.12\textwidth}{ \includegraphics[width = .12\textwidth]{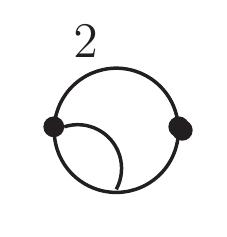}}} -{\parbox[c]{.17\textwidth}{ \includegraphics[width = .17\textwidth]{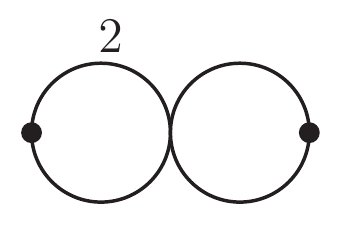}}}\Bigg]~,
\label{int1app}
\end{equation}
where
\vspace{-0.15cm}
\begin{subequations}
	\begin{align}
	& {\parbox[c]{.12\textwidth}{ \includegraphics[width = .12\textwidth]{D1.pdf}}} \equiv\,
	\int\! \frac{d^D k\,d^D q}{(2\pi)^{2D}}\,\frac{1}{(k^2)^2
		\,(k-q)^2\,(k-p)^2\,(q-p)^2}=\frac{I_{1,1}\, I_{2,1+\epsilon}}{(p^2)^{1+2\epsilon}}
	~,\\
	& {\parbox[c]{.17\textwidth}{ \includegraphics[width = .17\textwidth]{D2.pdf}}}
	\equiv\,\int\! \frac{d^D k\,d^D q}{(2\pi)^{2D}}\,\frac{1}{(k^2)^2\,(k-p)^2\,q^2\,(q-p)^2}
	=\frac{I_{2,1}\, I_{1,1}}{(p^2)^{1+2\epsilon}}~.
	\end{align}
\end{subequations}
The last steps in these equations follow 
from (\ref{int1}). Inserting these expressions into (\ref{int1app}) 
and expanding for $\epsilon\to 0$, we obtain
\begin{equation}
{\parbox[c]{.10\textwidth}{ \includegraphics[width = .10\textwidth]{Ydot1}}}
=\,\frac{I_{1,1}\,\big[I_{2,1+\epsilon}-I_{2,1}\big]}{\epsilon}\,\frac{1}{(p^2)^{1+2\epsilon}}=
\frac{6\zeta(3)}{(4\pi)^4}\,\frac{1}{(p^2)^{1+2\epsilon}}+\cdots~.
\end{equation}
After Fourier transforming and using (\ref{Pialphab}), we get
\begin{equation}
{\parbox[c]{.10\textwidth}{ \includegraphics[width = .10\textwidth]{Ydot1}}} \longrightarrow  ~~ \frac{6\zeta(3)}{(4\pi)^4}\,  \Pi_{1+2\epsilon} (x) +\cdots
~=~ \frac{6\zeta(3)}{(4\pi)^4}\,(\pi x^2)^{2\epsilon}\,\Delta(x)+\cdots~.
\label{int2x}
\end{equation}

The same procedure can be applied to express other two-loop integrals in terms of $I_{\alpha,\beta}$ defined in (\ref{Ialphabeta}). For example, we have
\begin{equation}
{\parbox[c]{.09\textwidth}{ \includegraphics[width = .09\textwidth]{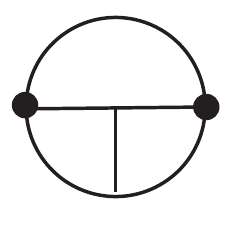}}} = ~
\frac{I_{1,1} I_{1,2}\,\big[ I_{1+\epsilon,1+\epsilon}-I_{1,1+2\epsilon}
	\big]}{\epsilon}\,\frac{1}{(p^2)^{3\epsilon}}
=\frac{2\zeta(3)}{\epsilon\,(4\pi)^6}\,\frac{1}{(p^2)^{3\epsilon}}+\cdots~.
\label{int3}
\end{equation}
Computing the Fourier transform and using (\ref{Pialpha1c}), we find
\begin{equation}
{\parbox[c]{.09\textwidth}{ \includegraphics[width = .09\textwidth]{Ydot5}}}
\longrightarrow  ~~ 
\frac{2\zeta(3)}{\epsilon\,(4\pi)^6}\,\Pi_{3\epsilon}(x)+\cdots=
\frac{6\zeta(3)}{(4\pi)^4}\,(\pi x^2)^{2\epsilon}\,\Delta(x)^2  +\cdots
~.
\label{int3x}
\end{equation}

Another scalar integral that will be needed is the one represented by the diagram
\begin{equation}
p^2\,{\parbox[c]{.08\textwidth}{ \includegraphics[width = .08\textwidth]{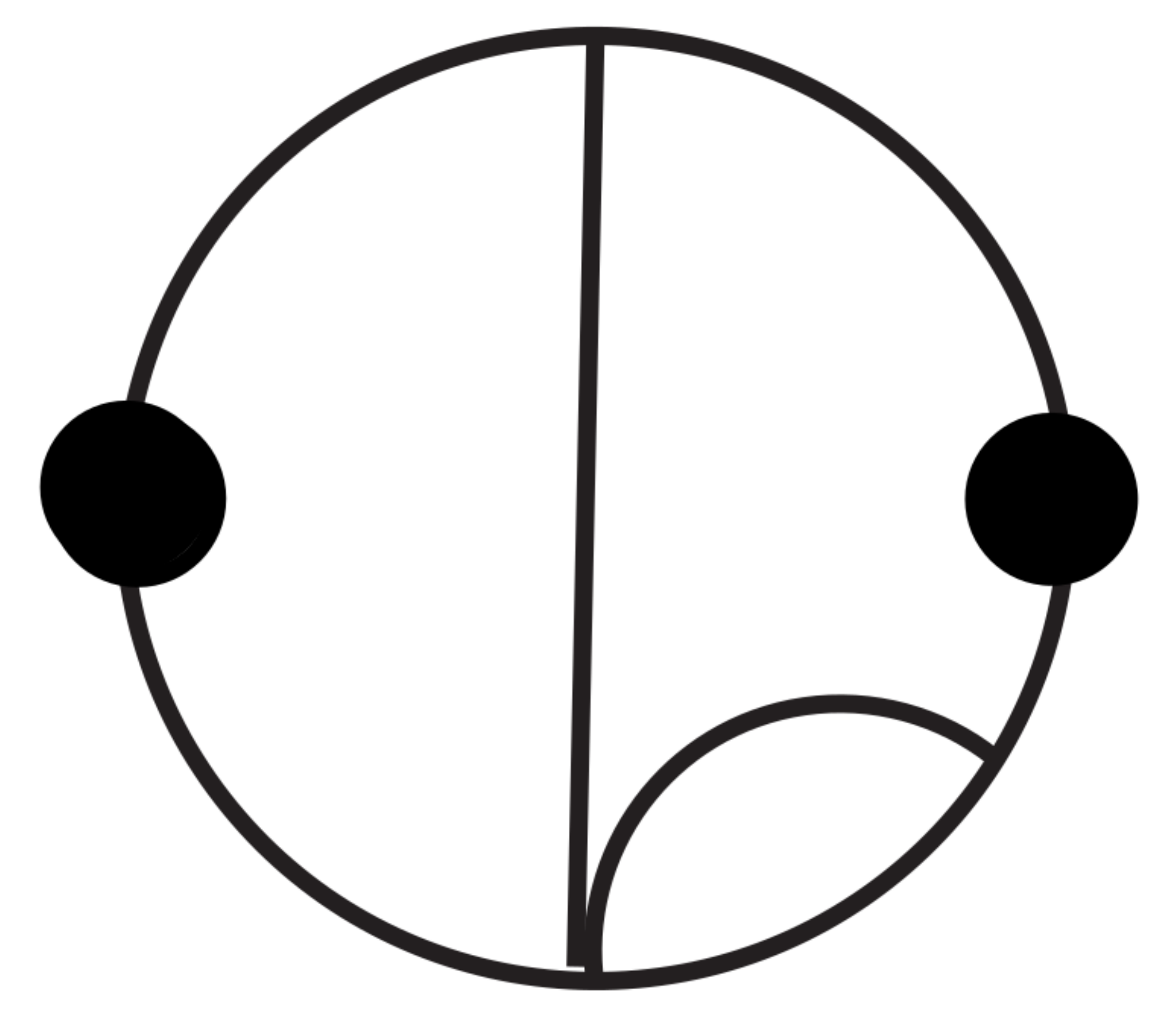}}}\,~.
\end{equation}
Using (\ref{int11}) and expanding for small $\epsilon$, one can prove that
\begin{equation}
{\parbox[c]{.08\textwidth}{ \includegraphics[width = .08\textwidth]{Ydot9}}} \,
=\,\frac{I_{1,1}}{(p^2)^{\epsilon}}{\parbox[c]{.10\textwidth}{ \includegraphics[width = .10\textwidth]{Ydot1}}} +\cdots\, =
\frac{6\zeta(3)}{\epsilon\,(4\pi)^6}\,\frac{1}{(p^2)^{1+3\epsilon}}+\cdots
\label{int4}
\end{equation}
where the ellipses stand for terms that vanish for $\epsilon\to 0$. Comparing with (\ref{int3}), 
we easily conclude that
\begin{equation}
p^2\,{\parbox[c]{.08\textwidth}{ \includegraphics[width = .08\textwidth]{Ydot9}}} \,=
3\,{\parbox[c]{.09\textwidth}{ \includegraphics[width = .09\textwidth]{Ydot5}}}+\cdots
\label{int4bis}
\end{equation}
so that, after Fourier transform, we have
\begin{equation}
p^2\,{\parbox[c]{.08\textwidth}{ \includegraphics[width = .08\textwidth]{Ydot9}}} ~
\longrightarrow  ~~
\frac{18\zeta(3)}{(4\pi)^4}\,(\pi x^2)^{2\epsilon}\,\Delta(x)^2  +\cdots ~.
\label{int4x}
\end{equation}

In a similar way one can derive the following relation
\begin{equation}
{\parbox[c]{.10\textwidth}{ \includegraphics[width = .10\textwidth]{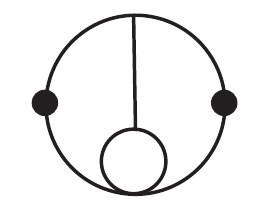}}} \,
=\,\frac{I_{1,1}}{(p^2)^{\epsilon}}{\parbox[c]{.10\textwidth}{ \includegraphics[width = .10\textwidth]{Ydot1}}} +\cdots\, =
\frac{6\zeta(3)}{\epsilon\,(4\pi)^6}\,\frac{1}{(p^2)^{1+3\epsilon}}+\cdots
\label{int50}
\end{equation}
from which we get
\begin{equation}
p^2 {\parbox[c]{.10\textwidth}{ \includegraphics[width = .10\textwidth]{Ydot8}}}  \,=
3\,{\parbox[c]{.09\textwidth}{ \includegraphics[width = .09\textwidth]{Ydot5}}}+\cdots~.
\label{int51app}
\end{equation}
Performing the Fourier transform we obtain
\begin{equation}
p^2 {\parbox[c]{.10\textwidth}{ \includegraphics[width = .10\textwidth]{Ydot8}}}  ~
\longrightarrow  ~~
\frac{18\zeta(3)}{(4\pi)^4}\,(\pi x^2)^{2\epsilon}\,\Delta(x)^2  +\cdots ~.
\label{int51x}
\end{equation}

The following divergent integrals  also appear in the two-loop calculation
\begin{subequations}
	\begin{align}
	\frac{1}{p^2}{\parbox[c]{.08\textwidth}{ \includegraphics[width = .08\textwidth]{Iab.jpg}}}  
	~~&~~=~  \frac{ I_{1,1}}{(p^2)^{1+\epsilon} } =\frac{\Gamma^2(1-\epsilon)\,
		\Gamma(\epsilon)}{(4\pi)^{2-\epsilon}\,\Gamma(2-2\epsilon)}\,\frac{1}{(p^2)^{1+\epsilon} }~,\label{int11app}\\[3mm]
	{\parbox[c]{.18\textwidth}{ \includegraphics[width = .18\textwidth]{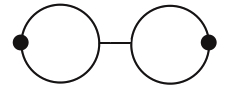}}}
	&~~=~  \frac{ I_{1,1}^2}{(p^2)^{1+2\epsilon} }=\bigg[\frac{\Gamma^2(1-\epsilon) \,
		\Gamma(\epsilon)}{(4\pi)^{2-\epsilon}\,\Gamma(2-2\epsilon)}\bigg]^2
	\frac{1}{(p^2)^{1+2\epsilon}}~,\label{int22app}\\[3mm]
	\frac{1}{p^2}{\parbox[c]{.12\textwidth}{ \includegraphics[width = .12\textwidth]{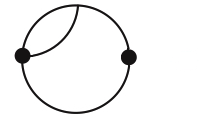}}}
	\!\!\!\!\!&~~=~\frac{ I_{1,1}\, I_{1,1+\epsilon}}{(p^2)^{1+2\epsilon}}=
	\frac{\Gamma^3(1-\epsilon)\,\Gamma(2\epsilon)}
	{(4\pi)^{4-2\epsilon}\,\epsilon (1-2\epsilon)\,\Gamma(2-3\epsilon)}
	\,\frac{1}{(p^2)^{1+2\epsilon}}~,
	\label{int33app}\\[3mm]
	{\parbox[c]{.085\textwidth}{ \includegraphics[width = .085\textwidth]{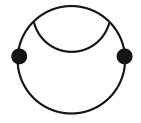}}}  
	\,\,&~~=~  \frac{ I_{1,1}\, I_{1,2+\epsilon}}{(p^2)^{1+2\epsilon}}=-
	\frac{\Gamma^3(1-\epsilon)\,\Gamma(2\epsilon)}
	{(4\pi)^{4-2\epsilon}\epsilon(1-2\epsilon)(1+\epsilon)\Gamma(1-3\epsilon)}
	\,\frac{1}{(p^2)^{1+2\epsilon}}~,
	\label{int44app}\\[3mm]
	{\parbox[c]{.10\textwidth}{ \includegraphics[width = .10\textwidth]{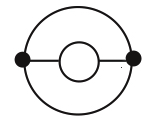}}}\!&~~=~
	\frac{I_{1,1}^2  I_{\epsilon,2+\epsilon}}{(p^2)^{3\epsilon}}=
	-\frac{\Gamma^4(1-\epsilon)\,\Gamma(3\epsilon)}
	{2(4\pi)^{6-3\epsilon}\,\epsilon^2(1-2\epsilon)\,(1+\epsilon)\,\Gamma(2-4\epsilon)}
	\,\frac{1}{(p^2)^{3\epsilon}}~.
	\label{int55app}
	\end{align}
\end{subequations}
After Fourier transforming, we get using (\ref{Pialpha1})  
\begin{subequations}
	\begin{align}
	\frac{1}{p^2}{\parbox[c]{.08\textwidth}{ \includegraphics[width = .08\textwidth]{Iab.jpg}}}\quad
	& ~~\longrightarrow  ~~ I_{1,1}\,  \Pi_{1+\epsilon} (x) = 
	\frac{(\pi x^2)^\epsilon\,\Gamma(1-\epsilon)}{(4\pi)^2\,(1-2\epsilon)}\,\Delta(x)~,
	\label{int11x}\\[3mm]
	{\parbox[c]{.18\textwidth}{ \includegraphics[width = .18\textwidth]{Ydot4.jpg}}}
	& ~~\longrightarrow ~~
	I_{1,1}^2  \Pi_{1+2\epsilon}(x)=\bigg[\frac{
		(\pi x^2)^\epsilon\,\Gamma(1-\epsilon)}{(4\pi)^2\,\epsilon(1-2\epsilon)}\bigg]^2\,\Delta(x)+O(\epsilon)~,\label{int22x}\\[3mm]
	\frac{1}{p^2}{\parbox[c]{.12\textwidth}{ \includegraphics[width = .12\textwidth]{Ydot2a.png}}}
	\!\!\!&~~ \longrightarrow ~~
	I_{1,1}\, I_{1,1+\epsilon}      \,  \Pi_{1+2\epsilon}(x) =
	\frac{(\pi x^2)^{2\epsilon}\,\Gamma^2(1-\epsilon)}
	{2(4\pi)^4\,\epsilon^2(1-2\epsilon)(1-3\epsilon)}\,\Delta(x)~,\label{int33x}\\[3mm]
	{\parbox[c]{.085\textwidth}{ \includegraphics[width = .085\textwidth]{Ydot3.jpg}}} 
	~~&~~ \longrightarrow ~~
	I_{1,1}\, I_{1,2+\epsilon}   \,  \Pi_{1+2\epsilon}(x)=-
	\frac{(\pi x^2)^{2\epsilon}\,\Gamma^2(1-\epsilon)}
	{2(4\pi)^4\,\epsilon^2(1-2\epsilon)(1+\epsilon)}\,\Delta(x)~,\label{int44x}\\[3mm]
	{\parbox[c]{.10\textwidth}{ \includegraphics[width = .10\textwidth]{Ydot7.jpg}}}
	& ~~\longrightarrow ~~
	I_{1,1}^2  I_{\epsilon,2+\epsilon}  \,  \Pi_{3\epsilon}(x)=-
	\frac{(\pi x^2)^{2\epsilon}\,\Gamma^2(1-\epsilon)}
	{2(4\pi)^4\,\epsilon^2(1-2\epsilon)(1+\epsilon)}\,\Delta(x)^2~.
	\label{int55x}
	\end{align}
\end{subequations}

\section{Evaluation of the relevant (super)diagrams}
\label{app:Feynman diagrams}
In this appendix we explicitly compute the diagrams discussed in Section~\ref{secn:pert}. 
We use dimensional regularization and the $\cN=1$ superspace formalism in the Feynman gauge
(we refer to \cite{Billo:2017glv} for more details).

\subsection{Feynman rules}
We first summarize the momentum-space Feynman rules in the chosen formalism. Let us start from the propagators for the chiral multiplets. We use a continuous line for the superfields $\Phi_I$ ($I=1,2,3$) of the $\cN=4$ gauge multiplet, which carry SU($N$) adjoint indices $a,b,\ldots$, a dashed line for the superfields $Q_A$ ($A=1,\ldots N_f$), which carry SU($N$) 
fundamental indices $u,v,\ldots$, and a dotted line for the superfields $\widetilde Q_A$, also 
carrying fundamental indices, which form a $\cN=2$ hypermultiplet together with $Q_A$.
We have\\
\begin{subequations}
	\begin{align}
	\label{chiralprops}
	\parbox[c]{.35\textwidth}{
\begingroup%
  \makeatletter%
  \providecommand\color[2][]{%
    \errmessage{(Inkscape) Color is used for the text in Inkscape, but the package 'color.sty' is not loaded}%
    \renewcommand\color[2][]{}%
  }%
  \providecommand\transparent[1]{%
    \errmessage{(Inkscape) Transparency is used (non-zero) for the text in Inkscape, but the package 'transparent.sty' is not loaded}%
    \renewcommand\transparent[1]{}%
  }%
  \providecommand\rotatebox[2]{#2}%
  \newcommand*\fsize{\dimexpr\f@size pt\relax}%
  \newcommand*\lineheight[1]{\fontsize{\fsize}{#1\fsize}\selectfont}%
  \ifx\svgwidth\undefined%
    \setlength{\unitlength}{180bp}%
    \ifx\svgscale\undefined%
      \relax%
    \else%
      \setlength{\unitlength}{\unitlength * \real{\svgscale}}%
    \fi%
  \else%
    \setlength{\unitlength}{\svgwidth}%
  \fi%
  \global\let\svgwidth\undefined%
  \global\let\svgscale\undefined%
  \makeatother%
  \begin{picture}(1,0.13343544)%
    \lineheight{1}%
    \setlength\tabcolsep{0pt}%
    \put(-0.00007697,-0.00218643){\color[rgb]{0,0,0}\makebox(0,0)[lt]{\lineheight{1.25}\smash{\begin{tabular}[t]{l}\textbf{$\theta_1,\bar\theta_1$}\end{tabular}}}}%
    \put(-0.00142977,0.10685608){\color[rgb]{0,0,0}\makebox(0,0)[lt]{\lineheight{1.25}\smash{\begin{tabular}[t]{l}\textbf{$a,I$ }\end{tabular}}}}%
    \put(0.58295695,0.10685608){\color[rgb]{0,0,0}\makebox(0,0)[lt]{\lineheight{1.25}\smash{\begin{tabular}[t]{l}\textbf{$b,J$}\end{tabular}}}}%
    \put(0.32917052,0.11168046){\color[rgb]{0,0,0}\makebox(0,0)[lt]{\lineheight{1.25}\smash{\begin{tabular}[t]{l}\textbf{$k$ }\end{tabular}}}}%
    \put(0,0){\includegraphics[width=\unitlength,page=1]{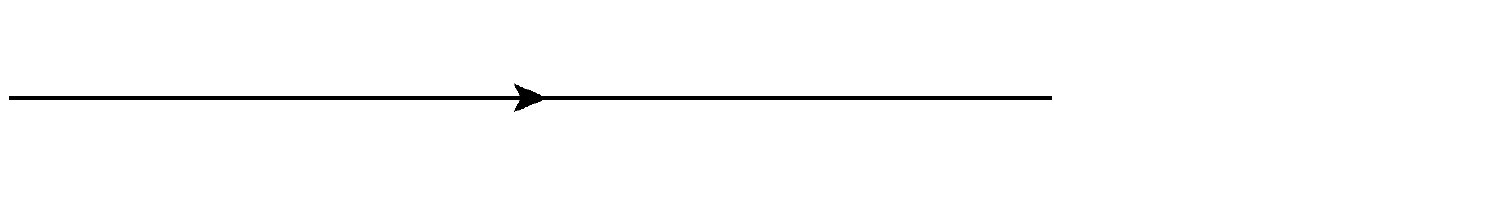}}%
    \put(0.58295695,-0.00218643){\color[rgb]{0,0,0}\makebox(0,0)[lt]{\lineheight{1.25}\smash{\begin{tabular}[t]{l}\textbf{$\theta_2,\bar\theta_2$}\end{tabular}}}}%
  \end{picture}%
\endgroup%
 } & \!\!\!\!=~~ 
	\delta^{ab}\,\delta^{IJ}\, \rme^{-\theta_1  k \bar \theta_1 -\theta_2  k \bar \theta_2
		+2\theta_1  k \bar \theta_2 } \, \frac{1}{k^2}~,\\ \nonumber \\
	\parbox[c]{.35\textwidth}{
\begingroup%
  \makeatletter%
  \providecommand\color[2][]{%
    \errmessage{(Inkscape) Color is used for the text in Inkscape, but the package 'color.sty' is not loaded}%
    \renewcommand\color[2][]{}%
  }%
  \providecommand\transparent[1]{%
    \errmessage{(Inkscape) Transparency is used (non-zero) for the text in Inkscape, but the package 'transparent.sty' is not loaded}%
    \renewcommand\transparent[1]{}%
  }%
  \providecommand\rotatebox[2]{#2}%
  \newcommand*\fsize{\dimexpr\f@size pt\relax}%
  \newcommand*\lineheight[1]{\fontsize{\fsize}{#1\fsize}\selectfont}%
  \ifx\svgwidth\undefined%
    \setlength{\unitlength}{180bp}%
    \ifx\svgscale\undefined%
      \relax%
    \else%
      \setlength{\unitlength}{\unitlength * \real{\svgscale}}%
    \fi%
  \else%
    \setlength{\unitlength}{\svgwidth}%
  \fi%
  \global\let\svgwidth\undefined%
  \global\let\svgscale\undefined%
  \makeatother%
  \begin{picture}(1,0.13343544)%
    \lineheight{1}%
    \setlength\tabcolsep{0pt}%
    \put(-0.00007697,-0.00218643){\color[rgb]{0,0,0}\makebox(0,0)[lt]{\lineheight{1.25}\smash{\begin{tabular}[t]{l}\textbf{$\theta_1,\bar\theta_1$}\end{tabular}}}}%
    \put(-0.00142977,0.10685608){\color[rgb]{0,0,0}\makebox(0,0)[lt]{\lineheight{1.25}\smash{\begin{tabular}[t]{l}\textbf{$u,A$ }\end{tabular}}}}%
    \put(0.58295695,0.10685608){\color[rgb]{0,0,0}\makebox(0,0)[lt]{\lineheight{1.25}\smash{\begin{tabular}[t]{l}\textbf{$v,B$}\end{tabular}}}}%
    \put(0.32917052,0.11168046){\color[rgb]{0,0,0}\makebox(0,0)[lt]{\lineheight{1.25}\smash{\begin{tabular}[t]{l}\textbf{$k$ }\end{tabular}}}}%
    \put(0.58295695,-0.00218643){\color[rgb]{0,0,0}\makebox(0,0)[lt]{\lineheight{1.25}\smash{\begin{tabular}[t]{l}\textbf{$\theta_2,\bar\theta_2$}\end{tabular}}}}%
    \put(0,0){\includegraphics[width=\unitlength,page=1]{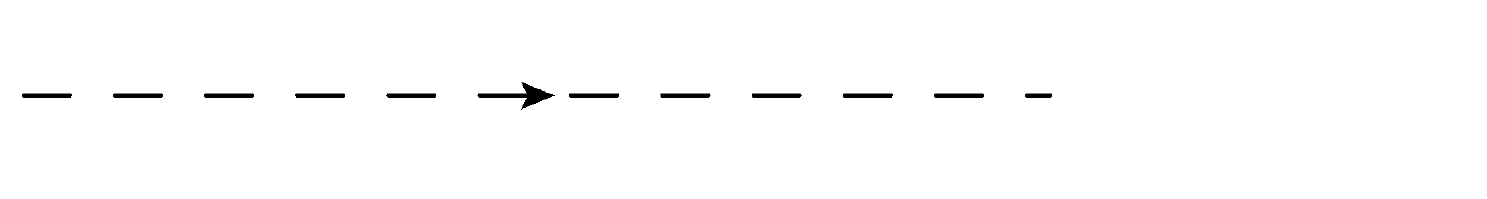}}%
  \end{picture}%
\endgroup%
 } & \!\!\!\!=~~ 
	\delta^{uv}\,\delta^{AB}\, \rme^{-\theta_1  k \bar \theta_1 -\theta_2  k \bar \theta_2
		+2\theta_1  k \bar \theta_2 } \, \frac{1}{k^2}~,\\ \nonumber\\
	\parbox[c]{.35\textwidth}{
\begingroup%
  \makeatletter%
  \providecommand\color[2][]{%
    \errmessage{(Inkscape) Color is used for the text in Inkscape, but the package 'color.sty' is not loaded}%
    \renewcommand\color[2][]{}%
  }%
  \providecommand\transparent[1]{%
    \errmessage{(Inkscape) Transparency is used (non-zero) for the text in Inkscape, but the package 'transparent.sty' is not loaded}%
    \renewcommand\transparent[1]{}%
  }%
  \providecommand\rotatebox[2]{#2}%
  \newcommand*\fsize{\dimexpr\f@size pt\relax}%
  \newcommand*\lineheight[1]{\fontsize{\fsize}{#1\fsize}\selectfont}%
  \ifx\svgwidth\undefined%
    \setlength{\unitlength}{180bp}%
    \ifx\svgscale\undefined%
      \relax%
    \else%
      \setlength{\unitlength}{\unitlength * \real{\svgscale}}%
    \fi%
  \else%
    \setlength{\unitlength}{\svgwidth}%
  \fi%
  \global\let\svgwidth\undefined%
  \global\let\svgscale\undefined%
  \makeatother%
  \begin{picture}(1,0.13343544)%
    \lineheight{1}%
    \setlength\tabcolsep{0pt}%
    \put(-0.00007697,-0.00218643){\color[rgb]{0,0,0}\makebox(0,0)[lt]{\lineheight{1.25}\smash{\begin{tabular}[t]{l}\textbf{$\theta_1,\bar\theta_1$}\end{tabular}}}}%
    \put(-0.00142977,0.10685608){\color[rgb]{0,0,0}\makebox(0,0)[lt]{\lineheight{1.25}\smash{\begin{tabular}[t]{l}\textbf{$u,A$ }\end{tabular}}}}%
    \put(0.58295695,0.10685608){\color[rgb]{0,0,0}\makebox(0,0)[lt]{\lineheight{1.25}\smash{\begin{tabular}[t]{l}\textbf{$v,B$}\end{tabular}}}}%
    \put(0.32917052,0.11168046){\color[rgb]{0,0,0}\makebox(0,0)[lt]{\lineheight{1.25}\smash{\begin{tabular}[t]{l}\textbf{$k$ }\end{tabular}}}}%
    \put(0.58295695,-0.00218643){\color[rgb]{0,0,0}\makebox(0,0)[lt]{\lineheight{1.25}\smash{\begin{tabular}[t]{l}\textbf{$\theta_2,\bar\theta_2$}\end{tabular}}}}%
    \put(0,0){\includegraphics[width=\unitlength,page=1]{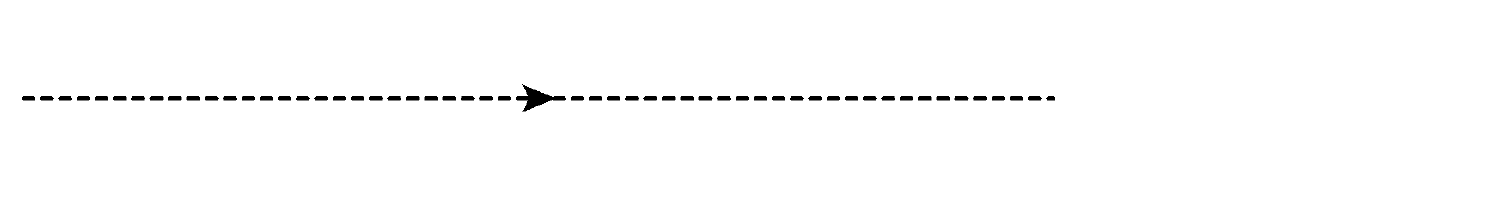}}%
  \end{picture}%
\endgroup%
 } & \!\!\!\!=~~ 
	\delta^{uv}\,\delta^{AB}\, \rme^{-\theta_1  k \bar \theta_1 -\theta_2  k \bar \theta_2
		+2\theta_1  k \bar \theta_2 } \, \frac{1}{k^2}~. \\ \nonumber
	\end{align}
\end{subequations}
Note that the arrow indicates both the orientation of the chiral propagator and the flow of the momentum. In (\ref{chiralprops}) we have used the notation
\begin{equation}
\label{convtheta}
\theta k \bar\theta = \theta^T\sigma^\mu \,\bar{\theta}\, k_\mu = \theta^{\alpha}\,
(\sigma^\mu)_{\alpha\dot{\beta}}\,\bar{\theta}^{\dot{\beta}}\,k_\mu~.
\end{equation}
Our conventions on spinor indices and Pauli matrices 
are the same as those explained in Appendix~A of \cite{Billo:2017glv}.

The propagator for the $\cN=1$ vector superfield is given by
\begin{equation}
\label{vectorprop}
\parbox[c]{.3\textwidth}{
\begingroup%
  \makeatletter%
  \providecommand\color[2][]{%
    \errmessage{(Inkscape) Color is used for the text in Inkscape, but the package 'color.sty' is not loaded}%
    \renewcommand\color[2][]{}%
  }%
  \providecommand\transparent[1]{%
    \errmessage{(Inkscape) Transparency is used (non-zero) for the text in Inkscape, but the package 'transparent.sty' is not loaded}%
    \renewcommand\transparent[1]{}%
  }%
  \providecommand\rotatebox[2]{#2}%
  \newcommand*\fsize{\dimexpr\f@size pt\relax}%
  \newcommand*\lineheight[1]{\fontsize{\fsize}{#1\fsize}\selectfont}%
  \ifx\svgwidth\undefined%
    \setlength{\unitlength}{180bp}%
    \ifx\svgscale\undefined%
      \relax%
    \else%
      \setlength{\unitlength}{\unitlength * \real{\svgscale}}%
    \fi%
  \else%
    \setlength{\unitlength}{\svgwidth}%
  \fi%
  \global\let\svgwidth\undefined%
  \global\let\svgscale\undefined%
  \makeatother%
  \begin{picture}(1,0.19869173)%
    \lineheight{1}%
    \setlength\tabcolsep{0pt}%
    \put(-0.0013758,-0.01499064){\color[rgb]{0,0,0}\makebox(0,0)[lt]{\lineheight{1.25}\smash{\begin{tabular}[t]{l}\textbf{$\theta_1,\bar\theta_1$}\end{tabular}}}}%
    \put(0.00913758,0.14577938){\color[rgb]{0,0,0}\makebox(0,0)[lt]{\lineheight{1.25}\smash{\begin{tabular}[t]{l}\textbf{$a$ }\end{tabular}}}}%
    \put(0.69492539,0.14447765){\color[rgb]{0,0,0}\makebox(0,0)[lt]{\lineheight{1.25}\smash{\begin{tabular}[t]{l}\textbf{$b$}\end{tabular}}}}%
    \put(0.33757197,0.177758){\color[rgb]{0,0,0}\makebox(0,0)[lt]{\lineheight{1.25}\smash{\begin{tabular}[t]{l}\textbf{$k$ }\end{tabular}}}}%
    \put(0.59492539,-0.01499064){\color[rgb]{0,0,0}\makebox(0,0)[lt]{\lineheight{1.25}\smash{\begin{tabular}[t]{l}\textbf{$\theta_2,\bar\theta_2$}\end{tabular}}}}%
    \put(0,0){\includegraphics[width=\unitlength,page=1]{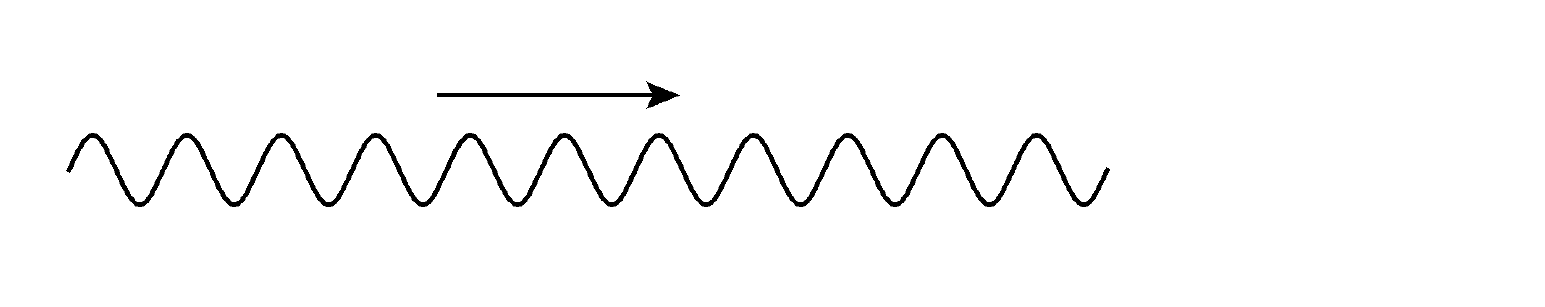}}%
  \end{picture}%
\endgroup%
 }~~~
=~\, - \frac{\delta^{ab}}{2} \,  \theta_{12}^2 \, \bar\theta_{12}^2  \, \frac{1}{k^2}\,,
\end{equation}
where $\theta_{12} \equiv \theta_1 - \theta_2$. 

The diagrams we have to compute only contain three-point vertices. These are given 
by the following rules:
\begin{subequations}
	\begin{align}
	\label{vertex1}
	&\parbox[c]{.24\textwidth}{
\begingroup%
  \makeatletter%
  \providecommand\color[2][]{%
    \errmessage{(Inkscape) Color is used for the text in Inkscape, but the package 'color.sty' is not loaded}%
    \renewcommand\color[2][]{}%
  }%
  \providecommand\transparent[1]{%
    \errmessage{(Inkscape) Transparency is used (non-zero) for the text in Inkscape, but the package 'transparent.sty' is not loaded}%
    \renewcommand\transparent[1]{}%
  }%
  \providecommand\rotatebox[2]{#2}%
  \newcommand*\fsize{\dimexpr\f@size pt\relax}%
  \newcommand*\lineheight[1]{\fontsize{\fsize}{#1\fsize}\selectfont}%
  \ifx\svgwidth\undefined%
    \setlength{\unitlength}{80bp}%
    \ifx\svgscale\undefined%
      \relax%
    \else%
      \setlength{\unitlength}{\unitlength * \real{\svgscale}}%
    \fi%
  \else%
    \setlength{\unitlength}{\svgwidth}%
  \fi%
  \global\let\svgwidth\undefined%
  \global\let\svgscale\undefined%
  \makeatother%
  \begin{picture}(1,0.81778956)%
    \lineheight{1}%
    \setlength\tabcolsep{0pt}%
    \put(0.0523874,0.78086603){\color[rgb]{0,0,0}\makebox(0,0)[lt]{\lineheight{1.25}\smash{\begin{tabular}[t]{l}\textbf{$a,I$ }\end{tabular}}}}%
    \put(0,0){\includegraphics[width=\unitlength,page=1]{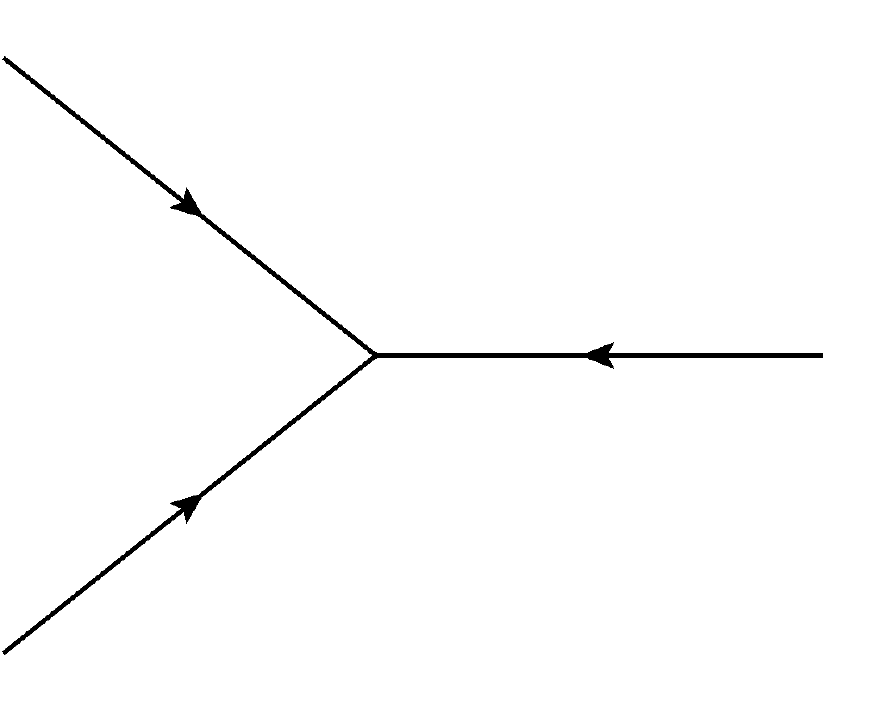}}%
    \put(0.0523874,0.00766068){\color[rgb]{0,0,0}\makebox(0,0)[lt]{\lineheight{1.25}\smash{\begin{tabular}[t]{l}\textbf{$b,J$ }\end{tabular}}}}%
    \put(0.8792796,0.29663151){\color[rgb]{0,0,0}\makebox(0,0)[lt]{\lineheight{1.25}\smash{\begin{tabular}[t]{l}\textbf{$c,K$ }\end{tabular}}}}%
  \end{picture}%
\endgroup%
 }
	= ~\frac{1}{3!} \,\epsilon_{IJK} \, \sqrt{2} g_0 \, \theta^2\, (T^a)^{bc}~,\\
	\notag\\
	\label{vertex2}
	&\parbox[c]{.24\textwidth}{
\begingroup%
  \makeatletter%
  \providecommand\color[2][]{%
    \errmessage{(Inkscape) Color is used for the text in Inkscape, but the package 'color.sty' is not loaded}%
    \renewcommand\color[2][]{}%
  }%
  \providecommand\transparent[1]{%
    \errmessage{(Inkscape) Transparency is used (non-zero) for the text in Inkscape, but the package 'transparent.sty' is not loaded}%
    \renewcommand\transparent[1]{}%
  }%
  \providecommand\rotatebox[2]{#2}%
  \newcommand*\fsize{\dimexpr\f@size pt\relax}%
  \newcommand*\lineheight[1]{\fontsize{\fsize}{#1\fsize}\selectfont}%
  \ifx\svgwidth\undefined%
    \setlength{\unitlength}{80bp}%
    \ifx\svgscale\undefined%
      \relax%
    \else%
      \setlength{\unitlength}{\unitlength * \real{\svgscale}}%
    \fi%
  \else%
    \setlength{\unitlength}{\svgwidth}%
  \fi%
  \global\let\svgwidth\undefined%
  \global\let\svgscale\undefined%
  \makeatother%
  \begin{picture}(1,0.81973424)%
    \lineheight{1}%
    \setlength\tabcolsep{0pt}%
    \put(0.05013401,0.78272291){\color[rgb]{0,0,0}\makebox(0,0)[lt]{\lineheight{1.25}\smash{\begin{tabular}[t]{l}\textbf{$a,I$ }\end{tabular}}}}%
    \put(0.05013401,0.0076789){\color[rgb]{0,0,0}\makebox(0,0)[lt]{\lineheight{1.25}\smash{\begin{tabular}[t]{l}\textbf{$b,J$ }\end{tabular}}}}%
    \put(0.87899253,0.29733689){\color[rgb]{0,0,0}\makebox(0,0)[lt]{\lineheight{1.25}\smash{\begin{tabular}[t]{l}\textbf{$c,K$ }\end{tabular}}}}%
    \put(0,0){\includegraphics[width=\unitlength,page=1]{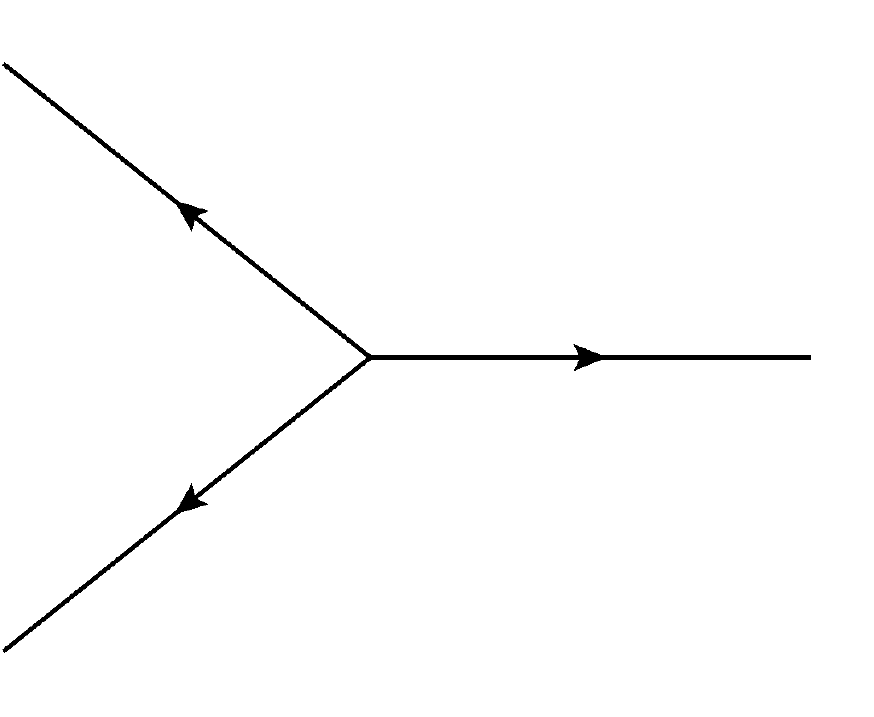}}%
  \end{picture}%
\endgroup%
 }
	= ~- \frac{1}{3!}\, \epsilon_{IJK} \, \sqrt{2}g_0 \, \bar\theta^2\, (T^a)^{bc}~, \\
	\notag\\
	\label{vertex3}
	&\parbox[c]{.24\textwidth}{
\begingroup%
  \makeatletter%
  \providecommand\color[2][]{%
    \errmessage{(Inkscape) Color is used for the text in Inkscape, but the package 'color.sty' is not loaded}%
    \renewcommand\color[2][]{}%
  }%
  \providecommand\transparent[1]{%
    \errmessage{(Inkscape) Transparency is used (non-zero) for the text in Inkscape, but the package 'transparent.sty' is not loaded}%
    \renewcommand\transparent[1]{}%
  }%
  \providecommand\rotatebox[2]{#2}%
  \newcommand*\fsize{\dimexpr\f@size pt\relax}%
  \newcommand*\lineheight[1]{\fontsize{\fsize}{#1\fsize}\selectfont}%
  \ifx\svgwidth\undefined%
    \setlength{\unitlength}{80bp}%
    \ifx\svgscale\undefined%
      \relax%
    \else%
      \setlength{\unitlength}{\unitlength * \real{\svgscale}}%
    \fi%
  \else%
    \setlength{\unitlength}{\svgwidth}%
  \fi%
  \global\let\svgwidth\undefined%
  \global\let\svgscale\undefined%
  \makeatother%
  \begin{picture}(1,0.83305539)%
    \lineheight{1}%
    \setlength\tabcolsep{0pt}%
    \put(0.04956104,0.7954426){\color[rgb]{0,0,0}\makebox(0,0)[lt]{\lineheight{1.25}\smash{\begin{tabular}[t]{l}\textbf{$u,A$ }\end{tabular}}}}%
    \put(0.04956104,0.00780368){\color[rgb]{0,0,0}\makebox(0,0)[lt]{\lineheight{1.25}\smash{\begin{tabular}[t]{l}\textbf{$v,B$ }\end{tabular}}}}%
    \put(0.89188899,0.30216878){\color[rgb]{0,0,0}\makebox(0,0)[lt]{\lineheight{1.25}\smash{\begin{tabular}[t]{l}\textbf{$a,I$ }\end{tabular}}}}%
    \put(0,0){\includegraphics[width=\unitlength,page=1]{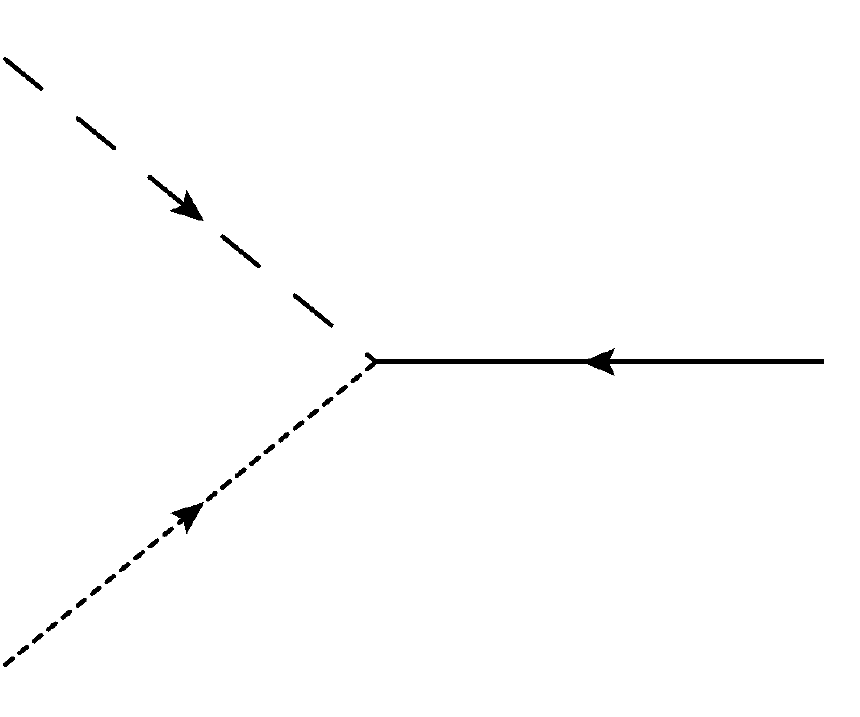}}%
  \end{picture}%
\endgroup%
 }
	= ~- \ii \,\delta_{AB}\, \delta_{I1}\, \sqrt{2} g_0\, \theta^2\, (T^a)_{uv}~, \\
	\notag\\
	\label{vertex4}
	&\parbox[c]{.24\textwidth}{
\begingroup%
  \makeatletter%
  \providecommand\color[2][]{%
    \errmessage{(Inkscape) Color is used for the text in Inkscape, but the package 'color.sty' is not loaded}%
    \renewcommand\color[2][]{}%
  }%
  \providecommand\transparent[1]{%
    \errmessage{(Inkscape) Transparency is used (non-zero) for the text in Inkscape, but the package 'transparent.sty' is not loaded}%
    \renewcommand\transparent[1]{}%
  }%
  \providecommand\rotatebox[2]{#2}%
  \newcommand*\fsize{\dimexpr\f@size pt\relax}%
  \newcommand*\lineheight[1]{\fontsize{\fsize}{#1\fsize}\selectfont}%
  \ifx\svgwidth\undefined%
    \setlength{\unitlength}{80bp}%
    \ifx\svgscale\undefined%
      \relax%
    \else%
      \setlength{\unitlength}{\unitlength * \real{\svgscale}}%
    \fi%
  \else%
    \setlength{\unitlength}{\svgwidth}%
  \fi%
  \global\let\svgwidth\undefined%
  \global\let\svgscale\undefined%
  \makeatother%
  \begin{picture}(1,0.82989822)%
    \lineheight{1}%
    \setlength\tabcolsep{0pt}%
    \put(0.05316308,0.79242798){\color[rgb]{0,0,0}\makebox(0,0)[lt]{\lineheight{1.25}\smash{\begin{tabular}[t]{l}\textbf{$v,B$ }\end{tabular}}}}%
    \put(0.05316308,0.00777411){\color[rgb]{0,0,0}\makebox(0,0)[lt]{\lineheight{1.25}\smash{\begin{tabular}[t]{l}\textbf{$u,A$ }\end{tabular}}}}%
    \put(0.89229872,0.3010236){\color[rgb]{0,0,0}\makebox(0,0)[lt]{\lineheight{1.25}\smash{\begin{tabular}[t]{l}\textbf{$a,I$ }\end{tabular}}}}%
    \put(0,0){\includegraphics[width=\unitlength,page=1]{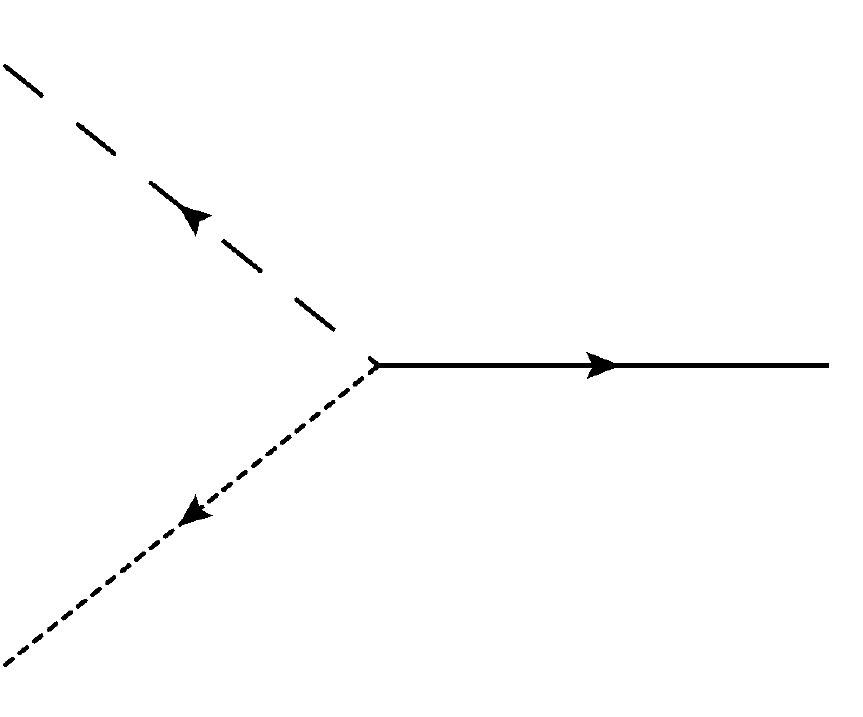}}%
  \end{picture}%
\endgroup%
 }
	=~ \ii \,\delta_{AB}\, \delta_{I1}\,  \sqrt{2} g_0\, \bar\theta^2\, (T^a)_{uv}~,\\
	\notag\\
	\label{vertex5}
	&\parbox[c]{.24\textwidth}{
\begingroup%
  \makeatletter%
  \providecommand\color[2][]{%
    \errmessage{(Inkscape) Color is used for the text in Inkscape, but the package 'color.sty' is not loaded}%
    \renewcommand\color[2][]{}%
  }%
  \providecommand\transparent[1]{%
    \errmessage{(Inkscape) Transparency is used (non-zero) for the text in Inkscape, but the package 'transparent.sty' is not loaded}%
    \renewcommand\transparent[1]{}%
  }%
  \providecommand\rotatebox[2]{#2}%
  \newcommand*\fsize{\dimexpr\f@size pt\relax}%
  \newcommand*\lineheight[1]{\fontsize{\fsize}{#1\fsize}\selectfont}%
  \ifx\svgwidth\undefined%
    \setlength{\unitlength}{80bp}%
    \ifx\svgscale\undefined%
      \relax%
    \else%
      \setlength{\unitlength}{\unitlength * \real{\svgscale}}%
    \fi%
  \else%
    \setlength{\unitlength}{\svgwidth}%
  \fi%
  \global\let\svgwidth\undefined%
  \global\let\svgscale\undefined%
  \makeatother%
  \begin{picture}(1,0.8529858)%
    \lineheight{1}%
    \setlength\tabcolsep{0pt}%
    \put(0.05464207,0.81447315){\color[rgb]{0,0,0}\makebox(0,0)[lt]{\lineheight{1.25}\smash{\begin{tabular}[t]{l}\textbf{$b,I$ }\end{tabular}}}}%
    \put(0.05464207,0.00799038){\color[rgb]{0,0,0}\makebox(0,0)[lt]{\lineheight{1.25}\smash{\begin{tabular}[t]{l}\textbf{$c,J$ }\end{tabular}}}}%
    \put(0.91712227,0.28939801){\color[rgb]{0,0,0}\makebox(0,0)[lt]{\lineheight{1.25}\smash{\begin{tabular}[t]{l}\textbf{$a$ }\end{tabular}}}}%
    \put(0,0){\includegraphics[width=\unitlength,page=1]{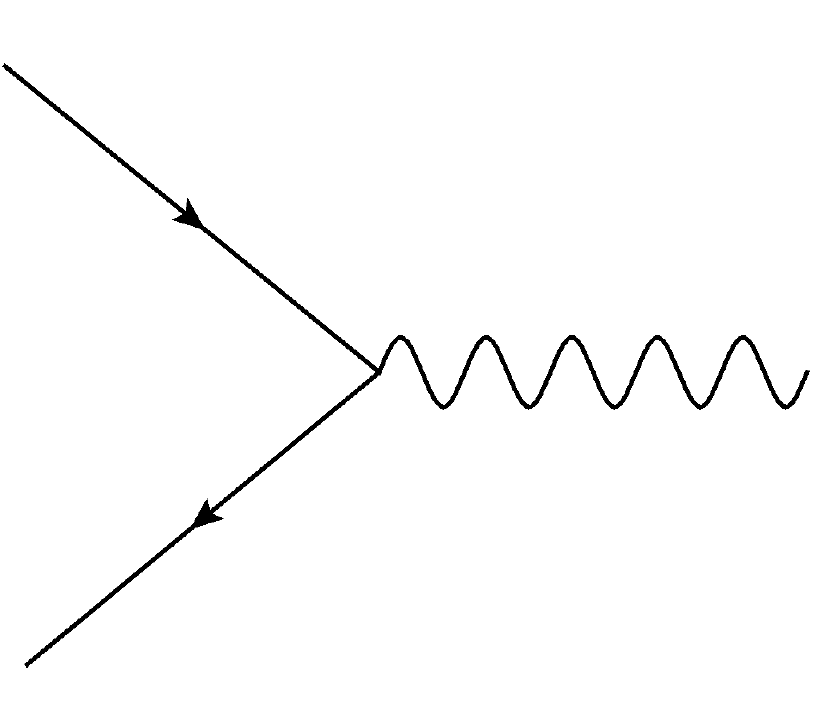}}%
  \end{picture}%
\endgroup%
 }
	~= ~ \delta_{IJ}\,  2g_0\, (T^a)^{bc}~,\\
	\notag\\
	\label{vertex6}
	&\parbox[c]{.24\textwidth}{
\begingroup%
  \makeatletter%
  \providecommand\color[2][]{%
    \errmessage{(Inkscape) Color is used for the text in Inkscape, but the package 'color.sty' is not loaded}%
    \renewcommand\color[2][]{}%
  }%
  \providecommand\transparent[1]{%
    \errmessage{(Inkscape) Transparency is used (non-zero) for the text in Inkscape, but the package 'transparent.sty' is not loaded}%
    \renewcommand\transparent[1]{}%
  }%
  \providecommand\rotatebox[2]{#2}%
  \newcommand*\fsize{\dimexpr\f@size pt\relax}%
  \newcommand*\lineheight[1]{\fontsize{\fsize}{#1\fsize}\selectfont}%
  \ifx\svgwidth\undefined%
    \setlength{\unitlength}{80bp}%
    \ifx\svgscale\undefined%
      \relax%
    \else%
      \setlength{\unitlength}{\unitlength * \real{\svgscale}}%
    \fi%
  \else%
    \setlength{\unitlength}{\svgwidth}%
  \fi%
  \global\let\svgwidth\undefined%
  \global\let\svgscale\undefined%
  \makeatother%
  \begin{picture}(1,0.8529858)%
    \lineheight{1}%
    \setlength\tabcolsep{0pt}%
    \put(0.05464207,0.81447315){\color[rgb]{0,0,0}\makebox(0,0)[lt]{\lineheight{1.25}\smash{\begin{tabular}[t]{l}\textbf{$u,A$ }\end{tabular}}}}%
    \put(0.05464207,0.00799038){\color[rgb]{0,0,0}\makebox(0,0)[lt]{\lineheight{1.25}\smash{\begin{tabular}[t]{l}\textbf{$v,B$ }\end{tabular}}}}%
    \put(0.91712227,0.28939801){\color[rgb]{0,0,0}\makebox(0,0)[lt]{\lineheight{1.25}\smash{\begin{tabular}[t]{l}\textbf{$a$ }\end{tabular}}}}%
    \put(0,0){\includegraphics[width=\unitlength,page=1]{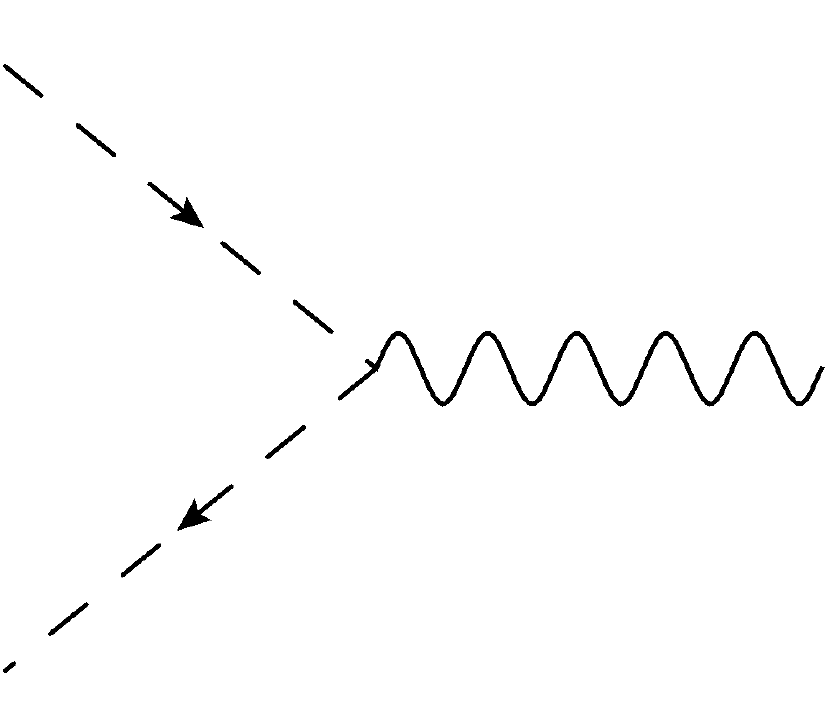}}%
  \end{picture}%
\endgroup%
 }
	~= ~\delta_{AB}\,  2g_0\, (T^a)_{uv}~,\\
	\notag\\
	\label{vertex7}
	&\parbox[c]{.24\textwidth}{
\begingroup%
  \makeatletter%
  \providecommand\color[2][]{%
    \errmessage{(Inkscape) Color is used for the text in Inkscape, but the package 'color.sty' is not loaded}%
    \renewcommand\color[2][]{}%
  }%
  \providecommand\transparent[1]{%
    \errmessage{(Inkscape) Transparency is used (non-zero) for the text in Inkscape, but the package 'transparent.sty' is not loaded}%
    \renewcommand\transparent[1]{}%
  }%
  \providecommand\rotatebox[2]{#2}%
  \newcommand*\fsize{\dimexpr\f@size pt\relax}%
  \newcommand*\lineheight[1]{\fontsize{\fsize}{#1\fsize}\selectfont}%
  \ifx\svgwidth\undefined%
    \setlength{\unitlength}{80bp}%
    \ifx\svgscale\undefined%
      \relax%
    \else%
      \setlength{\unitlength}{\unitlength * \real{\svgscale}}%
    \fi%
  \else%
    \setlength{\unitlength}{\svgwidth}%
  \fi%
  \global\let\svgwidth\undefined%
  \global\let\svgscale\undefined%
  \makeatother%
  \begin{picture}(1,0.8529858)%
    \lineheight{1}%
    \setlength\tabcolsep{0pt}%
    \put(0.05464207,0.81447315){\color[rgb]{0,0,0}\makebox(0,0)[lt]{\lineheight{1.25}\smash{\begin{tabular}[t]{l}\textbf{$v,B$ }\end{tabular}}}}%
    \put(0.05464207,0.00799038){\color[rgb]{0,0,0}\makebox(0,0)[lt]{\lineheight{1.25}\smash{\begin{tabular}[t]{l}\textbf{$u,A$ }\end{tabular}}}}%
    \put(0.91712227,0.29939801){\color[rgb]{0,0,0}\makebox(0,0)[lt]{\lineheight{1.25}\smash{\begin{tabular}[t]{l}\textbf{$a$ }\end{tabular}}}}%
    \put(0,0){\includegraphics[width=\unitlength,page=1]{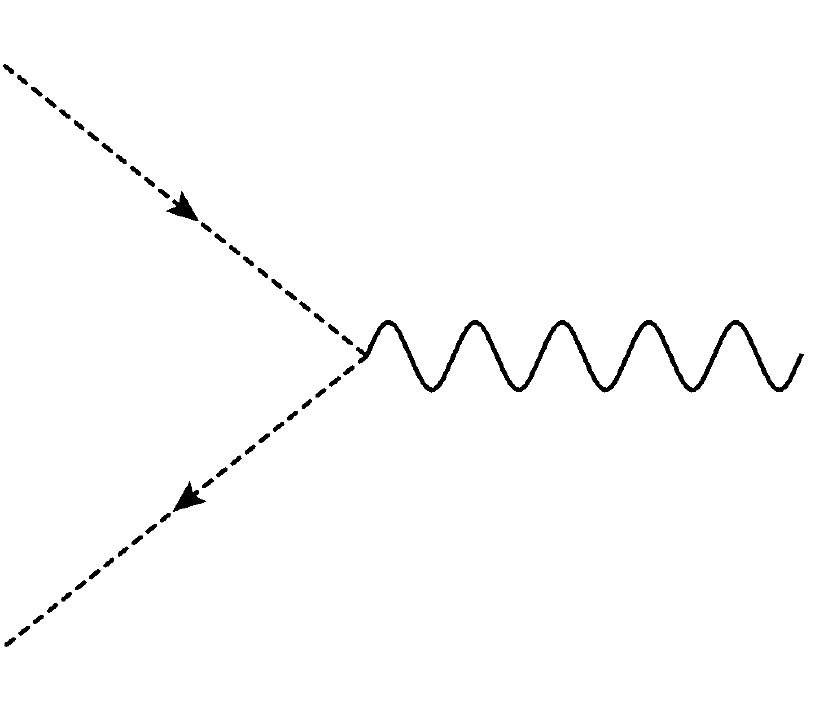}}%
  \end{picture}%
\endgroup%
 }
	~= ~-\delta_{AB}\,  2g_0\, (T^a)_{uv}~.
	\end{align}
\end{subequations}
Here $(T^a)^{bc}=-\ii\,f^{abc}$ are the generators in the adjoint representation, 
and $(T^a)_{uv}$ those in the fundamental representation. 
The $\theta$ variables appearing in the vertices are those associated to the vertex point.

\subsection{One-loop diagrams}
At one loop, we have to compute the diagrams in Fig.~\ref{fig:onelooppphi-iniz}. Denoting
\begin{equation}
\label{v21is}
{\parbox[c]{.25\textwidth}{\includegraphics[width = .25\textwidth]{v21}}} {}  = ~
W_1^{ab}(x)
\end{equation}
and isolating a prefactor containing the combinatorial and color factors, we have 
\begin{equation}
\label{W1is}
\begin{aligned}
W_1^{ab}(x) & =  (\sqrt{2} g_0)^2\,
\big(N_f\,\tr T^a T^b  - \tr_{\text{adj}} T^a T^b\big)\, W_1(x)\\[2mm]
& = g_0^2\, (N_f - 2N)\, W_1(x)\,\delta^{ab}~,
\end{aligned}
\end{equation}
where $W_1(x)$ is given by the one-loop Feynman diagram shown in (\ref{cW1is}) below. 
The first term in the color factor arises from the loop diagram of $N_f$ fundamental $Q,\tilde Q$ superfields, while the second term originates from the loop of the adjoint hypermultiplet $H$. 

It is convenient to Fourier transform $W_1(x)$ and write
\begin{equation}
\label{W1ft}
W_1(x) = \int \!\frac{d^Dp}{(2\pi)^D} \, \cW_1(p)\,\rme^{\ii p\cdot x}~,
\end{equation}
where $\cW_1(p)$ is described by the following diagram in the momentum space:
\begin{equation}
\label{cW1is}
\cW_1(p) \,=~ \parbox[c]{.40\textwidth}{
\begingroup%
  \makeatletter%
  \providecommand\color[2][]{%
    \errmessage{(Inkscape) Color is used for the text in Inkscape, but the package 'color.sty' is not loaded}%
    \renewcommand\color[2][]{}%
  }%
  \providecommand\transparent[1]{%
    \errmessage{(Inkscape) Transparency is used (non-zero) for the text in Inkscape, but the package 'transparent.sty' is not loaded}%
    \renewcommand\transparent[1]{}%
  }%
  \providecommand\rotatebox[2]{#2}%
  \newcommand*\fsize{\dimexpr\f@size pt\relax}%
  \newcommand*\lineheight[1]{\fontsize{\fsize}{#1\fsize}\selectfont}%
  \ifx\svgwidth\undefined%
    \setlength{\unitlength}{140bp}%
    \ifx\svgscale\undefined%
      \relax%
    \else%
      \setlength{\unitlength}{\unitlength * \real{\svgscale}}%
    \fi%
  \else%
    \setlength{\unitlength}{\svgwidth}%
  \fi%
  \global\let\svgwidth\undefined%
  \global\let\svgscale\undefined%
  \makeatother%
  \begin{picture}(1,0.41156878)%
    \lineheight{1}%
    \setlength\tabcolsep{0pt}%
    \put(-0.00143871,0.14003063){\color[rgb]{0,0,0}\makebox(0,0)[lt]{\lineheight{1.25}\smash{\begin{tabular}[t]{l}\textbf{$1$}\end{tabular}}}}%
    \put(0.15012083,0.25686486){\color[rgb]{0,0,0}\makebox(0,0)[lt]{\lineheight{1.25}\smash{\begin{tabular}[t]{l}\textbf{$p$ }\end{tabular}}}}%
    \put(0.78842056,0.25686486){\color[rgb]{0,0,0}\makebox(0,0)[lt]{\lineheight{1.25}\smash{\begin{tabular}[t]{l}\textbf{$p$}\end{tabular}}}}%
    \put(0.9594331,0.14003063){\color[rgb]{0,0,0}\makebox(0,0)[lt]{\lineheight{1.25}\smash{\begin{tabular}[t]{l}\textbf{$2$}\end{tabular}}}}%
    \put(0.35792518,0.16455946){\color[rgb]{0,0,0}\makebox(0,0)[lt]{\lineheight{1.25}\smash{\begin{tabular}[t]{l}\textbf{$3$}\end{tabular}}}}%
    \put(0.5947893,0.16455946){\color[rgb]{0,0,0}\makebox(0,0)[lt]{\lineheight{1.25}\smash{\begin{tabular}[t]{l}\textbf{$4$}\end{tabular}}}}%
    \put(0.44134338,0.37967776){\color[rgb]{0,0,0}\makebox(0,0)[lt]{\lineheight{1.25}\smash{\begin{tabular}[t]{l}\textbf{$-k$ }\end{tabular}}}}%
    \put(0.41985143,-0.00521886){\color[rgb]{0,0,0}\makebox(0,0)[lt]{\lineheight{1.25}\smash{\begin{tabular}[t]{l}\textbf{$k-p$ }\end{tabular}}}}%
    \put(0,0){\includegraphics[width=\unitlength,page=1]{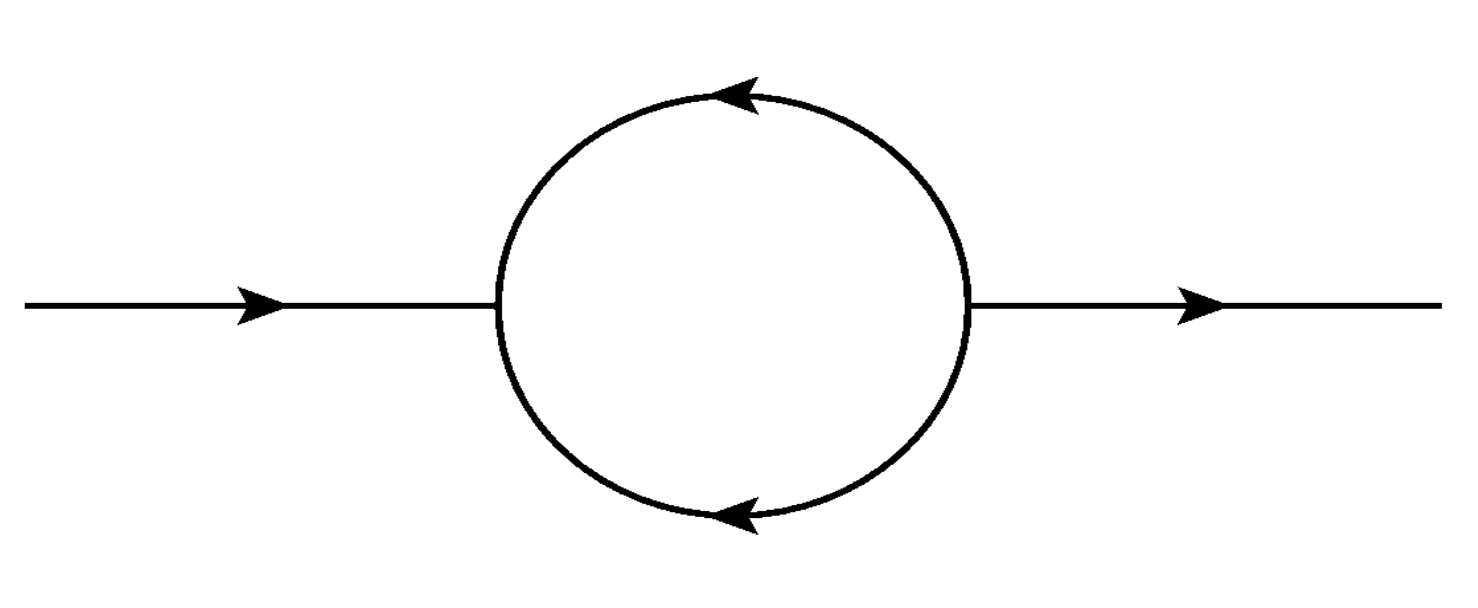}}%
  \end{picture}%
\endgroup%
 }~.
\end{equation}
Here the numbers label the external points and the interaction vertices. 
Note that the two external points $1$ and $2$ are connected to a bosonic scalar field, so that the propagators from $1$ to $3$ and from
$4$ to $2$ are in fact free scalar propagators with no $\theta$ dependence. 
Taking into account the Feynman rules given above, we get 
\begin{align}
\label{cW1nc}
\cW_1(p) & = \int \!\frac{d^Dk}{(2\pi)^D}\, \frac{1}{p^4\, k^2\,(k-p)^2} \int d^2\bar\theta_3\, d^2\theta_4\, \rme^{-2 \theta_4 p\, \bar{\theta_3} }\nonumber\\
& = - \frac{1}{p^2} \int \frac{d^Dk}{(2\pi)^D} \frac{1}{k^2\,(k-p)^2} = 
-\frac{1}{p^2}{\parbox[c]{.08\textwidth}{ \includegraphics[width = .08\textwidth]{Iab.jpg}}}  
~.
\end{align}
In the second step we used the Grassmann integral identity
\begin{equation}
\int d^2\theta_1 \,d^2\bar \theta_2 \,\rme^{\,\alpha\,\theta_1 k\,  \bar\theta_2}  = -
\frac{\alpha^2}{4}\, k^2~,
\end{equation}
and then we exploited the graphical representation introduced in Appendix~\ref{app:loop_integrals}.
Using (\ref{int11app}) and the Fourier transform (\ref{int11x}), we finally obtain 
\begin{equation}
\label{W1fin}
W_1(x) =- \frac{(\pi x^2)^\epsilon \, \Gamma(1-\epsilon)}{(4\pi)^2\,\epsilon(1 - 2\epsilon)}\,\Delta(x) ~,
\end{equation}
leading to the relation (\ref{v1loop}) in the main text:
\begin{equation}
\label{W1abfin}
W_1^{ab}(x) = \frac{g_0^2}{8\pi^2} (2N-N_f) \,\frac{(\pi x^2)^\epsilon\,\Gamma(1-\epsilon)}{2\epsilon(1 - 2\epsilon)}\,
\Delta(x)\,\delta^{ab} \equiv v_{2,1} \, \Delta(x)\,\delta^{ab}~.
\end{equation} 

\subsection{Two-loop diagrams}
At two loops we have to compute diagrams that correct either a two-point or a four-point vertex.  Such diagrams have been displayed in Section~\ref{secn:pert}. All of them have two external points, corresponding to the positions of the two operators $O_{\vec n}(x)$ and $\widebar{O}_{\vec m}(0)$, 
four internal points, corresponding to the interaction vertices, and either seven or eight 
propagators, for the corrections to the two-point or the four-point vertex respectively. 

\subsection*{Some notations}
It is useful to introduce some notation that allows us to write the various diagrams in a uniform way. 
In each two-loop diagram labeled by an index $I$, we label by $i=1,2$ the external points 
and by $i=3,\ldots, 6$ the internal ones. 
We denote by $\cE_I$ the set of propagators of the diagram, and label each propagator in $\cE_I$ 
by $s$. Any propagator connects a point $i$ to a point $j$, and in general there can be a number 
$r(i,j)$ of propagators connecting the same two points. A possible way of expressing the label $s$ of 
the propagators is thus
\begin{equation}
\label{slabelis}
s \to (i,j;r)
\end{equation}
where $r=1,\ldots, r(i,j)$; in the following we will omit the index $r$ if $r(i,j)=1$. 
The momentum $k_s$ associated to the  propagator will then be denoted as $k_{ij;r}$, with the convention that we take it to flow from $i$ to $j$. This is useful to write the delta-functions of 
momentum conservation at internal vertices, which take the form
\begin{equation}
\label{consk}
\delta_{\text{int}}(k) \equiv \prod_{i=3}^6\delta^D\Big(\sum_{j}\sum_{r=1}^{r(i,j)} k_{ij;r}\Big)~.
\end{equation}
Similarly, the relation between the internal momenta and the external momentum $p$ is enforced by
\begin{equation}
\label{pis}
\delta_{\text{ext}}(p,k) \equiv \delta^D\Big(p -\sum_{j}\sum_{r=1}^{r(1,j)} k_{1j}\Big)~.
\end{equation}  

Just as in the one-loop case, any two-loop diagram will be written as the product of a factor 
containing the weights in the vertices, the combinatorial and color factors, and of 
a colorless diagram $W_I(x)$, which we will obtain from its Fourier transform $\cW_I(p)$. 
The latter has the following structure
\begin{equation}
\label{WYD}
\cW_I(p) = \int \!\prod_{s\in\cE_I} \frac{dk_{s}}{(2\pi)^D}~\cY_I(p,k)~\cZ_I(p,k)~,
\end{equation}
where
\begin{equation}
\begin{aligned}
\cY_I(p,k) &= \prod_{s\in\cE_I}\frac{1}{k_s^2}\, \delta_{\rm int}(k)\,\delta_{\rm ext}(p,k) ~,\\
\cZ_I(p,k) &=\int \!\prod_{i=3}^6 d^2\theta_i\,d^2\bar\theta_i\,\, \mathfrak D_I (p,k,\theta,\bar\theta)~.
\end{aligned}
\label{Y}
\end{equation}
The factor $\cY_I$ contains the contribution of the propagators and the conditions for 
the momentum conservation at each vertex of the $I$-th diagram,
while the factor $\mathfrak D_I (p,k,\theta,\bar\theta)$ contains all $\theta$ and $\bar{\theta}$ terms coming from the vertices and from the superfield propagators. Some of the Grassmann integrations yielding might be obvious, in which case we will indicate only the non-trivial 
integrals and denote the integrand of $\cZ_I$ as $\widetilde{\mathfrak D}_I$.

\subsection*{Reducible diagrams}

The reducible two-loop diagrams are represented in Fig.~\ref{fig:2p1l} and Fig.~\ref{fig:pphi2l}.
In the diagrams of Fig.~\ref{fig:2p1l} there are two independent one-loop corrections to 
a propagator line. Hence the result follows simply from the one-loop computation presented in the 
previous subsection, and has been given in (\ref{2p1lres}) of the main text. 

Let us consider then the diagrams of Fig.~\ref{fig:pphi2l}. The overall factors are simply the square of those of (\ref{W1is}), and thus we get
\begin{equation}
\label{W2def}
\parbox[c]{.35\textwidth}{ \includegraphics[width = .35\textwidth]{v21square2.png}} = 
g_0^4 (N_f - 2N)^2  \, W_2(x) \, \delta^{ab}~.
\end{equation}
The Fourier transform of $W_2(x)$ is given by the diagram 
\begin{equation}
\label{W2pis}
\cW_2(p) =~~ \parbox[c]{.50\textwidth}{
\begingroup%
  \makeatletter%
  \providecommand\color[2][]{%
    \errmessage{(Inkscape) Color is used for the text in Inkscape, but the package 'color.sty' is not loaded}%
    \renewcommand\color[2][]{}%
  }%
  \providecommand\transparent[1]{%
    \errmessage{(Inkscape) Transparency is used (non-zero) for the text in Inkscape, but the package 'transparent.sty' is not loaded}%
    \renewcommand\transparent[1]{}%
  }%
  \providecommand\rotatebox[2]{#2}%
  \newcommand*\fsize{\dimexpr\f@size pt\relax}%
  \newcommand*\lineheight[1]{\fontsize{\fsize}{#1\fsize}\selectfont}%
  \ifx\svgwidth\undefined%
    \setlength{\unitlength}{180bp}%
    \ifx\svgscale\undefined%
      \relax%
    \else%
      \setlength{\unitlength}{\unitlength * \real{\svgscale}}%
    \fi%
  \else%
    \setlength{\unitlength}{\svgwidth}%
  \fi%
  \global\let\svgwidth\undefined%
  \global\let\svgscale\undefined%
  \makeatother%
  \begin{picture}(1,0.3453679)%
    \lineheight{1}%
    \setlength\tabcolsep{0pt}%
    \put(-0.00112054,0.12272266){\color[rgb]{0,0,0}\makebox(0,0)[lt]{\lineheight{1.25}\smash{\begin{tabular}[t]{l}\textbf{$1$}\end{tabular}}}}%
    \put(0.96373149,0.12272266){\color[rgb]{0,0,0}\makebox(0,0)[lt]{\lineheight{1.25}\smash{\begin{tabular}[t]{l}\textbf{$2$}\end{tabular}}}}%
    \put(0.20068509,0.14872266){\color[rgb]{0,0,0}\makebox(0,0)[lt]{\lineheight{1.25}\smash{\begin{tabular}[t]{l}\textbf{$3$}\end{tabular}}}}%
    \put(0.36624803,0.14872266){\color[rgb]{0,0,0}\makebox(0,0)[lt]{\lineheight{1.25}\smash{\begin{tabular}[t]{l}\textbf{$4$}\end{tabular}}}}%
    \put(0.59933766,0.14872266){\color[rgb]{0,0,0}\makebox(0,0)[lt]{\lineheight{1.25}\smash{\begin{tabular}[t]{l}\textbf{$5$}\end{tabular}}}}%
    \put(0.76340201,0.14872266){\color[rgb]{0,0,0}\makebox(0,0)[lt]{\lineheight{1.25}\smash{\begin{tabular}[t]{l}\textbf{$6$}\end{tabular}}}}%
    \put(0,0){\includegraphics[width=\unitlength,page=1]{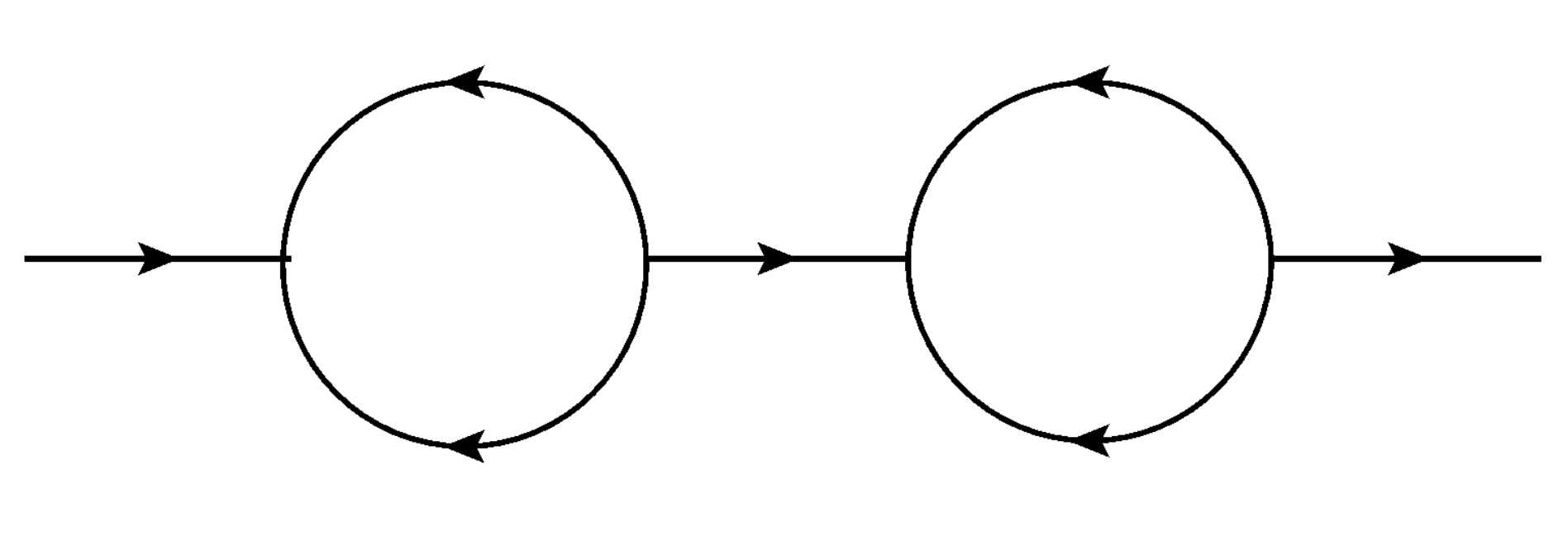}}%
    \put(0.07498926,0.21884458){\color[rgb]{0,0,0}\makebox(0,0)[lt]{\lineheight{1.25}\smash{\begin{tabular}[t]{l}\textbf{$k_{13}$ }\end{tabular}}}}%
    \put(0.27459794,0.32816433){\color[rgb]{0,0,0}\makebox(0,0)[lt]{\lineheight{1.25}\smash{\begin{tabular}[t]{l}\textbf{$k_{43;1}$}\end{tabular}}}}%
    \put(0.27459794,0.0040647){\color[rgb]{0,0,0}\makebox(0,0)[lt]{\lineheight{1.25}\smash{\begin{tabular}[t]{l}\textbf{$k_{43;2}$}\end{tabular}}}}%
    \put(0.67722783,0.32816433){\color[rgb]{0,0,0}\makebox(0,0)[lt]{\lineheight{1.25}\smash{\begin{tabular}[t]{l}\textbf{$k_{65;1}$}\end{tabular}}}}%
    \put(0.67722783,0.0040647){\color[rgb]{0,0,0}\makebox(0,0)[lt]{\lineheight{1.25}\smash{\begin{tabular}[t]{l}\textbf{$k_{65;2}$}\end{tabular}}}}%
    \put(0.46792387,0.21884458){\color[rgb]{0,0,0}\makebox(0,0)[lt]{\lineheight{1.25}\smash{\begin{tabular}[t]{l}\textbf{$k_{45}$ }\end{tabular}}}}%
    \put(0.87701367,0.21884458){\color[rgb]{0,0,0}\makebox(0,0)[lt]{\lineheight{1.25}\smash{\begin{tabular}[t]{l}\textbf{$k_{62}$ }\end{tabular}}}}%
  \end{picture}%
\endgroup%
 }~.
\vspace{0.2cm}
\end{equation}
According to the conventions described earlier, the labeling of the momenta is determined by that of 
the vertices. In the following, therefore, in drawing momentum space diagrams we will only exhibit 
the labeling of the vertices. This diagram can be expressed in the form (\ref{WYD}), with 
\begin{equation}
\label{w2D}
\cZ_2(p,k)=
\int \! d^2\bar\theta_3\, d^2\theta_4\, d^2\bar\theta_5\, d^2\theta_6\, \widetilde{\mathfrak D }_2~,
\end{equation}
where
\begin{equation}
\label{w2D1}
\begin{aligned}
\widetilde{\mathfrak D }_2
& = \exp \left( 2 \theta_4 (k_{43;1} +k_{43;2}) \bar\theta_3
+ 2 \theta_6 (k_{65;1} +k_{65;2}) \bar\theta_5
+ 2 \theta_4 k_{45} \bar\theta_5 \right) \\
& = \exp \left(\!-2 \theta_4 p\, \bar\theta_3 - 2 \theta_6 p \,\bar\theta_5
+ 2 \theta_4 p\, \bar\theta_{5} \right)~.
\end{aligned}
\end{equation}
In the second step we used the momentum conservation $\delta$-functions that are present 
in the factor $\cY$ defined in (\ref{WYD}). Performing the Grassmann integrals, one finds 
$\cZ_2(p,k)= p^4$.
This factor cancels the two ``external'' propagators in $\cY_2$ and the 
integral over internal momenta can be represented in the graphical notation of Appendix~\ref{app:loop_integrals} as follows:
\begin{equation}
\label{w2D3}
\cW_2(p) = \parbox[c]{.18\textwidth}{ \includegraphics[width = .18\textwidth]{Ydot4.jpg}}~.
\end{equation}
Taking the Fourier transform of this expression via (\ref{int22x}) and inserting it into (\ref{W2def}),
we finally find 
\begin{eqnarray}
\parbox[c]{.35\textwidth}{ \includegraphics[width = .35\textwidth]{v21square2.png}} &=&
g_0^4 (2N-N_f )^2\, \bigg[\frac{(\pi x^2)^\epsilon\,\Gamma(1-\epsilon)}{(4\pi)^2\,\epsilon(1 - 2\epsilon)}
\bigg]^2 \Delta(x) \, \delta^{ab}+  O(\epsilon)\nonumber\\
&=& v_{2,1}^2\, \Delta(x)\, \delta^{ab} +  O(\epsilon)~, \label{w2final}
\end{eqnarray}
in agreement with the formula (\ref{2loopv212}) in the main text.

\subsection*{Irreducible diagrams: the $v_{2,2}$ part}
Let us now consider the irreducible two-loop corrections to the scalar propagator, namely the diagrams represented in Fig.~\ref{fig:pphi2l-irred}. 
We start from
\begin{equation}
\label{w3x}
\begin{aligned}
W_3^{ab}(x) & \equiv ~\parbox[c]{.35\textwidth}{
\begingroup%
  \makeatletter%
  \providecommand\color[2][]{%
    \errmessage{(Inkscape) Color is used for the text in Inkscape, but the package 'color.sty' is not loaded}%
    \renewcommand\color[2][]{}%
  }%
  \providecommand\transparent[1]{%
    \errmessage{(Inkscape) Transparency is used (non-zero) for the text in Inkscape, but the package 'transparent.sty' is not loaded}%
    \renewcommand\transparent[1]{}%
  }%
  \providecommand\rotatebox[2]{#2}%
  \newcommand*\fsize{\dimexpr\f@size pt\relax}%
  \newcommand*\lineheight[1]{\fontsize{\fsize}{#1\fsize}\selectfont}%
  \ifx\svgwidth\undefined%
    \setlength{\unitlength}{130bp}%
    \ifx\svgscale\undefined%
      \relax%
    \else%
      \setlength{\unitlength}{\unitlength * \real{\svgscale}}%
    \fi%
  \else%
    \setlength{\unitlength}{\svgwidth}%
  \fi%
  \global\let\svgwidth\undefined%
  \global\let\svgscale\undefined%
  \makeatother%
  \begin{picture}(1,0.39033469)%
    \lineheight{1}%
    \setlength\tabcolsep{0pt}%
    \put(-0.00232559,0.25693898){\color[rgb]{0,0,0}\makebox(0,0)[lt]{\lineheight{1.25}\smash{\begin{tabular}[t]{l}\textbf{$a$}\end{tabular}}}}%
    \put(-0.00232559,0.12132969){\color[rgb]{0,0,0}\makebox(0,0)[lt]{\lineheight{1.25}\smash{\begin{tabular}[t]{l}\textbf{$x$}\end{tabular}}}}%
    \put(0.92348062,0.12132969){\color[rgb]{0,0,0}\makebox(0,0)[lt]{\lineheight{1.25}\smash{\begin{tabular}[t]{l}\textbf{$0$}\end{tabular}}}}%
    \put(0.92348062,0.25693898){\color[rgb]{0,0,0}\makebox(0,0)[lt]{\lineheight{1.25}\smash{\begin{tabular}[t]{l}\textbf{$b$}\end{tabular}}}}%
    \put(0,0){\includegraphics[width=\unitlength,page=1]{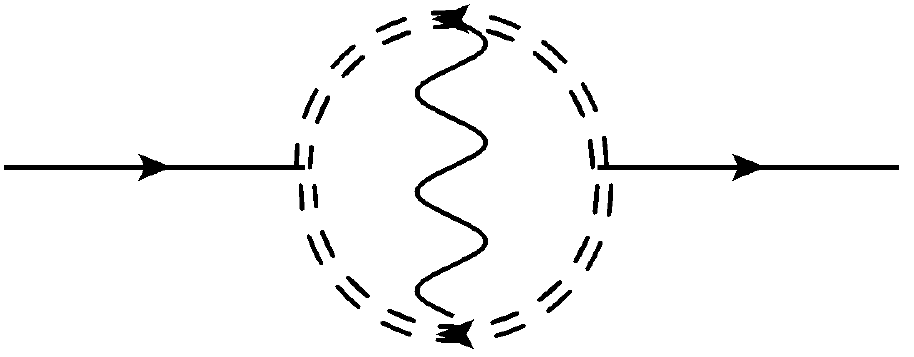}}%
  \end{picture}%
\endgroup%
} \\ 
& = \frac{1}{2} (\sqrt{2} g_0)^2 (2 g_0)^2 
\big(N_f\,\tr T^a T^c T^b T^c - \tr_{\text{adj}}T^a T^c T^b T^c\big)\, W_3(x)\\[2mm]
& = - 2 g_0^4\,\Big(\frac{N_f}{2N}+ N^2\Big)\,W_3(x)\,\delta^{ab}~.
\end{aligned}
\end{equation}
In momentum space, we have to compute
\begin{equation}
\label{w3}
\cW_3(p)  ~=~ \parbox[c]{.45\textwidth}{
\begingroup%
  \makeatletter%
  \providecommand\color[2][]{%
    \errmessage{(Inkscape) Color is used for the text in Inkscape, but the package 'color.sty' is not loaded}%
    \renewcommand\color[2][]{}%
  }%
  \providecommand\transparent[1]{%
    \errmessage{(Inkscape) Transparency is used (non-zero) for the text in Inkscape, but the package 'transparent.sty' is not loaded}%
    \renewcommand\transparent[1]{}%
  }%
  \providecommand\rotatebox[2]{#2}%
  \newcommand*\fsize{\dimexpr\f@size pt\relax}%
  \newcommand*\lineheight[1]{\fontsize{\fsize}{#1\fsize}\selectfont}%
  \ifx\svgwidth\undefined%
    \setlength{\unitlength}{135bp}%
    \ifx\svgscale\undefined%
      \relax%
    \else%
      \setlength{\unitlength}{\unitlength * \real{\svgscale}}%
    \fi%
  \else%
    \setlength{\unitlength}{\svgwidth}%
  \fi%
  \global\let\svgwidth\undefined%
  \global\let\svgscale\undefined%
  \makeatother%
  \begin{picture}(1,0.47906544)%
    \lineheight{1}%
    \setlength\tabcolsep{0pt}%
    \put(-0.00154666,0.18625171){\color[rgb]{0,0,0}\makebox(0,0)[lt]{\lineheight{1.25}\smash{\begin{tabular}[t]{l}\textbf{$1$}\end{tabular}}}}%
    \put(0.94993926,0.18625171){\color[rgb]{0,0,0}\makebox(0,0)[lt]{\lineheight{1.25}\smash{\begin{tabular}[t]{l}\textbf{$2$}\end{tabular}}}}%
    \put(0.25726988,0.18625171){\color[rgb]{0,0,0}\makebox(0,0)[lt]{\lineheight{1.25}\smash{\begin{tabular}[t]{l}\textbf{$3$}\end{tabular}}}}%
    \put(0.67120308,0.18625171){\color[rgb]{0,0,0}\makebox(0,0)[lt]{\lineheight{1.25}\smash{\begin{tabular}[t]{l}\textbf{$4$}\end{tabular}}}}%
    \put(0.50124199,0.45553197){\color[rgb]{0,0,0}\makebox(0,0)[lt]{\lineheight{1.25}\smash{\begin{tabular}[t]{l}\textbf{$5$}\end{tabular}}}}%
    \put(0.50124199,0.0033056){\color[rgb]{0,0,0}\makebox(0,0)[lt]{\lineheight{1.25}\smash{\begin{tabular}[t]{l}\textbf{$6$}\end{tabular}}}}%
    \put(0,0){\includegraphics[width=\unitlength,page=1]{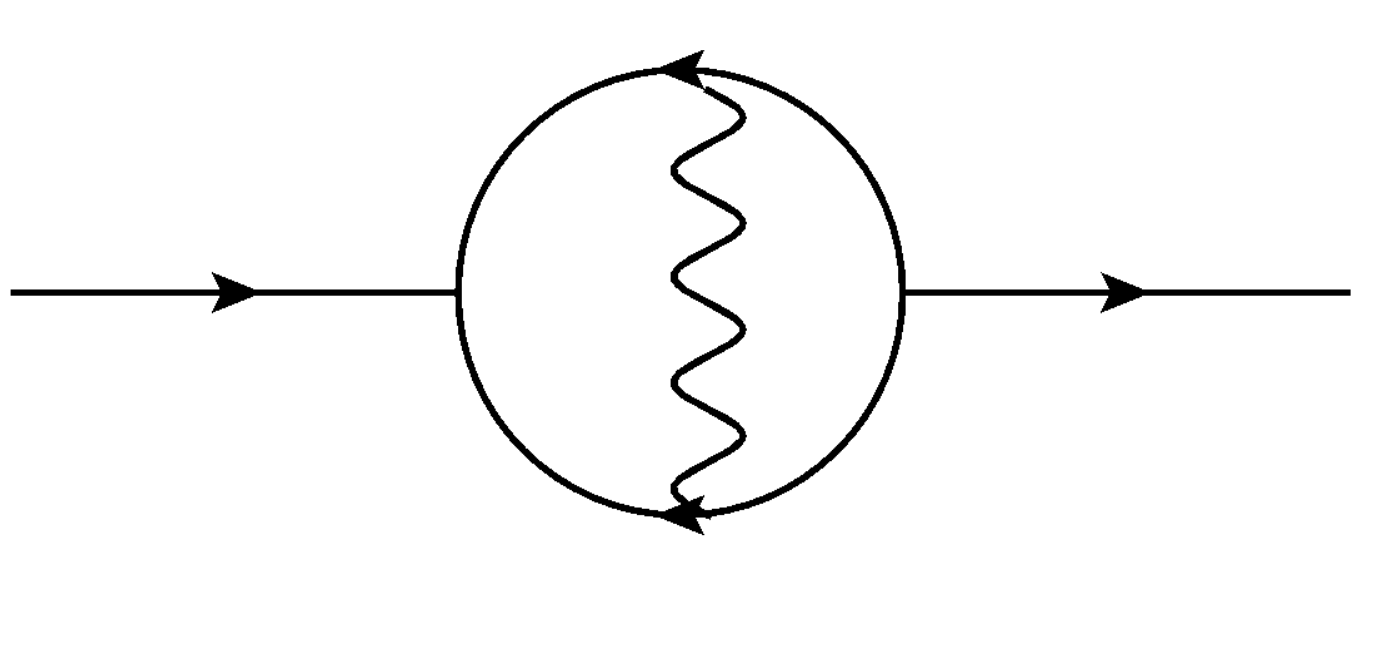}}%
  \end{picture}%
\endgroup%
}
\end{equation}
which has the general form (\ref{WYD}).
The $\theta$-factors present in the two chiral vertices saturate the integrations over 
$\theta_3$ and $\bar{\theta}_4$, while those in the gluon propagator set $\theta_{6}$ 
and $\bar\theta_{6}$ equal to $\theta_{5}$ and $\bar\theta_{5}$ respectively. 
The remaining Grassmann integrations are
\begin{equation}
\label{w3-1}
\cZ_3(p,k)=\int   d^2\bar\theta_3\, d^2\theta_4\, d^2\theta_5\, d^2\bar\theta_5\, \,
\widetilde{\mathfrak D}_3 
\end{equation}
with
\begin{align}
\label{w3-2}
\widetilde{\mathfrak D}_3
& = \exp\left(\!-\theta_5 (k_{53} + k_{63} + k_{45} + k_{46})\,\bar{\theta}_5
+ 2 \theta_5 (k_{53} + k_{63}) \bar{\theta_3} + 2 \theta_4 (k_{45} + k_{46}) 
\,\bar{\theta_5}\right)
\nonumber\\
& = \exp\left(2 \theta_5 p\, \bar{\theta}_5 - 2 \theta_5 p\, \bar\theta_3 - 2 \theta_4 p \,\bar{\theta}_5\right)~, 
\end{align}
where in the second step we used momentum conservation.
Performing the $\theta$-integrals, we get $\cZ_3(p,k)=p^4$,
which cancels the two ``external'' propagators in $\cY_3$; the remaining
integral over internal momenta can be represented in the graphical notation of Appendix~\ref{app:loop_integrals} as follows:
\begin{equation}
\label{w3-4}
\cW_3(p) = {\parbox[c]{.10\textwidth}{ \includegraphics[width = .10\textwidth]{Ydot1}}}
~.
\end{equation}
Taking the Fourier transform via (\ref{int2x}), and inserting the result into (\ref{w3x}), 
we obtain
\begin{equation}
\label{w3-5}
W_3^{ab}(x) = 
-\Big(\frac{g_0^2}{8\pi^2}\Big)^2\,3\,\zeta(3)\,\Big(\frac{N_f}{2N}+N^2\Big)\,(\pi x^2)^{2\epsilon}\,\Delta(x)
\, \delta^{ab} + \cdots 
\end{equation}

Let us now consider the diagram
\begin{equation}
\begin{aligned}
\label{w4x}
W_4^{ab}(x) & ~\equiv~~ \parbox[c]{.29\textwidth}{
\begingroup%
  \makeatletter%
  \providecommand\color[2][]{%
    \errmessage{(Inkscape) Color is used for the text in Inkscape, but the package 'color.sty' is not loaded}%
    \renewcommand\color[2][]{}%
  }%
  \providecommand\transparent[1]{%
    \errmessage{(Inkscape) Transparency is used (non-zero) for the text in Inkscape, but the package 'transparent.sty' is not loaded}%
    \renewcommand\transparent[1]{}%
  }%
  \providecommand\rotatebox[2]{#2}%
  \newcommand*\fsize{\dimexpr\f@size pt\relax}%
  \newcommand*\lineheight[1]{\fontsize{\fsize}{#1\fsize}\selectfont}%
  \ifx\svgwidth\undefined%
    \setlength{\unitlength}{130bp}%
    \ifx\svgscale\undefined%
      \relax%
    \else%
      \setlength{\unitlength}{\unitlength * \real{\svgscale}}%
    \fi%
  \else%
    \setlength{\unitlength}{\svgwidth}%
  \fi%
  \global\let\svgwidth\undefined%
  \global\let\svgscale\undefined%
  \makeatother%
  \begin{picture}(1,0.46386359)%
    \lineheight{1}%
    \setlength\tabcolsep{0pt}%
    \put(0.03227698,0.25327315){\color[rgb]{0,0,0}\makebox(0,0)[lt]{\lineheight{1.25}\smash{\begin{tabular}[t]{l}\textbf{$a$}\end{tabular}}}}%
    \put(0.03227698,0.12049827){\color[rgb]{0,0,0}\makebox(0,0)[lt]{\lineheight{1.25}\smash{\begin{tabular}[t]{l}\textbf{$x$}\end{tabular}}}}%
    \put(0.96417868,0.12049827){\color[rgb]{0,0,0}\makebox(0,0)[lt]{\lineheight{1.25}\smash{\begin{tabular}[t]{l}\textbf{$0$}\end{tabular}}}}%
    \put(0.96417868,0.25327315){\color[rgb]{0,0,0}\makebox(0,0)[lt]{\lineheight{1.25}\smash{\begin{tabular}[t]{l}\textbf{$b$}\end{tabular}}}}%
    \put(0,0){\includegraphics[width=\unitlength,page=1]{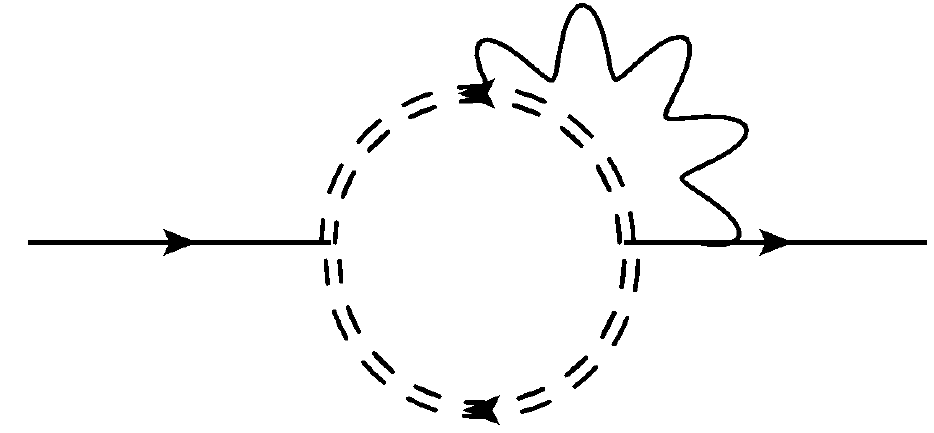}}%
  \end{picture}%
\endgroup%
}\\ 
& = 4\times \frac{1}{2} (\sqrt{2} g_0)^2 (2 g_0)^2 
\big(N_f \,\tr T^a T^d T^c - \tr_{\text{adj}} T^a T^d T^c\big)\,
(T^c)^{db}\, W_4(x)~.
\end{aligned}
\end{equation}
Using the relations
\begin{equation}
\label{3traces}
\tr T^{a} T^{d} T^c = \frac{1}{4} \,\big(d^{a d c} + \ii\, f^{a d c}\big)~,~~~
\tr_{\text{adj}} T^{a} T^{d} T^c= \ii\,\frac{N}{2}\, f^{a d c}
\end{equation}
we get
\begin{equation}
\begin{aligned}
\label{w4xbis}
W_4^{ab}(x) & = 4 g_0^4 \,\big(N_f \big(d^{adc}+ \ii \,f^{adc}\big) - 2 N \ii f^{adc}\big)
\,\ii \,f^{cdb}\, W_4(x) \\[2mm]
& = - 4 g_0^4\, N(2N-N_f)\, W_4(x)\,\delta^{ab}~,  
\end{aligned}
\end{equation}
where in the second step we took advantage of the identities
\begin{equation}
d^{adc}\, f^{cdb} = 0~, ~~~ f^{adc}\, f^{cdb} = - \tr_{\text{adj}}T^a T^b = - N \,\delta^{ab}~.
\end{equation} 
In momentum space, we have to compute
\begin{equation}
\label{w4}
\cW_4(p)  \,= \!\!\!\!\!\!\!\!\!\!\!\parbox[c]{.45\textwidth}{      
\begingroup%
  \makeatletter%
  \providecommand\color[2][]{%
    \errmessage{(Inkscape) Color is used for the text in Inkscape, but the package 'color.sty' is not loaded}%
    \renewcommand\color[2][]{}%
  }%
  \providecommand\transparent[1]{%
    \errmessage{(Inkscape) Transparency is used (non-zero) for the text in Inkscape, but the package 'transparent.sty' is not loaded}%
    \renewcommand\transparent[1]{}%
  }%
  \providecommand\rotatebox[2]{#2}%
  \newcommand*\fsize{\dimexpr\f@size pt\relax}%
  \newcommand*\lineheight[1]{\fontsize{\fsize}{#1\fsize}\selectfont}%
  \ifx\svgwidth\undefined%
    \setlength{\unitlength}{160bp}
    \ifx\svgscale\undefined%
      \relax%
    \else%
      \setlength{\unitlength}{\unitlength * \real{\svgscale}}%
    \fi%
  \else%
    \setlength{\unitlength}{\svgwidth}%
  \fi%
  \global\let\svgwidth\undefined%
  \global\let\svgscale\undefined%
  \makeatother%
  \begin{picture}(1,0.43503623)%
    \lineheight{1}%
    \setlength\tabcolsep{0pt}%
    \put(0.46081262,0.17535969){\color[rgb]{0,0,0}\makebox(0,0)[lt]{\lineheight{1.25}\smash{\begin{tabular}[t]{l}\textbf{$3$}\end{tabular}}}}%
    \put(0.1386326,0.15135969){\color[rgb]{0,0,0}\makebox(0,0)[lt]{\lineheight{1.25}\smash{\begin{tabular}[t]{l}\textbf{$1$}\end{tabular}}}}%
    \put(0.9361669,0.15135969){\color[rgb]{0,0,0}\makebox(0,0)[lt]{\lineheight{1.25}\smash{\begin{tabular}[t]{l}\textbf{$2$}\end{tabular}}}}%
    \put(0.63645449,0.17535969){\color[rgb]{0,0,0}\makebox(0,0)[lt]{\lineheight{1.25}\smash{\begin{tabular}[t]{l}\textbf{$4$}\end{tabular}}}}%
    \put(0,0.07){\includegraphics[width=\unitlength,page=1]{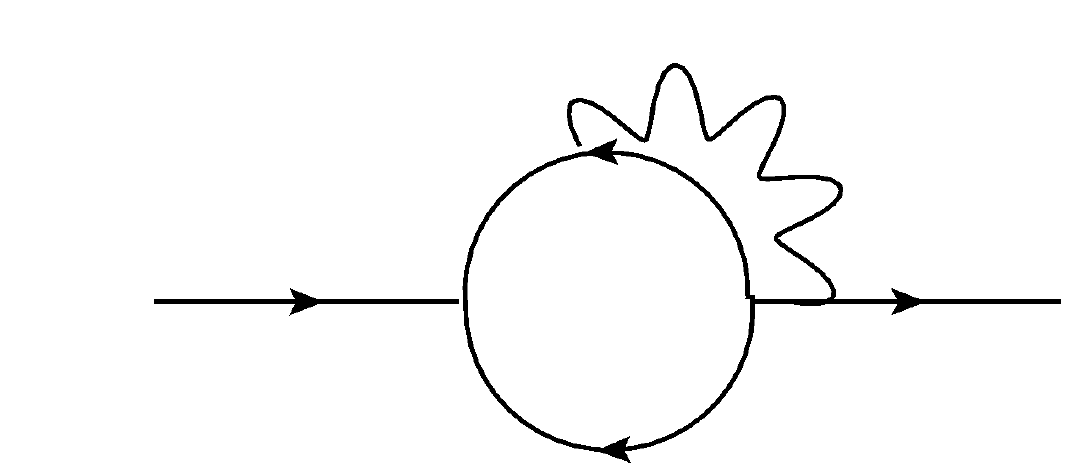}}%
    \put(0.5226316,0.28468961){\color[rgb]{0,0,0}\makebox(0,0)[lt]{\lineheight{1.25}\smash{\begin{tabular}[t]{l}\textbf{$5$}\end{tabular}}}}%
    \put(0.75207943,0.15135969){\color[rgb]{0,0,0}\makebox(0,0)[lt]{\lineheight{1.25}\smash{\begin{tabular}[t]{l}\textbf{$6$}\end{tabular}}}}%
  \end{picture}%
\endgroup%
}
\vspace{-0.2cm}
\end{equation}
which has the general form (\ref{WYD}). Again, 
the $\theta$-factors of the vertices saturate the integrations over $\theta_3$ and 
$\bar{\theta}_4$, while the gluon propagator sets $\theta_{6}$ and $\bar\theta_{6}$ equal to 
$\theta_{5}$ and $\bar\theta_{5}$. The remaining Grassmann integrations are as in
(\ref{w3-1}), but now with 
\begin{align}
\label{w4-2}
\widetilde{\mathfrak D}_4 & = \exp\left(-\theta_5 (k_{53} + k_{45} + k_{46} + k_{62})\,\bar{\theta}_5
+ 2 \theta_5 k_{53}\, \bar{\theta_3} + 2 \theta_4 (k_{45} + k_{46})\, \bar{\theta_5}
+ 2 \theta_4 k_{43}\,\bar{\theta}_3\right)
\nonumber\\[2mm]
& = \exp\left(- 2 \theta_5 p\, \bar{\theta}_5 - 2 \theta_5 k_{53}\, \bar\theta_{53} - 2 \theta_4 k_{43}\, \bar{\theta}_{53}\right)
\end{align}
where in the second step we used momentum conservation.
The Grassmann integrations yield $\cZ_4(p,k)= p^2\, k_{43}^2$.
This factor cancels one external and one internal propagator in $\cY_4$ and we remain with
\begin{equation}
\label{w4-4}
\cW_4(p) = 
\frac{1}{p^2}{\parbox[c]{.12\textwidth}{ \includegraphics[width = .12\textwidth]{Ydot2a.png}}}
\!\!\!\!\!\!.
\end{equation}
Thus, the momentum space contribution corresponding to (\ref{w4xbis}) is
\begin{equation}
\label{w4-5}
\cW_4^{ab}(p) = - 4 g_0^4\, N(2N-N_f)\,\frac{1}{p^2}{\parbox[c]{.12\textwidth}{ \includegraphics[width = .12\textwidth]{Ydot2a.png}}}  \!\!\!\!\!\!\delta^{ab} ~.
\end{equation}

We now consider the diagram
\begin{equation}
\begin{aligned}
\label{w5x}
W_5^{ab}(x) & \equiv ~~\parbox[c]{.25\textwidth}{
\begingroup%
  \makeatletter%
  \providecommand\color[2][]{%
    \errmessage{(Inkscape) Color is used for the text in Inkscape, but the package 'color.sty' is not loaded}%
    \renewcommand\color[2][]{}%
  }%
  \providecommand\transparent[1]{%
    \errmessage{(Inkscape) Transparency is used (non-zero) for the text in Inkscape, but the package 'transparent.sty' is not loaded}%
    \renewcommand\transparent[1]{}%
  }%
  \providecommand\rotatebox[2]{#2}%
  \newcommand*\fsize{\dimexpr\f@size pt\relax}%
  \newcommand*\lineheight[1]{\fontsize{\fsize}{#1\fsize}\selectfont}%
  \ifx\svgwidth\undefined%
    \setlength{\unitlength}{130bp}%
    \ifx\svgscale\undefined%
      \relax%
    \else%
      \setlength{\unitlength}{\unitlength * \real{\svgscale}}%
    \fi%
  \else%
    \setlength{\unitlength}{\svgwidth}%
  \fi%
  \global\let\svgwidth\undefined%
  \global\let\svgscale\undefined%
  \makeatother%
  \begin{picture}(1,0.55736077)%
    \lineheight{1}%
    \setlength\tabcolsep{0pt}%
    \put(-0.00182171,0.27218555){\color[rgb]{0,0,0}\makebox(0,0)[lt]{\lineheight{1.25}\smash{\begin{tabular}[t]{l}\textbf{$a$}\end{tabular}}}}%
    \put(-0.00182171,0.13664444){\color[rgb]{0,0,0}\makebox(0,0)[lt]{\lineheight{1.25}\smash{\begin{tabular}[t]{l}\textbf{$x$}\end{tabular}}}}%
    \put(0.92351908,0.13664444){\color[rgb]{0,0,0}\makebox(0,0)[lt]{\lineheight{1.25}\smash{\begin{tabular}[t]{l}\textbf{$0$}\end{tabular}}}}%
    \put(0.92351908,0.27218555){\color[rgb]{0,0,0}\makebox(0,0)[lt]{\lineheight{1.25}\smash{\begin{tabular}[t]{l}\textbf{$b$}\end{tabular}}}}%
    \put(0,0.02){\includegraphics[width=\unitlength,page=1]{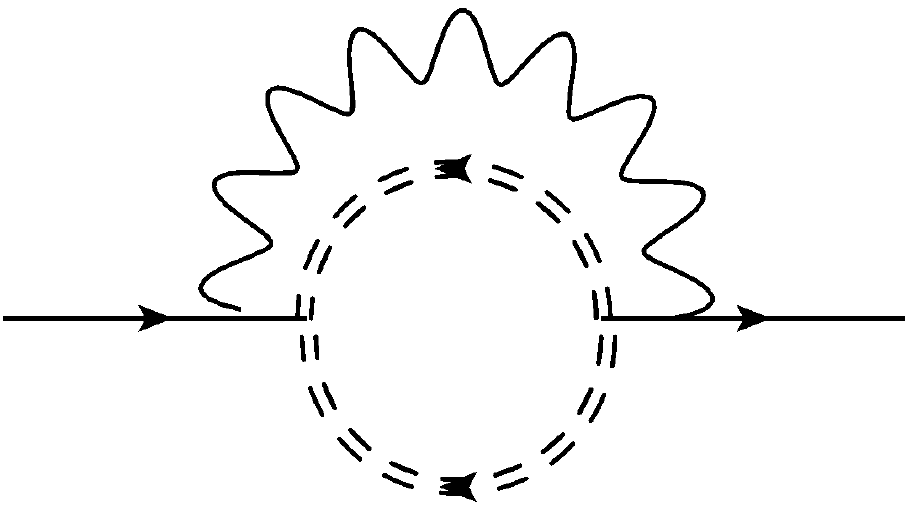}}%
  \end{picture}%
\endgroup%
}\\ 
& = \, - \frac{1}{2} (\sqrt{2} g_0)^2 (2 g_0)^2 
\big(T^a T^b\big)^{cd} \,\big(N_f \,\tr T^c T^d - \tr_{\text{adj}} T^c T^d\big) \,
W_5(x)\\[2mm]
& =\, 2 g_0^4 \,N(2N-N_f)\,W_5(x)\,\delta^{ab} ~.
\end{aligned}
\end{equation}
In momentum space, we have to compute 
\begin{equation}
\label{w5}
\vspace{0.1cm}
\cW_5(p)  = \parbox[c]{.45\textwidth}{
\begingroup%
  \makeatletter%
  \providecommand\color[2][]{%
    \errmessage{(Inkscape) Color is used for the text in Inkscape, but the package 'color.sty' is not loaded}%
    \renewcommand\color[2][]{}%
  }%
  \providecommand\transparent[1]{%
    \errmessage{(Inkscape) Transparency is used (non-zero) for the text in Inkscape, but the package 'transparent.sty' is not loaded}%
    \renewcommand\transparent[1]{}%
  }%
  \providecommand\rotatebox[2]{#2}%
  \newcommand*\fsize{\dimexpr\f@size pt\relax}%
  \newcommand*\lineheight[1]{\fontsize{\fsize}{#1\fsize}\selectfont}%
  \ifx\svgwidth\undefined%
    \setlength{\unitlength}{140bp}%
    \ifx\svgscale\undefined%
      \relax%
    \else%
      \setlength{\unitlength}{\unitlength * \real{\svgscale}}%
    \fi%
  \else%
    \setlength{\unitlength}{\svgwidth}%
  \fi%
  \global\let\svgwidth\undefined%
  \global\let\svgscale\undefined%
  \makeatother%
  \begin{picture}(1,0.55551226)%
    \lineheight{1}%
    \setlength\tabcolsep{0pt}%
    \put(0.41302803,0.20530449){\color[rgb]{0,0,0}\makebox(0,0)[lt]{\lineheight{1.25}\smash{\begin{tabular}[t]{l}\textbf{$3$}\end{tabular}}}}%
    \put(0.05984859,0.16530449){\color[rgb]{0,0,0}\makebox(0,0)[lt]{\lineheight{1.25}\smash{\begin{tabular}[t]{l}\textbf{$1$}\end{tabular}}}}%
    \put(0.9293087,0.16530449){\color[rgb]{0,0,0}\makebox(0,0)[lt]{\lineheight{1.25}\smash{\begin{tabular}[t]{l}\textbf{$2$}\end{tabular}}}}%
    \put(0.6066229,0.20530449){\color[rgb]{0,0,0}\makebox(0,0)[lt]{\lineheight{1.25}\smash{\begin{tabular}[t]{l}\textbf{$4$}\end{tabular}}}}%
    \put(0,0.08){\includegraphics[width=\unitlength,page=1]{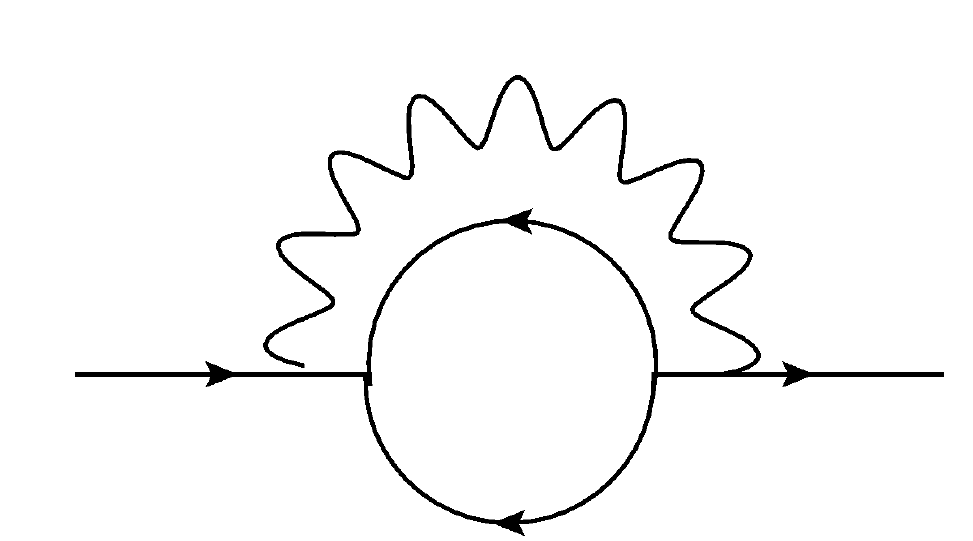}}%
    \put(0.27879254,0.16530449){\color[rgb]{0,0,0}\makebox(0,0)[lt]{\lineheight{1.25}\smash{\begin{tabular}[t]{l}\textbf{$5$}\end{tabular}}}}%
    \put(0.73647125,0.16530449){\color[rgb]{0,0,0}\makebox(0,0)[lt]{\lineheight{1.25}\smash{\begin{tabular}[t]{l}\textbf{$6$}\end{tabular}}}}%
  \end{picture}%
\endgroup%
}
\vspace{-0.4cm}
\end{equation}
which again has the form (\ref{WYD}) with
\begin{equation}
\begin{aligned}
\label{w5-2}
\widetilde{\mathfrak D}_5 & = \exp\left(
-2 \theta_5 (p + k_{53})\,\bar{\theta}_5 + 2 \theta_5 k_{53} \,\bar{\theta}_3 
- 2 \theta_4 k_{53}\, \bar{\theta}_3+ 2 \theta_4 k_{53}\, \bar{\theta}_5 \right)\\[2mm]
& = \exp\left(- 2 \theta_5 p\, \bar{\theta}_5 + 2 \theta_5 k_{53} \,\bar{\theta}_{35} - 2 \theta_4 k_{53} \,\bar{\theta}_{35}\right)~.
\end{aligned}
\end{equation}
The Grassmann integration leads to $\cZ_5(p,k)=p^2\, k_{53}^2$, which 
cancels one external and one internal propagator in $\cY_5$. We then remain with 
\begin{equation}
\label{w5-4}
\cW_5(p) = 
\frac{1}{p^2}{\parbox[c]{.12\textwidth}{ \includegraphics[width = .12\textwidth]{Ydot2a.png}}}\!\!
\!\!\!\!.
\end{equation}
Using this in (\ref{w5x}), we find that the total diagram in momentum 
space is given by  
\begin{equation}
\label{w5-5}
\cW_5^{ab}(p) = 2 g_0^4\, N(2N-N_f)\,\frac{1}{p^2}{\parbox[c]{.12\textwidth}{ \includegraphics[width = .12\textwidth]{Ydot2a.png}}}  \!\!\!\!\!\!\delta^{ab} ~.
\end{equation}

Finally, we consider the diagram
\begin{equation}
\begin{aligned}
\label{w6x}
W_6^{ab}(x) & \equiv ~~\parbox[c]{.29\textwidth}{
\begingroup%
  \makeatletter%
  \providecommand\color[2][]{%
    \errmessage{(Inkscape) Color is used for the text in Inkscape, but the package 'color.sty' is not loaded}%
    \renewcommand\color[2][]{}%
  }%
  \providecommand\transparent[1]{%
    \errmessage{(Inkscape) Transparency is used (non-zero) for the text in Inkscape, but the package 'transparent.sty' is not loaded}%
    \renewcommand\transparent[1]{}%
  }%
  \providecommand\rotatebox[2]{#2}%
  \newcommand*\fsize{\dimexpr\f@size pt\relax}%
  \newcommand*\lineheight[1]{\fontsize{\fsize}{#1\fsize}\selectfont}%
  \ifx\svgwidth\undefined%
    \setlength{\unitlength}{130bp}%
    \ifx\svgscale\undefined%
      \relax%
    \else%
      \setlength{\unitlength}{\unitlength * \real{\svgscale}}%
    \fi%
  \else%
    \setlength{\unitlength}{\svgwidth}%
  \fi%
  \global\let\svgwidth\undefined%
  \global\let\svgscale\undefined%
  \makeatother%
  \begin{picture}(1,0.55935832)%
    \lineheight{1}%
    \setlength\tabcolsep{0pt}%
    \put(-0.00231934,0.14020202){\color[rgb]{0,0,0}\makebox(0,0)[lt]{\lineheight{1.25}\smash{\begin{tabular}[t]{l}\textbf{$a$}\end{tabular}}}}%
    \put(-0.00231934,0.00495703){\color[rgb]{0,0,0}\makebox(0,0)[lt]{\lineheight{1.25}\smash{\begin{tabular}[t]{l}\textbf{$x$}\end{tabular}}}}%
    \put(0.92099982,0.00495703){\color[rgb]{0,0,0}\makebox(0,0)[lt]{\lineheight{1.25}\smash{\begin{tabular}[t]{l}\textbf{$0$}\end{tabular}}}}%
    \put(0.92099982,0.14020202){\color[rgb]{0,0,0}\makebox(0,0)[lt]{\lineheight{1.25}\smash{\begin{tabular}[t]{l}\textbf{$b$}\end{tabular}}}}%
    \put(0,0){\includegraphics[width=\unitlength,page=1]{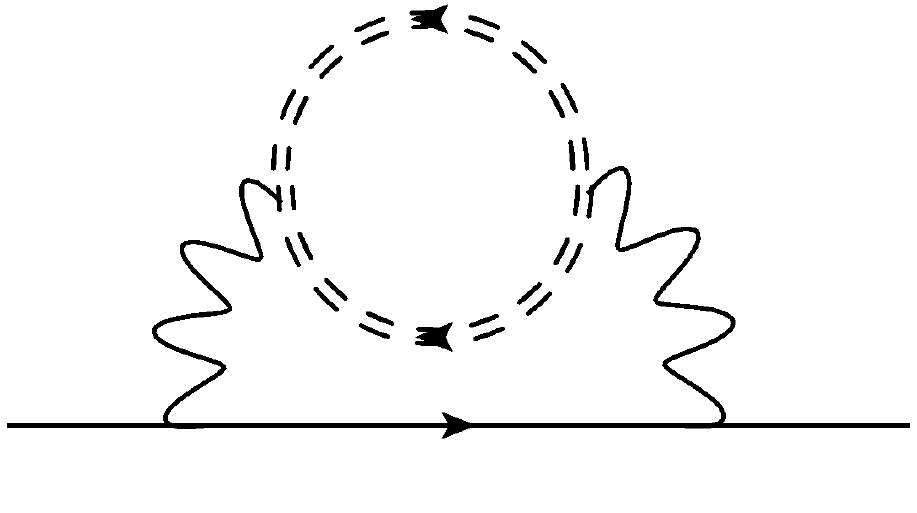}}%
  \end{picture}%
\endgroup%
}\\ 
& =  2\times \Big(\!-\frac{1}{2}\Big)^2 (2 g_0)^4 
\big(T^a T^b\big)^{cd}\,\big(N_f \,\tr T^c T^d - \tr_{\text{adj}}T^c T^d\big) \,W_6(x)
\\[2mm]
& =  -4 g_0^2\, N(2N-N_f)\,W_6(x)\,\delta^{ab}~.
\end{aligned}
\end{equation}
In momentum space, we have to compute
\begin{equation}
\label{w6}
\cW_6(p)  = \parbox[c]{.35\textwidth}{
\begingroup%
  \makeatletter%
  \providecommand\color[2][]{%
    \errmessage{(Inkscape) Color is used for the text in Inkscape, but the package 'color.sty' is not loaded}%
    \renewcommand\color[2][]{}%
  }%
  \providecommand\transparent[1]{%
    \errmessage{(Inkscape) Transparency is used (non-zero) for the text in Inkscape, but the package 'transparent.sty' is not loaded}%
    \renewcommand\transparent[1]{}%
  }%
  \providecommand\rotatebox[2]{#2}%
  \newcommand*\fsize{\dimexpr\f@size pt\relax}%
  \newcommand*\lineheight[1]{\fontsize{\fsize}{#1\fsize}\selectfont}%
  \ifx\svgwidth\undefined%
    \setlength{\unitlength}{140bp}%
    \ifx\svgscale\undefined%
      \relax%
    \else%
      \setlength{\unitlength}{\unitlength * \real{\svgscale}}%
    \fi%
  \else%
    \setlength{\unitlength}{\svgwidth}%
  \fi%
  \global\let\svgwidth\undefined%
  \global\let\svgscale\undefined%
  \makeatother%
  \begin{picture}(1,0.53047109)%
    \lineheight{1}%
    \setlength\tabcolsep{0pt}%
    \put(0.16815657,0.00827991){\color[rgb]{0,0,0}\makebox(0,0)[lt]{\lineheight{1.25}\smash{\begin{tabular}[t]{l}\textbf{$3$}\end{tabular}}}}%
    \put(-0.00223033,0.00827991){\color[rgb]{0,0,0}\makebox(0,0)[lt]{\lineheight{1.25}\smash{\begin{tabular}[t]{l}\textbf{$1$}\end{tabular}}}}%
    \put(0.92781078,0.00827991){\color[rgb]{0,0,0}\makebox(0,0)[lt]{\lineheight{1.25}\smash{\begin{tabular}[t]{l}\textbf{$2$}\end{tabular}}}}%
    \put(0.75039761,0.00827991){\color[rgb]{0,0,0}\makebox(0,0)[lt]{\lineheight{1.25}\smash{\begin{tabular}[t]{l}\textbf{$4$}\end{tabular}}}}%
    \put(0,0){\includegraphics[width=\unitlength,page=1]{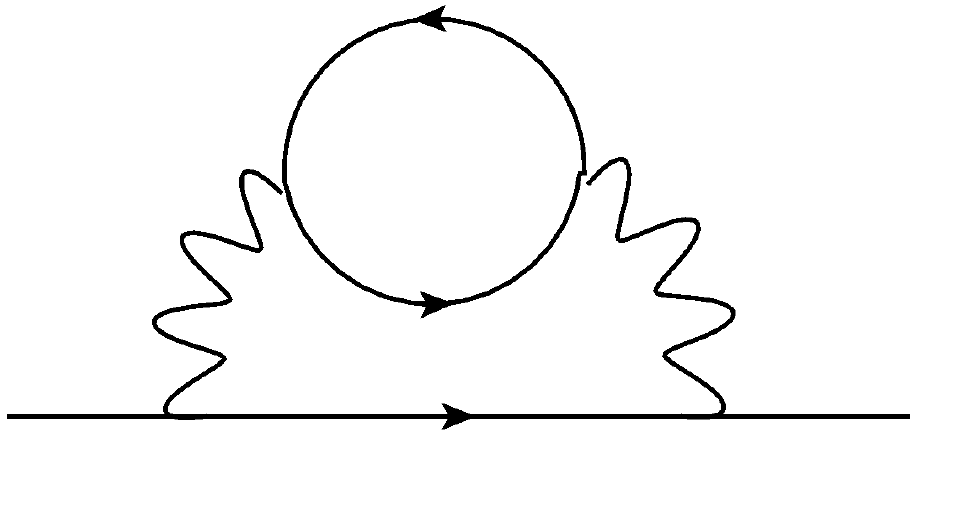}}%
    \put(0.33170693,0.31091128){\color[rgb]{0,0,0}\makebox(0,0)[lt]{\lineheight{1.25}\smash{\begin{tabular}[t]{l}\textbf{$5$}\end{tabular}}}}%
    \put(0.53546878,0.31091128){\color[rgb]{0,0,0}\makebox(0,0)[lt]{\lineheight{1.25}\smash{\begin{tabular}[t]{l}\textbf{$6$}\end{tabular}}}}%
  \end{picture}%
\endgroup%
}
\end{equation}
which, once again, is of the form (\ref{WYD}) with
\begin{equation}
\label{w6-2}
\begin{aligned}
\widetilde{\mathfrak D}_6 & = \exp\left(
-\theta_3(p + k_{56} + k_{65}+ k_{34})\,\bar{\theta}_3 
-\theta_4(p + k_{56} + k_{65}+ k_{34})\,\bar{\theta}_4 \right.\\[2mm]
& \phantom{+ \exp (}~ \left. + 2 \theta_3(k_{34} + k_{56})\,\bar{\theta}_4 + 2 \theta_4 k_{65}
\,\bar{\theta}_3\right)\\[2mm]
& = \exp\left(- 2 \theta_3(k_{34}+ k_{56})\,\bar{\theta}_3 + 2 \theta_{34} (k_{34}+ k_{56})
\,\bar{\theta}_4 + 2 \theta_4 k_{65}\, \bar{\theta}_3
\right) ~,
\end{aligned}
\end{equation}
having used momentum conservation in the second step. 
The Grassmann integral we have to compute in this case is
\begin{equation}
\label{w6-2bis}
\cZ_6(p,k)=\int d^2\theta_3 \,d^2\bar{\theta}_2\, d^2\theta_4\, d^2\bar{\theta}_4\,\, 
\widetilde{\mathfrak D}_6~.
\end{equation}
Integrating over $\bar{\theta}_4$ produces a factor of $ (k_{34}+ k_{56})^2 \, \theta_{34}^2$
which sets $\theta_4 = \theta_3$, so that in the end we remain with
\begin{equation}
\label{w6-2quater}
\cZ_6(p,k)= (k_{34}+ k_{56})^2 \int d^2\theta_3\, d^2\bar{\theta}_3 
\,\exp\left(-2 \theta_3 p \,\bar{\theta}_3\right) = (k_{34}+ k_{56})^2\, p^2
\sim 2 \left(k_{34}\cdot k_{56}\right)\, p^2~.
\end{equation}
The last step follows from the fact that when we integrate this expression over momenta, 
both the $k_{34}^2$ and the $k_{56}^2$ terms, canceling the corresponding propagator, give 
rise to tadpole-like integrals, which vanish in dimensional regularization. 
The symmetry of the diagram under $k_{56} \leftrightarrow - k_{65}$ allows us to rewrite 
the above result as 
\begin{equation}
\label{w6-3}
\cZ_6(p,k)=\big(k_{34}\cdot(k_{56}- k_{65})\big) \,p^2  = (k_{34}\cdot k_{35})\, p^2  
= \frac 12 \,(p^2 - k_{34}^2 - k_{35}^2)\, p^2 
\sim \frac 12 \,(p^2 - k_{35}^2) \, p^2~,
\end{equation}
where again we discarded the tadpole originating from $k_{34}^2$. 
The first term cancels the two external propagators of $\cY_6$, 
while the second term cancels one external and one internal propagator. 
Using the graphical notation of Appendix~\ref{app:loop_integrals}, we can write
\begin{equation}
\cW_6(p) =  \frac{1}{2} \,\bigg[  \parbox[c]{.085\textwidth}{ \includegraphics[width = .085\textwidth]{Ydot3.jpg}}
-  \frac{1}{p^2}{\parbox[c]{.12\textwidth}{ \includegraphics[width = .12\textwidth]{Ydot2a.png}}}
\!\!\!\!\!\!\!\bigg]~,
\label{w6-4}
\end{equation}
so that, from (\ref{w6x}) we see that the total diagram in momentum space is
\begin{equation}
\label{w6-5}
\cW_6^{ab}(p) = -2  g_0^4 \,N(2N-N_f) \,
\bigg[  \parbox[c]{.085\textwidth}{ \includegraphics[width = .085\textwidth]{Ydot3.jpg}}
-  \frac{1}{p^2}{\parbox[c]{.12\textwidth}{ \includegraphics[width = .12\textwidth]{Ydot2a.png}}}
\!\!\!\!\!\!\!\bigg]\,\delta^{ab}~. 
\end{equation}

Summing the three diagrams (\ref{w4-5}), (\ref{w5-5}) and
(\ref{w6-5}), a simplification takes place and we are left with
\begin{equation}
\label{w456}
\sum_{I=4}^6 \cW_I^{ab}(p) = - 2 g_0^4 \,N(2N-N_f) \,
\parbox[c]{.085\textwidth}{ \includegraphics[width = .085\textwidth]{Ydot3.jpg}}\,\delta^{ab} ~.
\end{equation} 
Taking the Fourier transform via (\ref{int44app}), we then have
\begin{equation}
\label{w456x}
\sum_{I=4}^6 W_I^{ab}(x) =\Big(\frac{g_0^2}{8\pi^2}\Big)^2\,N(2N-N_f)
\frac{\Gamma^2(1-\epsilon)}{4\epsilon^2(1-2\epsilon)(1+\epsilon)}\,(\pi x^2)^{2\epsilon}
\,\Delta(x)\, \delta^{ab}~. 
\end{equation}
If we include also the $W_3^{ab}(x)$ diagram given in (\ref{w3-5}), we obtain the 
two-loop irreducible corrections to the propagator:
\begin{equation}
\label{w3456}
\sum_{I=3}^6 W_I^{ab}(x)\equiv \,v_{2,2}\, \Delta(x)\, \delta^{ab}
\end{equation}
with 
\begin{equation}
\label{v22isapp}
v_{2,2}=-\Big(\frac{g_0^2}{8\pi^2}\Big)^2\left[3\,\zeta(3)
\Big(\frac{N_f}{2N}+N^2\Big)-N(2N-N_f)\frac{\Gamma^2(1-\epsilon)}{4\epsilon^2(1-2\epsilon)
	(1+\epsilon)}\right](\pi x^2)^{2\epsilon}~,
\end{equation}
as reported in the formula (\ref{v2}) of the main text.

\subsection*{Irreducible diagrams: the $v_{4,2}$ part}
We now evaluate the irreducible two-loop diagrams that give rise to the
contribution $(c)$ in Fig.~\ref{fig:diagrams}. 
We start from the diagrams represented in Fig.~\ref{fig:vh42diag}. 
The first of these is
\begin{align}
W_7^{a_1a_2b_1b_2}(x)  & \equiv ~~\parbox[c]{.25\textwidth}{
\begingroup%
  \makeatletter%
  \providecommand\color[2][]{%
    \errmessage{(Inkscape) Color is used for the text in Inkscape, but the package 'color.sty' is not loaded}%
    \renewcommand\color[2][]{}%
  }%
  \providecommand\transparent[1]{%
    \errmessage{(Inkscape) Transparency is used (non-zero) for the text in Inkscape, but the package 'transparent.sty' is not loaded}%
    \renewcommand\transparent[1]{}%
  }%
  \providecommand\rotatebox[2]{#2}%
  \newcommand*\fsize{\dimexpr\f@size pt\relax}%
  \newcommand*\lineheight[1]{\fontsize{\fsize}{#1\fsize}\selectfont}%
  \ifx\svgwidth\undefined%
    \setlength{\unitlength}{110bp}%
    \ifx\svgscale\undefined%
      \relax%
    \else%
      \setlength{\unitlength}{\unitlength * \real{\svgscale}}%
    \fi%
  \else%
    \setlength{\unitlength}{\svgwidth}%
  \fi%
  \global\let\svgwidth\undefined%
  \global\let\svgscale\undefined%
  \makeatother%
  \begin{picture}(1,0.92799752)%
    \lineheight{1}%
    \setlength\tabcolsep{0pt}%
    \put(0.01432714,0.89481285){\color[rgb]{0,0,0}\makebox(0,0)[lt]{\lineheight{1.25}\smash{\begin{tabular}[t]{l}\textbf{$a_1$}\end{tabular}}}}%
    \put(0.01432714,0.7676378){\color[rgb]{0,0,0}\makebox(0,0)[lt]{\lineheight{1.25}\smash{\begin{tabular}[t]{l}\textbf{$x$}\end{tabular}}}}%
    \put(0.88255269,0.7676378){\color[rgb]{0,0,0}\makebox(0,0)[lt]{\lineheight{1.25}\smash{\begin{tabular}[t]{l}\textbf{$0$}\end{tabular}}}}%
    \put(0.88255269,0.89481285){\color[rgb]{0,0,0}\makebox(0,0)[lt]{\lineheight{1.25}\smash{\begin{tabular}[t]{l}\textbf{$b_1$}\end{tabular}}}}%
    \put(0,0){\includegraphics[width=\unitlength,page=1]{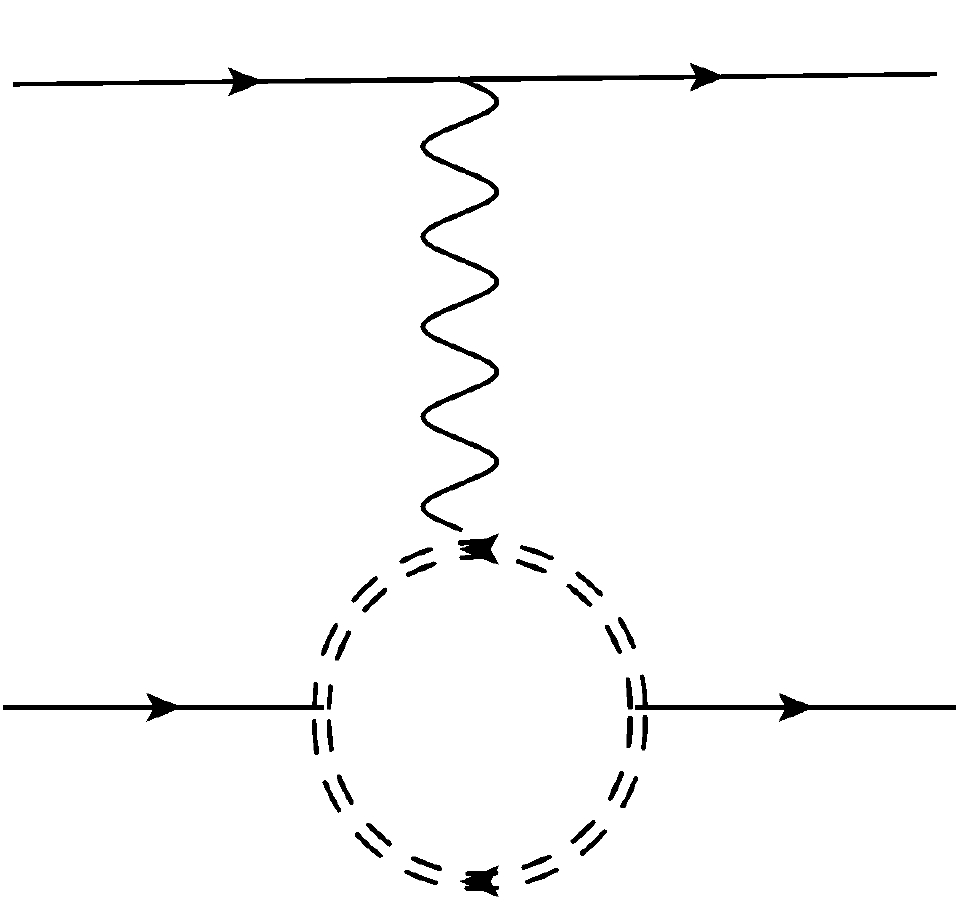}}%
    \put(0.01432714,0.25579237){\color[rgb]{0,0,0}\makebox(0,0)[lt]{\lineheight{1.25}\smash{\begin{tabular}[t]{l}\textbf{$a_2$}\end{tabular}}}}%
    \put(0.88255269,0.25579237){\color[rgb]{0,0,0}\makebox(0,0)[lt]{\lineheight{1.25}\smash{\begin{tabular}[t]{l}\textbf{$b_2$}\end{tabular}}}}%
    \put(0.01432714,0.11402905){\color[rgb]{0,0,0}\makebox(0,0)[lt]{\lineheight{1.25}\smash{\begin{tabular}[t]{l}\textbf{$x$}\end{tabular}}}}%
    \put(0.88255269,0.11402905){\color[rgb]{0,0,0}\makebox(0,0)[lt]{\lineheight{1.25}\smash{\begin{tabular}[t]{l}\textbf{$0$}\end{tabular}}}}%
  \end{picture}%
\endgroup%
}\label{w7x}\\
& = 2\!\times\! \Big(\!-\frac{1}{2}\Big) (\sqrt{2} g_0)^2 (2 g_0)^2
\big(N_f \, \tr  T^c T^{a_2} T^{b_2} -
\tr_{\text{adj}}T^c  T^{a_2} T^{b_2} \big)\,(T^c)^{a_1 b_1}\, W_7(x)~.\nonumber
\end{align}
Using the relations (\ref{3traces}), we find
\begin{equation}
\label{w7xbis}
W_7^{a_1a_2b_1b_2}(x) =  -2 g_0^4\,
\big(\ii \,f^{c a_2 b_2}
(N_f - 2 N)+ N_f \,d^{c a_2 b_2} \big)\big(-\ii\, f^{c a_1 b_1}\big)\, W_7(x)~.
\end{equation}
Defining the tensor (see (\ref{C4Ais}))
\begin{equation}
\label{defC4A}
C_{4}^{(A)\,a_1 a_2 b_1 b_2}
= -\frac{1}{N}\,f^{c\,a_1 b_1}\, f^{c\,a_2 b_2} ~,
\end{equation}
we can write
\begin{equation}
\label{w7xtris}
W_7^{a_1a_2b_1b_2}(x) =- 2 g_0^4 \big( N(2N-N_f)\,C_4^{(A)a_1 a_2 b_1 b_2} +
\ii\, N_f\, d^{a_2 b_2 c}\, f^{a_1 b_1 c} \big)\, W_7(x)~.
\end{equation}
Note that the term proportional to $d^{a_2 b_2 c} f^{a_1 b_1 c}$  is 
actually anti-symmetric in $(a_1,a_2)$ and in $(b_1, b_2)$, and thus 
it vanishes when we insert this sub-diagram in a chiral/anti-chiral correlator.
Therefore in the following we discard this term.

In momentum space, we have to compute
\begin{equation}
\label{w7}
\cW_7(p)  = ~~\parbox[c]{.40\textwidth}{
\begingroup%
  \makeatletter%
  \providecommand\color[2][]{%
    \errmessage{(Inkscape) Color is used for the text in Inkscape, but the package 'color.sty' is not loaded}%
    \renewcommand\color[2][]{}%
  }%
  \providecommand\transparent[1]{%
    \errmessage{(Inkscape) Transparency is used (non-zero) for the text in Inkscape, but the package 'transparent.sty' is not loaded}%
    \renewcommand\transparent[1]{}%
  }%
  \providecommand\rotatebox[2]{#2}%
  \newcommand*\fsize{\dimexpr\f@size pt\relax}%
  \newcommand*\lineheight[1]{\fontsize{\fsize}{#1\fsize}\selectfont}%
  \ifx\svgwidth\undefined%
    \setlength{\unitlength}{110bp}%
    \ifx\svgscale\undefined%
      \relax%
    \else%
      \setlength{\unitlength}{\unitlength * \real{\svgscale}}%
    \fi%
  \else%
    \setlength{\unitlength}{\svgwidth}%
  \fi%
  \global\let\svgwidth\undefined%
  \global\let\svgscale\undefined%
  \makeatother%
  \begin{picture}(1,0.85147601)%
    \lineheight{1}%
    \setlength\tabcolsep{0pt}%
    \put(0.3580719,0.16586013){\color[rgb]{0,0,0}\makebox(0,0)[lt]{\lineheight{1.25}\smash{\begin{tabular}[t]{l}\textbf{$3$}\end{tabular}}}}%
    \put(-0.00170167,0.7275656){\color[rgb]{0,0,0}\makebox(0,0)[lt]{\lineheight{1.25}\smash{\begin{tabular}[t]{l}\textbf{$1$}\end{tabular}}}}%
    \put(0.90266728,0.7275656){\color[rgb]{0,0,0}\makebox(0,0)[lt]{\lineheight{1.25}\smash{\begin{tabular}[t]{l}\textbf{$2$}\end{tabular}}}}%
    \put(0.58741534,0.16586013){\color[rgb]{0,0,0}\makebox(0,0)[lt]{\lineheight{1.25}\smash{\begin{tabular}[t]{l}\textbf{$4$}\end{tabular}}}}%
    \put(0.5371848,0.7275656){\color[rgb]{0,0,0}\makebox(0,0)[lt]{\lineheight{1.25}\smash{\begin{tabular}[t]{l}\textbf{$5$}\end{tabular}}}}%
    \put(0.53735183,0.36130454){\color[rgb]{0,0,0}\makebox(0,0)[lt]{\lineheight{1.25}\smash{\begin{tabular}[t]{l}\textbf{$6$}\end{tabular}}}}%
    \put(-0.00170167,0.07686013){\color[rgb]{0,0,0}\makebox(0,0)[lt]{\lineheight{1.25}\smash{\begin{tabular}[t]{l}\textbf{$1$}\end{tabular}}}}%
    \put(0.90266728,0.07686013){\color[rgb]{0,0,0}\makebox(0,0)[lt]{\lineheight{1.25}\smash{\begin{tabular}[t]{l}\textbf{$2$}\end{tabular}}}}%
    \put(0,0){\includegraphics[width=\unitlength,page=1]{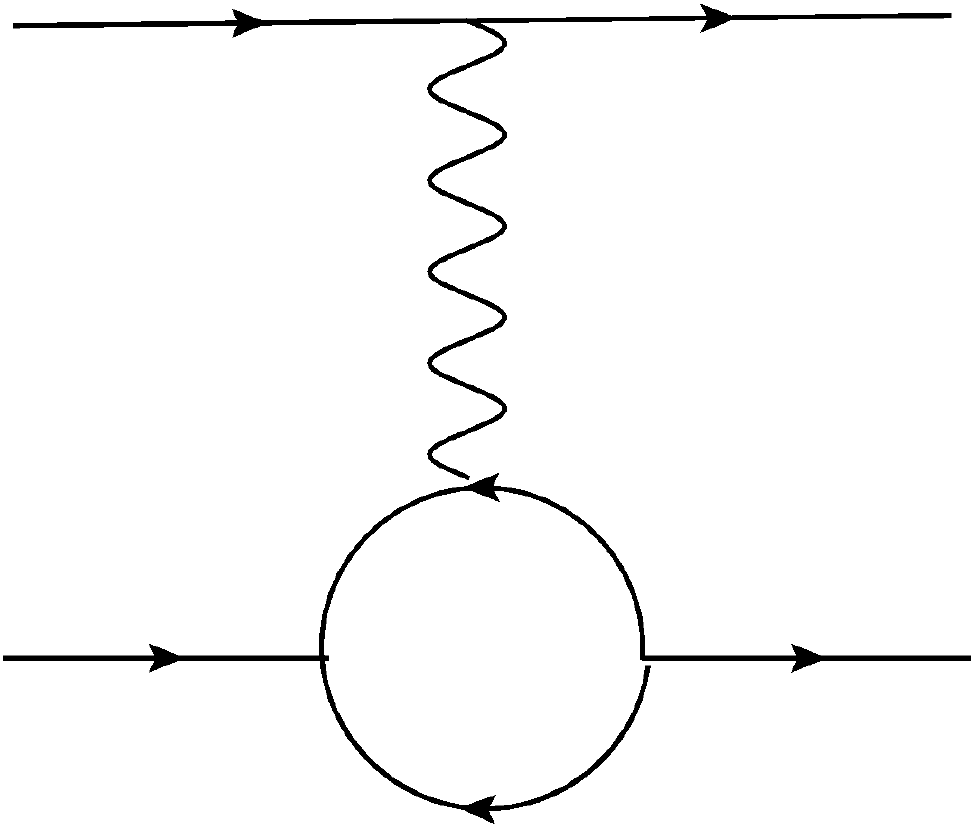}}%
  \end{picture}%
\endgroup%
}
\end{equation}
which has the canonical form (\ref{WYD}). In this case we have
\begin{equation}
\label{w7-2}
\begin{aligned}
\widetilde{\mathfrak D}_7 & = \exp\left(
-\theta_6(k_{63} + k_{46} + k_{15}+ k_{52})\,\bar{\theta}_6
+ 2  \theta_6 k_{63}\, \bar{\theta}_3 + 2 \theta_4 k_{46}\, \bar{\theta}_6 
+ 2 \theta_4 k_{43}\,\bar{\theta}_3 \right)\\[2mm]
& = \exp\left(- 2 \theta_6(k_{46}+ k_{15})\,\bar{\theta}_6 
+ 2  \theta_6 k_{63}\, \bar{\theta}_3 + 2 \theta_4 k_{46} \,
\bar{\theta}_6 + 2 \theta_4 k_{43}\,\bar{\theta}_3 \right) ~,
\end{aligned}
\end{equation}
where in the second step we used momentum conservation. To perform the Grassmann integral
\begin{equation}
\label{w7-3}
\cZ_7(p,k)=
\int d^2\bar{\theta}_3\, d^2\theta_4 \,d^2\theta_6\, d^2\bar{\theta}_6 \,\, 
\widetilde{\mathfrak D}_7 ~,
\end{equation}
we use the formula
\begin{equation}
\label{4trace}
\int d^2\theta_i \,d^2\bar{\theta}_j \,d^2\theta_k \,d^2\bar{\theta}_l \,
\,\eee^{2\theta_i A \bar{\theta}_j + 2 \theta_i B \bar{\theta}_l + 2 \theta_k C \bar{\theta}_l
	+ 2 \theta_k D \bar{\theta}_j} 
=  A^2 C^2 + B^2 D^2 - \tr \big(A D C B\big)
\end{equation}
where
\begin{equation}
\begin{aligned}
\tr \big(A D C B\big) &= \tr \big(\sigma^\mu\,\bar{\sigma}^\nu\,\sigma^\lambda\,\bar{\sigma}^\rho\big)\,A_\mu\,D_\nu\,C_\lambda\,B_\rho\\
&=2\,A\!\cdot\! D\,\,C\!\cdot\! B-2A\!\cdot\! C\,\,D\!\cdot\! B+2\,A\!\cdot\! B\,\,D\!\cdot\! C-\varepsilon^{\mu\nu\lambda\rho}\,
A_\mu\,D_\nu\,C_\lambda\,B_\rho~.
\end{aligned}
\label{trace4sigma}
\end{equation}
In this way we obtain
\begin{align}
\cZ_7(p,k)
& =
(k_{46} + k_{15})^2\, k_{43}^2 + k_{63}^2\, k_{46}^2 + 
\tr\big((k_{46} + k_{15})\, k_{46}\, k_{43}\, k_{63}\big)\nonumber\\[2mm]
& = (k_{46} + k_{15})^2 \,k_{43}^2 + 2 \big((k_{46} + k_{15})\cdot k_{46}\big)
\,(k_{43}\cdot k_{63})\label{w7-3bis}
\\[2mm]
& ~~~- 2 \big((k_{46} + k_{15})\cdot k_{43}\big)\,(k_{46}\cdot k_{63})
+ 2 \big((k_{46} + k_{15})\cdot k_{63}\big)\,(k_{43}\cdot k_{46})~,
\nonumber
\end{align}
where we have discarded a term proportional to the anti-symmetric 
$\varepsilon$-tensor coming from the trace of four Pauli matrices given in (\ref{trace4sigma}), 
that will not contribute to the correlator for symmetry reasons.
Using momentum conservation, after some algebra the polynomial $\cZ_7(p,k)$
can be rewritten as
\begin{align}
\label{w7-4}
&~(p\cdot k_{43})\,(k_{56}^2 - k_{13}^2 - k_{42}^2) 
-\frac{p^2}{2} \,(k_{13}^2 + k_{42}^2)
\nonumber\\
&+ \frac 12 \,(k_{15}^2 k_{43}^2 - k_{13}^2 k_{43}^2) 
+ \frac 12 \,(k_{52}^2 k_{43}^2 - k_{42}^2 k_{43}^2)
+ \frac 12 \,(k_{13}^2 k_{46}^2 - k_{15}^2 k_{46}^2)+
\frac 12 \,(k_{42}^2 k_{63}^2 - k_{52}^2 k_{63}^2)\nonumber\\
& + 
\frac{p^2}{2}\,(k_{63}^2 + k_{46}^2)+ \frac 12 k_{52}^2 k_{13}^2  + \frac 12 k_{15}^2 k_{42}^2~.
\end{align}
This polynomial has to be multiplied by the factor $\cY_7$ containing all propagators and then integrated over the momenta. It is not difficult to show that the terms
in the first line, proportional to $p\cdot k_{43}$ and to $p^2 k_{13}^2$ or $p^2 k_{42}^2$, yield contributions that vanish for $\epsilon \to 0$ after Fourier 
transform. The terms in the second line in each brackets cancel each other 
owing to the symmetries of the diagram. The remaining terms in the third line of (\ref{w7-4}) give
a non-vanishing contributions. Thus, we can effectively use
\begin{equation}
\cZ_7(p,k)=\frac{p^2}{2}\,(k_{63}^2 + k_{46}^2)+ \frac 12 k_{52}^2 k_{13}^2  + \frac 12 k_{15}^2 k_{42}^2~.
\end{equation}
All these terms lead to cancellations of some of the propagators of $\cY_7$ and the result
can be written in the graphical notation of Appendix~\ref{app:loop_integrals}.
Altogether we find,
\begin{equation}
\label{w7-5}
\cW_7(p)  =\,
p^2\,{\parbox[c]{.08\textwidth}{ \includegraphics[width = .08\textwidth]{Ydot9}}} \,+
\,{\parbox[c]{.09\textwidth}{ \includegraphics[width = .09\textwidth]{Ydot5}}}
=\,4\,{\parbox[c]{.09\textwidth}{ \includegraphics[width = .09\textwidth]{Ydot5}}}+\cdots
\end{equation}
where the last step follows from (\ref{int4bis}). Using this result, we find that the momentum
space expression corresponding to $W_7^{a_1a_2b_1b_2}(x)$ given in (\ref{w7xtris}) is
\begin{equation}
\cW_7^{a_1a_2b_1b_2}(p) = 
-8 g_0^4 \, N(2N-N_f) \,{\parbox[c]{.09\textwidth}{ \includegraphics[width = .09\textwidth]{Ydot5}}}
\,C_4^{(A)a_1 a_2 b_1 b_2}+\cdots
\label{w7-6}
\end{equation}
where the dots stand for terms that do not contribute in the correlators due to their colour factors or
that vanish for $\epsilon\to 0$.

The second diagram we have to consider is 
\begin{equation}
\label{w8x}
\begin{aligned}
W_8^{a_1a_2b_1b_2}(x)  & \equiv ~~\parbox[c]{.25\textwidth}{
\begingroup%
  \makeatletter%
  \providecommand\color[2][]{%
    \errmessage{(Inkscape) Color is used for the text in Inkscape, but the package 'color.sty' is not loaded}%
    \renewcommand\color[2][]{}%
  }%
  \providecommand\transparent[1]{%
    \errmessage{(Inkscape) Transparency is used (non-zero) for the text in Inkscape, but the package 'transparent.sty' is not loaded}%
    \renewcommand\transparent[1]{}%
  }%
  \providecommand\rotatebox[2]{#2}%
  \newcommand*\fsize{\dimexpr\f@size pt\relax}%
  \newcommand*\lineheight[1]{\fontsize{\fsize}{#1\fsize}\selectfont}%
  \ifx\svgwidth\undefined%
    \setlength{\unitlength}{110bp}%
    \ifx\svgscale\undefined%
      \relax%
    \else%
      \setlength{\unitlength}{\unitlength * \real{\svgscale}}%
    \fi%
  \else%
    \setlength{\unitlength}{\svgwidth}%
  \fi%
  \global\let\svgwidth\undefined%
  \global\let\svgscale\undefined%
  \makeatother%
  \begin{picture}(1,0.94286205)%
    \lineheight{1}%
    \setlength\tabcolsep{0pt}%
    \put(-0.00221756,0.90912036){\color[rgb]{0,0,0}\makebox(0,0)[lt]{\lineheight{1.25}\smash{\begin{tabular}[t]{l}\textbf{$a_1$}\end{tabular}}}}%
    \put(-0.00221756,0.77981066){\color[rgb]{0,0,0}\makebox(0,0)[lt]{\lineheight{1.25}\smash{\begin{tabular}[t]{l}\textbf{$x$}\end{tabular}}}}%
    \put(0.88058132,0.77981066){\color[rgb]{0,0,0}\makebox(0,0)[lt]{\lineheight{1.25}\smash{\begin{tabular}[t]{l}\textbf{$0$}\end{tabular}}}}%
    \put(0.88058132,0.90912036){\color[rgb]{0,0,0}\makebox(0,0)[lt]{\lineheight{1.25}\smash{\begin{tabular}[t]{l}\textbf{$b_1$}\end{tabular}}}}%
    \put(-0.00221756,0.25937381){\color[rgb]{0,0,0}\makebox(0,0)[lt]{\lineheight{1.25}\smash{\begin{tabular}[t]{l}\textbf{$a_2$}\end{tabular}}}}%
    \put(0.88058132,0.25937381){\color[rgb]{0,0,0}\makebox(0,0)[lt]{\lineheight{1.25}\smash{\begin{tabular}[t]{l}\textbf{$b_2$}\end{tabular}}}}%
    \put(-0.00221756,0.11523096){\color[rgb]{0,0,0}\makebox(0,0)[lt]{\lineheight{1.25}\smash{\begin{tabular}[t]{l}\textbf{$x$}\end{tabular}}}}%
    \put(0.88058132,0.11523096){\color[rgb]{0,0,0}\makebox(0,0)[lt]{\lineheight{1.25}\smash{\begin{tabular}[t]{l}\textbf{$0$}\end{tabular}}}}%
    \put(0,0){\includegraphics[width=\unitlength,page=1]{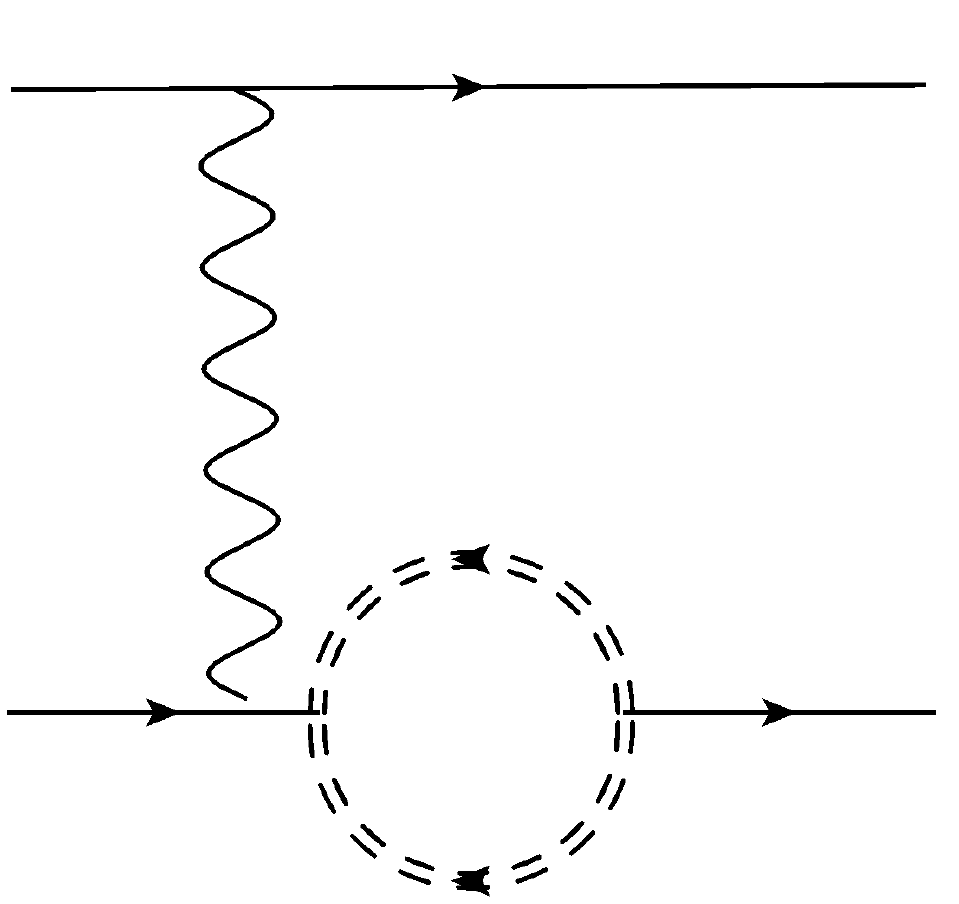}}%
  \end{picture}%
\endgroup%
}\\
& = 2\times \Big(\!-\frac{1}{2}\Big) (2 g_0)^2\, g_0^2\, (N_f - 2 N) \,(T^c)^{a_1 b_1} 
(T^c)^{a_2 b_2}\, W_8(x)\\[2mm]
& = 4 g_0^2\, N(2N-N_f)\, C_4^{(A)a_1 a_2 b_1 b_2}\,  W_8(x)~.
\end{aligned}
\end{equation}
In momentum space, we must compute
\begin{equation}
\label{w8}
\cW_8(p)  =~~ \parbox[c]{.40\textwidth}{
\begingroup%
  \makeatletter%
  \providecommand\color[2][]{%
    \errmessage{(Inkscape) Color is used for the text in Inkscape, but the package 'color.sty' is not loaded}%
    \renewcommand\color[2][]{}%
  }%
  \providecommand\transparent[1]{%
    \errmessage{(Inkscape) Transparency is used (non-zero) for the text in Inkscape, but the package 'transparent.sty' is not loaded}%
    \renewcommand\transparent[1]{}%
  }%
  \providecommand\rotatebox[2]{#2}%
  \newcommand*\fsize{\dimexpr\f@size pt\relax}%
  \newcommand*\lineheight[1]{\fontsize{\fsize}{#1\fsize}\selectfont}%
  \ifx\svgwidth\undefined%
    \setlength{\unitlength}{110bp}%
    \ifx\svgscale\undefined%
      \relax%
    \else%
      \setlength{\unitlength}{\unitlength * \real{\svgscale}}%
    \fi%
  \else%
    \setlength{\unitlength}{\svgwidth}%
  \fi%
  \global\let\svgwidth\undefined%
  \global\let\svgscale\undefined%
  \makeatother%
  \begin{picture}(1,0.85608863)%
    \lineheight{1}%
    \setlength\tabcolsep{0pt}%
    \put(0.30289471,0.73872364){\color[rgb]{0,0,0}\makebox(0,0)[lt]{\lineheight{1.25}\smash{\begin{tabular}[t]{l}\textbf{$5$}\end{tabular}}}}%
    \put(-0.00177644,0.73872364){\color[rgb]{0,0,0}\makebox(0,0)[lt]{\lineheight{1.25}\smash{\begin{tabular}[t]{l}\textbf{$1$}\end{tabular}}}}%
    \put(0.95057186,0.73872364){\color[rgb]{0,0,0}\makebox(0,0)[lt]{\lineheight{1.25}\smash{\begin{tabular}[t]{l}\textbf{$2$}\end{tabular}}}}%
    \put(0.22564056,0.07942694){\color[rgb]{0,0,0}\makebox(0,0)[lt]{\lineheight{1.25}\smash{\begin{tabular}[t]{l}\textbf{$6$}\end{tabular}}}}%
    \put(0.37271524,0.12942694){\color[rgb]{0,0,0}\makebox(0,0)[lt]{\lineheight{1.25}\smash{\begin{tabular}[t]{l}\textbf{$3$}\end{tabular}}}}%
    \put(0.59146947,0.12942694){\color[rgb]{0,0,0}\makebox(0,0)[lt]{\lineheight{1.25}\smash{\begin{tabular}[t]{l}\textbf{$4$}\end{tabular}}}}%
    \put(-0.00177644,0.07942694){\color[rgb]{0,0,0}\makebox(0,0)[lt]{\lineheight{1.25}\smash{\begin{tabular}[t]{l}\textbf{$1$}\end{tabular}}}}%
    \put(0.95057186,0.07942694){\color[rgb]{0,0,0}\makebox(0,0)[lt]{\lineheight{1.25}\smash{\begin{tabular}[t]{l}\textbf{$2$}\end{tabular}}}}%
    \put(0,0){\includegraphics[width=\unitlength,page=1]{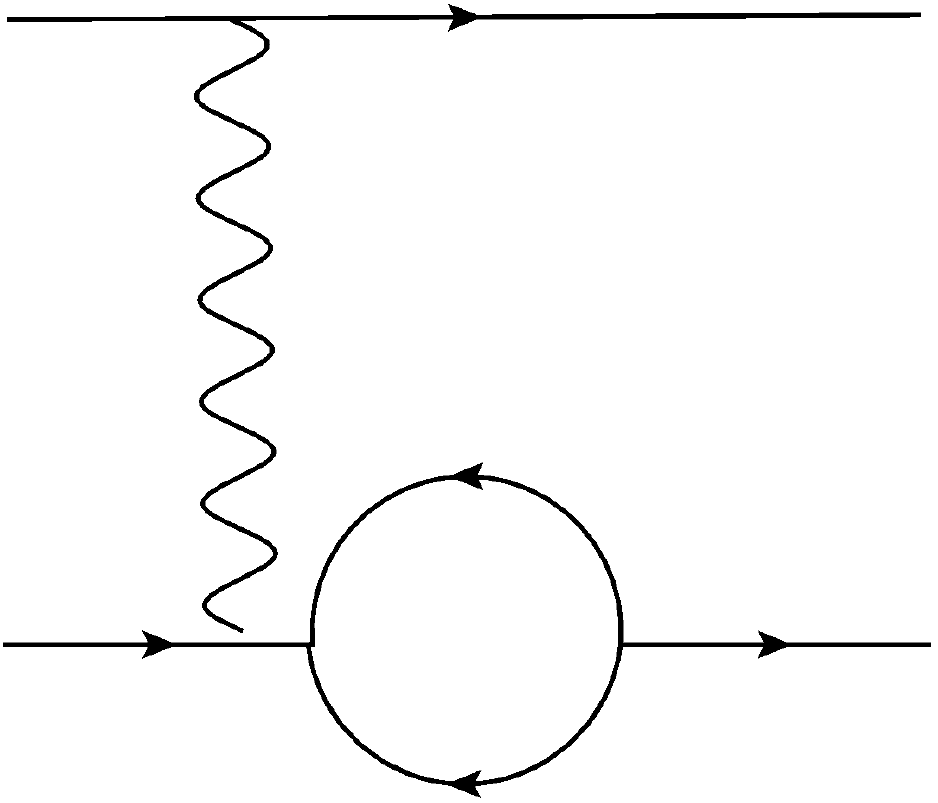}}%
  \end{picture}%
\endgroup%
}
\end{equation}
which again is of the form (\ref{WYD}).
In this case we have
\begin{equation}
\begin{aligned}
\label{w8-2}
\widetilde{\mathfrak D} & = \exp\left(
-\theta_6\left(k_{15} + k_{16} + k_{52} + k_{63}\right)\bar{\theta}_6
+ 2 \theta_6\, k_{63}\, \bar{\theta}_3 + 2 \theta_4\, k_{63}\, \bar{\theta}_3\right)\\[2mm]
& = \exp\left(- 2 \theta_6 \,p\, \bar{\theta}_6 + 2 \theta_6\, k_{63}\, \bar{\theta}_3 + 2 \theta_4 \,k_{63}\, \bar{\theta}_3\right)~,
\end{aligned}
\end{equation}
while the Grassmann integration yields $\cZ_8(p,k)= p^2 \,k_{63}^2$.
Inserting this into the momentum integrals, we remain with
\begin{equation}
\label{w8-5}
\cW_8(p) = p^2\,{\parbox[c]{.08\textwidth}{ \includegraphics[width = .08\textwidth]{Ydot9}}} =
\,3\,{\parbox[c]{.09\textwidth}{ \includegraphics[width = .09\textwidth]{Ydot5}}}+\cdots
\end{equation}
where the last step follows from (\ref{int4bis}). Using this result, we find that the momentum
space expression corresponding to $W_8^{a_1a_2b_1b_2}(x)$ given in (\ref{w8x}) is
\begin{equation}
\cW_8^{a_1a_2b_1b_2}(p) = 
12 g_0^4 \, N(2N-N_f) \,{\parbox[c]{.09\textwidth}{ \includegraphics[width = .09\textwidth]{Ydot5}}}
\,C_4^{(A)a_1 a_2 b_1 b_2}+\cdots
\label{w8-6}
\end{equation}
where the dots stand for terms that vanish for $\epsilon\to 0$.

The third diagram we need to consider is
\begin{equation}
\begin{aligned}
\label{w9x}
W_9^{a_1a_2b_1b_2}(x)  & \equiv ~~\parbox[c]{.29\textwidth}{
\begingroup%
  \makeatletter%
  \providecommand\color[2][]{%
    \errmessage{(Inkscape) Color is used for the text in Inkscape, but the package 'color.sty' is not loaded}%
    \renewcommand\color[2][]{}%
  }%
  \providecommand\transparent[1]{%
    \errmessage{(Inkscape) Transparency is used (non-zero) for the text in Inkscape, but the package 'transparent.sty' is not loaded}%
    \renewcommand\transparent[1]{}%
  }%
  \providecommand\rotatebox[2]{#2}%
  \newcommand*\fsize{\dimexpr\f@size pt\relax}%
  \newcommand*\lineheight[1]{\fontsize{\fsize}{#1\fsize}\selectfont}%
  \ifx\svgwidth\undefined%
    \setlength{\unitlength}{110bp}%
    \ifx\svgscale\undefined%
      \relax%
    \else%
      \setlength{\unitlength}{\unitlength * \real{\svgscale}}%
    \fi%
  \else%
    \setlength{\unitlength}{\svgwidth}%
  \fi%
  \global\let\svgwidth\undefined%
  \global\let\svgscale\undefined%
  \makeatother%
  \begin{picture}(1,0.83197155)%
    \lineheight{1}%
    \setlength\tabcolsep{0pt}%
    \put(-0.00173712,0.79824605){\color[rgb]{0,0,0}\makebox(0,0)[lt]{\lineheight{1.25}\smash{\begin{tabular}[t]{l}\textbf{$a_1$}\end{tabular}}}}%
    \put(-0.00173712,0.66899833){\color[rgb]{0,0,0}\makebox(0,0)[lt]{\lineheight{1.25}\smash{\begin{tabular}[t]{l}\textbf{$x$}\end{tabular}}}}%
    \put(0.88063856,0.66899833){\color[rgb]{0,0,0}\makebox(0,0)[lt]{\lineheight{1.25}\smash{\begin{tabular}[t]{l}\textbf{$0$}\end{tabular}}}}%
    \put(0.88063856,0.79824605){\color[rgb]{0,0,0}\makebox(0,0)[lt]{\lineheight{1.25}\smash{\begin{tabular}[t]{l}\textbf{$b_1$}\end{tabular}}}}%
    \put(-0.00173712,0.14881096){\color[rgb]{0,0,0}\makebox(0,0)[lt]{\lineheight{1.25}\smash{\begin{tabular}[t]{l}\textbf{$a_2$}\end{tabular}}}}%
    \put(0.88063856,0.14881096){\color[rgb]{0,0,0}\makebox(0,0)[lt]{\lineheight{1.25}\smash{\begin{tabular}[t]{l}\textbf{$b_2$}\end{tabular}}}}%
    \put(-0.00173712,0.00473722){\color[rgb]{0,0,0}\makebox(0,0)[lt]{\lineheight{1.25}\smash{\begin{tabular}[t]{l}\textbf{$x$}\end{tabular}}}}%
    \put(0.88063856,0.00473722){\color[rgb]{0,0,0}\makebox(0,0)[lt]{\lineheight{1.25}\smash{\begin{tabular}[t]{l}\textbf{$0$}\end{tabular}}}}%
    \put(0,0){\includegraphics[width=\unitlength,page=1]{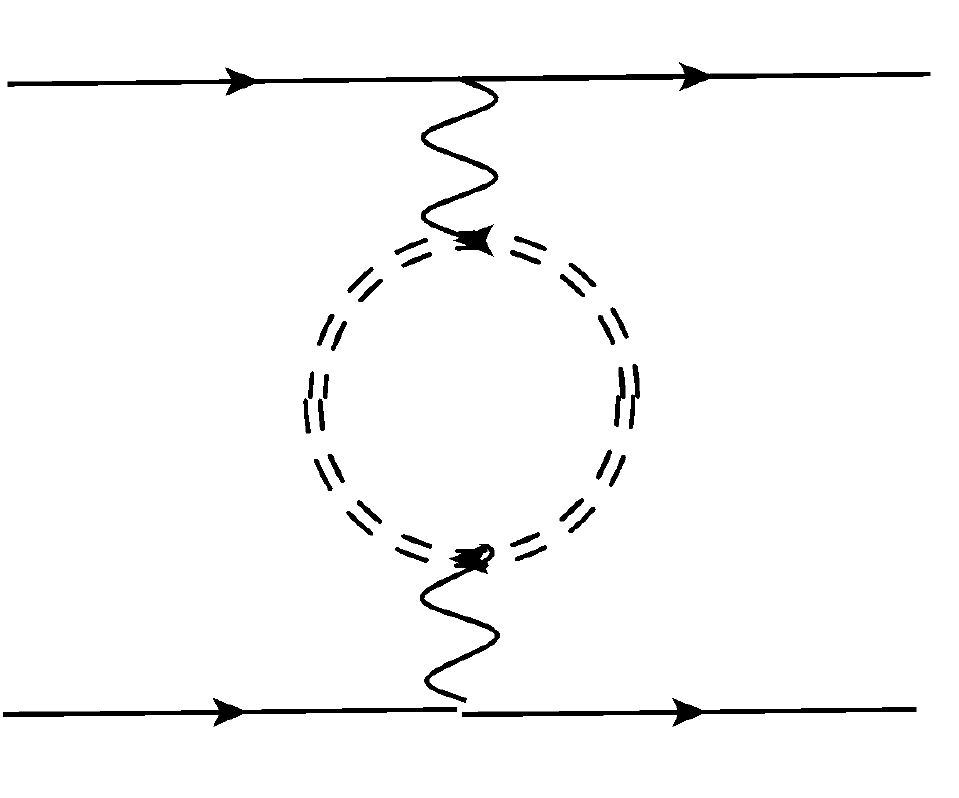}}%
  \end{picture}%
\endgroup%
}\nonumber\\
& = 2\!\times\!\Big(\!-\frac{1}{2}\Big)^2 (2 g_0)^2 \,g_0^2\,(N_f - 2 N) (T^c)_{a_1 b_1} (T^c)_{a_2 b_2}\, W_9(x)\nonumber\\[2mm]
& = -2 g_0^4\, N(2N-N_f)\, C_4^{(A)a_1 a_2 b_1 b_2} \, W_9(x)~.
\end{aligned}
\end{equation}
In momentum space, we have to compute
\begin{equation}
\label{w9}
\cW_9(p)  = ~~\parbox[c]{.40\textwidth}{
\begingroup%
  \makeatletter%
  \providecommand\color[2][]{%
    \errmessage{(Inkscape) Color is used for the text in Inkscape, but the package 'color.sty' is not loaded}%
    \renewcommand\color[2][]{}%
  }%
  \providecommand\transparent[1]{%
    \errmessage{(Inkscape) Transparency is used (non-zero) for the text in Inkscape, but the package 'transparent.sty' is not loaded}%
    \renewcommand\transparent[1]{}%
  }%
  \providecommand\rotatebox[2]{#2}%
  \newcommand*\fsize{\dimexpr\f@size pt\relax}%
  \newcommand*\lineheight[1]{\fontsize{\fsize}{#1\fsize}\selectfont}%
  \ifx\svgwidth\undefined%
    \setlength{\unitlength}{110bp}%
    \ifx\svgscale\undefined%
      \relax%
    \else%
      \setlength{\unitlength}{\unitlength * \real{\svgscale}}%
    \fi%
  \else%
    \setlength{\unitlength}{\svgwidth}%
  \fi%
  \global\let\svgwidth\undefined%
  \global\let\svgscale\undefined%
  \makeatother%
  \begin{picture}(1,0.79959963)%
    \lineheight{1}%
    \setlength\tabcolsep{0pt}%
    \put(0.44162375,0.52794249){\color[rgb]{0,0,0}\makebox(0,0)[lt]{\lineheight{1.25}\smash{\begin{tabular}[t]{l}\textbf{$5$}\end{tabular}}}}%
    \put(-0.00177644,0.68667906){\color[rgb]{0,0,0}\makebox(0,0)[lt]{\lineheight{1.25}\smash{\begin{tabular}[t]{l}\textbf{$1$}\end{tabular}}}}%
    \put(0.90057186,0.68667906){\color[rgb]{0,0,0}\makebox(0,0)[lt]{\lineheight{1.25}\smash{\begin{tabular}[t]{l}\textbf{$2$}\end{tabular}}}}%
    \put(0.44162375,0.33742692){\color[rgb]{0,0,0}\makebox(0,0)[lt]{\lineheight{1.25}\smash{\begin{tabular}[t]{l}\textbf{$6$}\end{tabular}}}}%
    \put(0.55992188,0.68667906){\color[rgb]{0,0,0}\makebox(0,0)[lt]{\lineheight{1.25}\smash{\begin{tabular}[t]{l}\textbf{$3$}\end{tabular}}}}%
    \put(0.46174503,-0.00484444){\color[rgb]{0,0,0}\makebox(0,0)[lt]{\lineheight{1.25}\smash{\begin{tabular}[t]{l}\textbf{$4$}\end{tabular}}}}%
    \put(-0.00177644,-0.00484444){\color[rgb]{0,0,0}\makebox(0,0)[lt]{\lineheight{1.25}\smash{\begin{tabular}[t]{l}\textbf{$1$}\end{tabular}}}}%
    \put(0.90057186,-0.00484444){\color[rgb]{0,0,0}\makebox(0,0)[lt]{\lineheight{1.25}\smash{\begin{tabular}[t]{l}\textbf{$2$}\end{tabular}}}}%
    \put(0,0){\includegraphics[width=\unitlength,page=1]{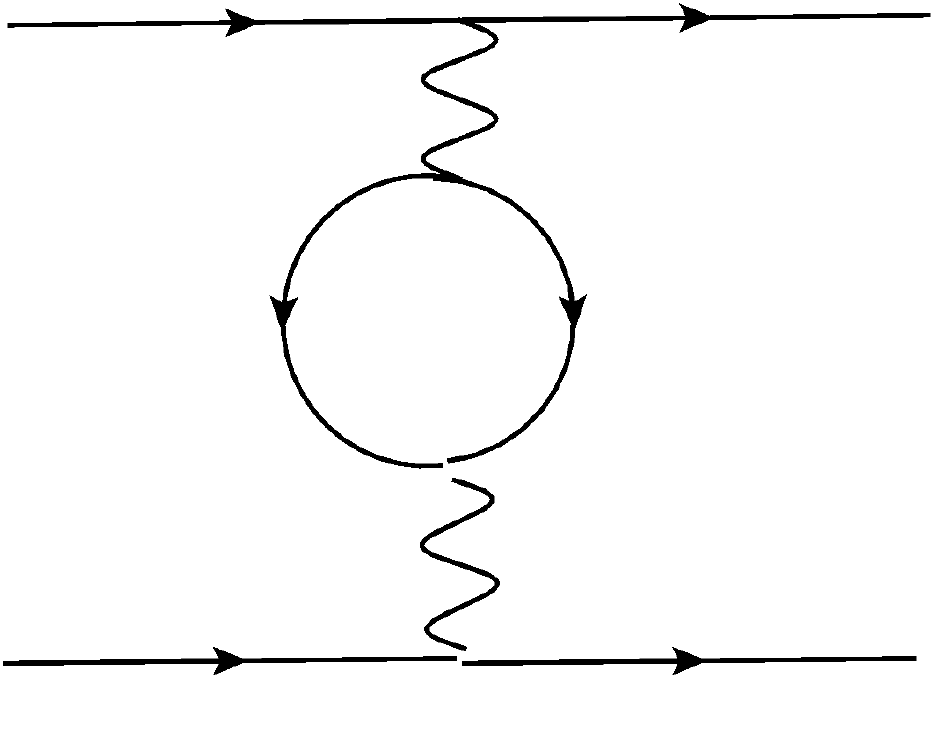}}%
  \end{picture}%
\endgroup%
}
\end{equation}
which has again the form (\ref{WYD}) with
\begin{equation}\label{w9-2}
\begin{aligned}
\widetilde{\mathfrak D}_9 & = \exp \bigl(
-\theta_3\left(k_{13} + k_{32} + k_{56} + k_{65}\right)\bar{\theta}_3
-\theta_4\left(k_{14} + k_{42} + k_{56} + k_{65}\right)\bar{\theta}_4
\\[1mm]
& \phantom{= \exp( }\,
+ 2 \theta_3\, k_{56} \,\bar{\theta}_4 + 2 \theta_4 \,k_{65}\, \bar{\theta_3}
\bigr)
\\[1mm]
& = \exp\left(- 2 \theta_3\left(k_{13} + k_{65}\right)\bar{\theta}_3
- 2 \theta_4\left(k_{14} + k_{56}\right)\bar{\theta}_4
+ 2 \theta_3\, k_{56}\, \bar{\theta}_4 + 2 \theta_4\, k_{65}\, \bar{\theta_3}
\right)~.
\end{aligned}
\end{equation}
The Grassmann integration is carried out using (\ref{4trace}) and gives
\begin{equation}
\label{w9-3}
\cZ_9(p,k)  = (k_{13} + k_{65})^2 (k_{14} + k_{56})^2 + k_{56}^2 k_{65}^2
- \tr \left((k_{13} + k_{65}) k_{65} (k_{14} + k_{56}) k_{56}\right)~. 
\end{equation}
We expand the trace according to (\ref{trace4sigma}), and take into account the part 
proportional to the $\varepsilon$-tensor does not contribute. The terms proportional to
$k_{56}^2$ and/or to $k_{65}^2$, as well as the term $k_{13}^2 k_{14}^2$, are tadpole-like and vanish in dimensional regularization, and thus we remain with
\begin{equation}
\begin{aligned}
\label{w9-4}
\cZ_9(p,k)=&\,
2 k_{14}^2 (k_{65}\cdot k_{13}) + 2 k_{13}^2 (k_{56}\cdot k_{14})
- 2 (k_{13}\cdot k_{65})(k_{14}\cdot k_{56}) 
\\[1mm]
&+ 2 (k_{13}\cdot k_{56})(k_{14}\cdot k_{65})+ 2 (k_{13}\cdot  k_{14})(k_{56}\cdot k_{65})~.
\end{aligned}
\end{equation}
The diagram is symmetric under the exchange $k_{56} \leftrightarrow k_{65}$. Symmetrizing 
$\cZ_9(p,k)$ with respect to this exchange, exploiting momentum conservation and discarding tadpole-like
terms proportional to $k_{56}^2$, $k_{65}^2$ or $k_{13}^2 k_{14}^2$ we can rewrite (\ref{w9-4}) 
as
\begin{equation}
\begin{aligned}
\label{w9-5}
\cZ_9(p,k)&=
- k_{14}^2 (k_{35}\cdot k_{13}) + k_{13}^2 (k_{64}\cdot k_{14}) - k_{64}^2  (k_{13}\cdot k_{14})
\\
&= \frac 12 k_{14}^2\, k_{32}^2 + \frac 12 k_{13}^2 \,k_{42}^2 - \frac 12 p^2\, k_{64}^2~.
\end{aligned}
\end{equation}
The first two terms in the last expression give the same result and cancel two internal propagators, while the last term cancels one external and one internal propagator of $\cY_9$. In the end, adopting
the graphical notation of Appendix~\ref{app:loop_integrals}, we have
\begin{align}
\label{w9-6}
\cW_9(p) =  {\parbox[c]{.10\textwidth}{ \includegraphics[width = .10\textwidth]{Ydot7.jpg}}}
- \,\frac 12\, p^2 {\parbox[c]{.10\textwidth}{ \includegraphics[width = .10\textwidth]{Ydot8}}} =
{\parbox[c]{.10\textwidth}{ \includegraphics[width = .10\textwidth]{Ydot7.jpg}}}
- \,\frac 32\,{\parbox[c]{.09\textwidth}{ \includegraphics[width = .09\textwidth]{Ydot5}}}
+\cdots
\end{align}
where the second step follows from (\ref{int51app}). Inserting this result in (\ref{w9x}), we see that
the momentum space expression corresponding to $W_9^{a_1a_2b_1b_2}(x)$ is
\begin{equation}
\label{w9-9}
\cW_9^{a_1a_2b_1b_2}(p) = g_0^4\, N(2N-N_f)
\bigg[-2  {\parbox[c]{.10\textwidth}{ \includegraphics[width = .10\textwidth]{Ydot7.jpg}}}
+3\, {\parbox[c]{.09\textwidth}{ \includegraphics[width = .09\textwidth]{Ydot5}}}
\bigg] C_4^{(A)a_1 a_2 b_1 b_2}+\cdots~.
\end{equation}
Summing the three diagrams (\ref{w7-6}), (\ref{w8-6}) and
(\ref{w9-9}), we find
\begin{equation}
\label{w789p}
\sum_{I=7}^9 \cW_I^{a_1a_2b_1b_2}(p) = g_0^4\, N(2N-N_f)
\bigg[7\, {\parbox[c]{.09\textwidth}{ \includegraphics[width = .09\textwidth]{Ydot5}}}
-2  {\parbox[c]{.10\textwidth}{ \includegraphics[width = .10\textwidth]{Ydot7.jpg}}}
\bigg] C_4^{(A)a_1 a_2 b_1 b_2}+\cdots~.
\end{equation}
Performing the Fourier transform using (\ref{int2x}) and (\ref{int51x}), we finally obtain
\begin{equation}
\sum_{I=7}^9 W_I^{a_1a_2b_1b_2}(x) =  
v_{4,2}^{(A)} \,\Delta(x)^2 \, C_{4}^{(A)\,a_1 a_2 b_1 b_2}+ \cdots
\label{w789x}
\end{equation}
with 
\begin{equation}
v_{4,2}^{(A)} =\left(\frac{g_0^2}{8\pi^2}\right)^2 N(2N-N_f)
\left[\frac{21}{2} \zeta(3) + \frac{\Gamma^2(1-\epsilon)}{4\epsilon^2(1-2\epsilon)(1 + \epsilon)}\right]
(\pi x^2)^{2\epsilon}+\ldots
\label{v42Aapp}
\end{equation}
in agreement with the formula (\ref{v42Ares}) of the main text.

The last two-loop diagram we have to compute is
\begin{align}
\label{w10x}
W_{10}^{a_1a_2b_1b_2}(x)  & \equiv~~ \parbox[c]{.29\textwidth}{
\begingroup%
  \makeatletter%
  \providecommand\color[2][]{%
    \errmessage{(Inkscape) Color is used for the text in Inkscape, but the package 'color.sty' is not loaded}%
    \renewcommand\color[2][]{}%
  }%
  \providecommand\transparent[1]{%
    \errmessage{(Inkscape) Transparency is used (non-zero) for the text in Inkscape, but the package 'transparent.sty' is not loaded}%
    \renewcommand\transparent[1]{}%
  }%
  \providecommand\rotatebox[2]{#2}%
  \newcommand*\fsize{\dimexpr\f@size pt\relax}%
  \newcommand*\lineheight[1]{\fontsize{\fsize}{#1\fsize}\selectfont}%
  \ifx\svgwidth\undefined%
    \setlength{\unitlength}{110bp}%
    \ifx\svgscale\undefined%
      \relax%
    \else%
      \setlength{\unitlength}{\unitlength * \real{\svgscale}}%
    \fi%
  \else%
    \setlength{\unitlength}{\svgwidth}%
  \fi%
  \global\let\svgwidth\undefined%
  \global\let\svgscale\undefined%
  \makeatother%
  \begin{picture}(1,0.80997038)%
    \lineheight{1}%
    \setlength\tabcolsep{0pt}%
    \put(0.02475344,0.77713673){\color[rgb]{0,0,0}\makebox(0,0)[lt]{\lineheight{1.25}\smash{\begin{tabular}[t]{l}\textbf{$a_1$}\end{tabular}}}}%
    \put(0.02475344,0.65130692){\color[rgb]{0,0,0}\makebox(0,0)[lt]{\lineheight{1.25}\smash{\begin{tabular}[t]{l}\textbf{$x$}\end{tabular}}}}%
    \put(0.88379503,0.65130692){\color[rgb]{0,0,0}\makebox(0,0)[lt]{\lineheight{1.25}\smash{\begin{tabular}[t]{l}\textbf{$0$}\end{tabular}}}}%
    \put(0.88379503,0.77713673){\color[rgb]{0,0,0}\makebox(0,0)[lt]{\lineheight{1.25}\smash{\begin{tabular}[t]{l}\textbf{$b_1$}\end{tabular}}}}%
    \put(0.02475344,0.14487571){\color[rgb]{0,0,0}\makebox(0,0)[lt]{\lineheight{1.25}\smash{\begin{tabular}[t]{l}\textbf{$b_2$}\end{tabular}}}}%
    \put(0.88379503,0.14487571){\color[rgb]{0,0,0}\makebox(0,0)[lt]{\lineheight{1.25}\smash{\begin{tabular}[t]{l}\textbf{$a_2$}\end{tabular}}}}%
    \put(0.02475344,0.00461194){\color[rgb]{0,0,0}\makebox(0,0)[lt]{\lineheight{1.25}\smash{\begin{tabular}[t]{l}\textbf{$0$}\end{tabular}}}}%
    \put(0.88379503,0.00461194){\color[rgb]{0,0,0}\makebox(0,0)[lt]{\lineheight{1.25}\smash{\begin{tabular}[t]{l}\textbf{$x$}\end{tabular}}}}%
    \put(0,0){\includegraphics[width=\unitlength,page=1]{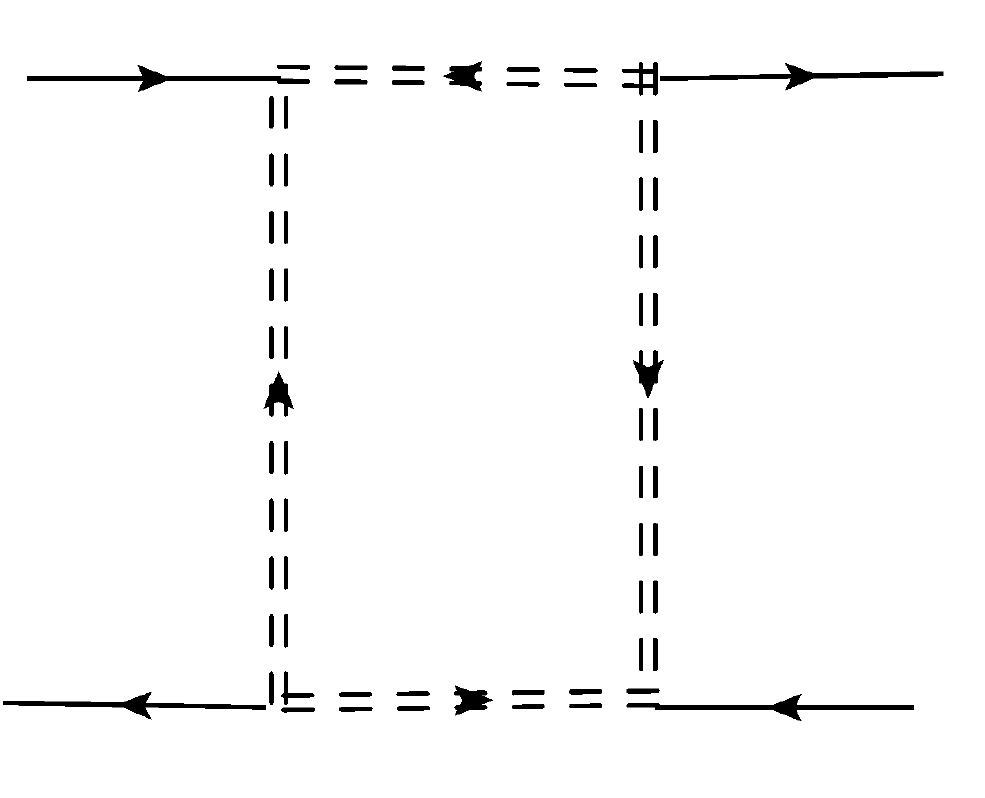}}%
  \end{picture}%
\endgroup%
}\\[1mm]
& = \frac{2}{2!^2} (\sqrt{2}g_0)^4\,
\big(N_f\,\tr T^{a_1} T^{b_1} T^{a_2} T^{b_2}
- \tr_{\text{adj}}T^{a_1} T^{b_1} T^{a_2} T^{b_2}\big)\, W_{10}(x)~.
\nonumber
\end{align}
This diagram was already computed in \cite{Billo:2017glv} in configuration space. For completeness
we report here its evaluation in momentum space. Using the relation 
\begin{align}
\label{tr4adj}
\tr_{\text{adj}}T^{a_1} T^{b_1} T^{a_2} T^{b_2} & =
2 N \,\tr T^{a_1} T^{b_1} T^{a_2} T^{b_2} + \frac 12 
\big(\delta^{a_1b_1}\delta^{a_2 b_2}+\delta^{a_1a_2}\delta^{b_1 b_2}
+\delta^{a_1b_2}\delta^{a_2 b_1}\big)\nonumber\\
& + \frac{\ii\,N}{4}\big(f^{a_1b_1c}\,d^{a_2b_2c}+f^{a_2b_2c}\,d^{a_1b_1c}\big)~,
\end{align}
and introducing the tensor (see (\ref{C4}))
\begin{equation}
{C}_{4}^{(B)\,a_1a_2b_1b_2}= -(2N-N_f)\, \tr T^{a_1}T^{b_1}T^{a_2}T^{b_2}\,-\,\frac{1}{2}\big(
\delta^{a_1b_1}\delta^{a_2 b_2}+\delta^{a_1a_2}\delta^{b_1 b_2}
+\delta^{a_1b_2}\delta^{a_2 b_1}\big)~,
\label{C4B}
\end{equation}
we can rewrite (\ref{w10x}) as 
\begin{equation}
\label{w10xbis}
W_{10}^{a_1a_2b_1b_2}(x) = 2 g_0^4 \,\Big[
{C}_{4}^{(B)\,a_1a_2b_1b_2} - \frac{\ii\,N}{4}\,\big(f^{a_1b_1c}\,d^{a_2b_2c}+f^{a_2b_2c}\,d^{a_1b_1c}\big)\Big]\, W_{10}(x)~.
\end{equation}
As noted after (\ref{w7xtris}), the last two terms in the square brackets 
are anti-symmetric in $(a_1,a_2)$ and $(b_1,b_2)$.
Therefore they vanish when inserted in a chiral/anti-chiral two-point function and can be discarded.
The momentum space diagram corresponding to $W_{10}(x)$ is
\begin{equation}
\label{w10}
\cW_{10}(p)  =~~\parbox[c]{.40\textwidth}{
\begingroup%
  \makeatletter%
  \providecommand\color[2][]{%
    \errmessage{(Inkscape) Color is used for the text in Inkscape, but the package 'color.sty' is not loaded}%
    \renewcommand\color[2][]{}%
  }%
  \providecommand\transparent[1]{%
    \errmessage{(Inkscape) Transparency is used (non-zero) for the text in Inkscape, but the package 'transparent.sty' is not loaded}%
    \renewcommand\transparent[1]{}%
  }%
  \providecommand\rotatebox[2]{#2}%
  \newcommand*\fsize{\dimexpr\f@size pt\relax}%
  \newcommand*\lineheight[1]{\fontsize{\fsize}{#1\fsize}\selectfont}%
  \ifx\svgwidth\undefined%
    \setlength{\unitlength}{110bp}%
    \ifx\svgscale\undefined%
      \relax%
    \else%
      \setlength{\unitlength}{\unitlength * \real{\svgscale}}%
    \fi%
  \else%
    \setlength{\unitlength}{\svgwidth}%
  \fi%
  \global\let\svgwidth\undefined%
  \global\let\svgscale\undefined%
  \makeatother%
  \begin{picture}(1,0.79119933)%
    \lineheight{1}%
    \setlength\tabcolsep{0pt}%
    \put(0.33748508,0.67816393){\color[rgb]{0,0,0}\makebox(0,0)[lt]{\lineheight{1.25}\smash{\begin{tabular}[t]{l}\textbf{$3$}\end{tabular}}}}%
    \put(0.03128118,0.67816393){\color[rgb]{0,0,0}\makebox(0,0)[lt]{\lineheight{1.25}\smash{\begin{tabular}[t]{l}\textbf{$1$}\end{tabular}}}}%
    \put(0.92574585,0.67816393){\color[rgb]{0,0,0}\makebox(0,0)[lt]{\lineheight{1.25}\smash{\begin{tabular}[t]{l}\textbf{$2$}\end{tabular}}}}%
    \put(0.59017994,0.67816393){\color[rgb]{0,0,0}\makebox(0,0)[lt]{\lineheight{1.25}\smash{\begin{tabular}[t]{l}\textbf{$4$}\end{tabular}}}}%
    \put(0.59017994,0.00480212){\color[rgb]{0,0,0}\makebox(0,0)[lt]{\lineheight{1.25}\smash{\begin{tabular}[t]{l}\textbf{$5$}\end{tabular}}}}%
    \put(0.33748508,0.00480212){\color[rgb]{0,0,0}\makebox(0,0)[lt]{\lineheight{1.25}\smash{\begin{tabular}[t]{l}\textbf{$6$}\end{tabular}}}}%
    \put(0.03128118,0.00480212){\color[rgb]{0,0,0}\makebox(0,0)[lt]{\lineheight{1.25}\smash{\begin{tabular}[t]{l}\textbf{$2$}\end{tabular}}}}%
    \put(0.92574585,0.00480212){\color[rgb]{0,0,0}\makebox(0,0)[lt]{\lineheight{1.25}\smash{\begin{tabular}[t]{l}\textbf{$1$}\end{tabular}}}}%
    \put(0,0){\includegraphics[width=\unitlength,page=1]{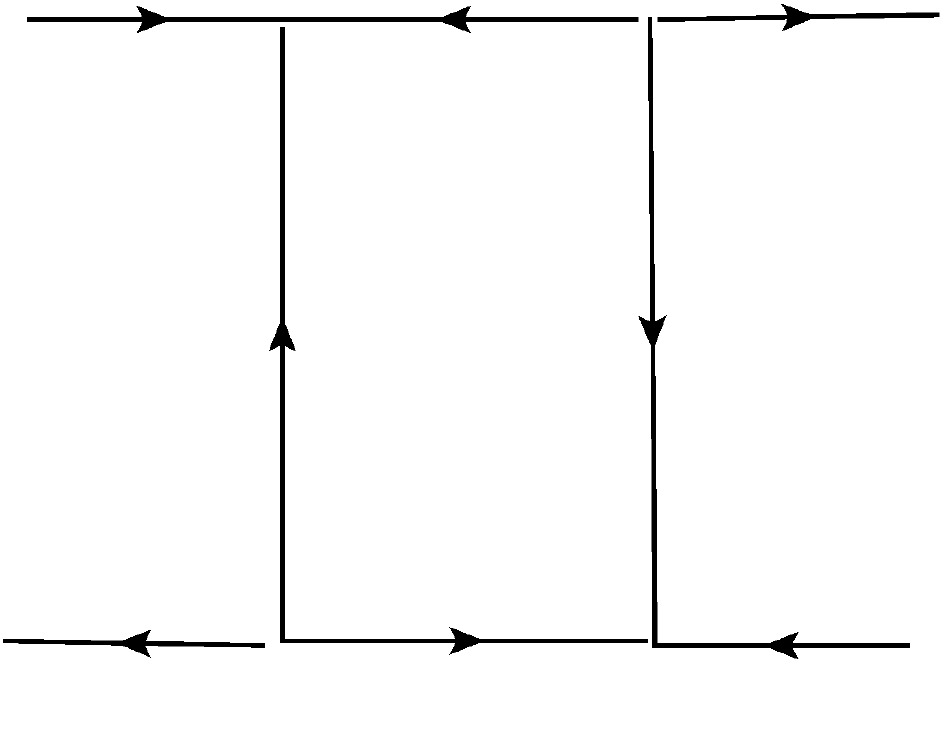}}%
  \end{picture}%
\endgroup%
}
\end{equation}
which has the form (\ref{WYD}) with
\begin{equation}
\label{w10-2}
\widetilde{\mathfrak D}_{10} = 
\exp \bigl(
2 \theta_4\, k_{43}\, \bar{\theta}_3 + 2 \theta_4 \,k_{45}\, \bar{\theta}_5 + 2 \theta_6\, k_{65} 
\,\bar{\theta}_5 + 2 \theta_6 \,k_{63}\, \bar{\theta}_3
\bigr)~.
\end{equation}
The corresponding Grassmann integration is easily carried out using (\ref{4trace}):
\begin{align}
\label{w10-3}
\mathcal{Z}_{10}(p,k) & = \int d^2\bar{\theta}_3\,d^2\theta_4\,  d^2\bar{\theta}_5 \,
d^2\theta_6\, \,\widetilde{\mathfrak D}_{10}
= k_{43}^2\, k_{65}^2 + k_{45}^2\, k_{63}^2 
\\ 
\notag
&- 2 \left(k_{43}\cdot k_{63}\right) \left(k_{45}\cdot k_{65}\right)
+ 2  \left(k_{43}\cdot k_{65}\right) \left(k_{45}\cdot k_{63}\right)
- 2 \left(k_{43}\cdot k_{45}\right) \left(k_{63}\cdot k_{65}\right)~.
\end{align}
Here we have neglected a term proportional to the to the anti-symmetric 
$\varepsilon$-tensor coming from the trace of four Pauli matrices given in (\ref{trace4sigma}) 
which does not contribute for symmetry reasons.
Recalling that $p = k_{13}+ k_{15}= k_{42}+ k_{62}$, we can exploit momentum conservation and discard terms involving $p^2$ and $p\cdot k_{43}$, which give contributions that
vanish for $\epsilon\to 0$ after Fourier transform. After some algebra we are left with 
\begin{equation}
\label{w10-4}
\cZ_{10}(p,k) = k_{13}^2\, k_{62}^2~.  
\end{equation}
When inserted in the momentum integral $\cY_{10}$, this cancels two propagators so that
\begin{equation}
\label{w10-5}
\cW_{10}(p) = {\parbox[c]{.09\textwidth}{ \includegraphics[width = .09\textwidth]{Ydot5}}}~.
\end{equation}
Going back to configuration space using (\ref{int3x}) and inserting the result in (\ref{w10xbis}), 
up to terms that do not contribute for their color structure or that vanish in the limit 
$\epsilon\to 0$, we get
\begin{equation}
W_{10}^{a_1a_2b_1b_2}(x) ={v}_{4,2}^{(B)}\, \Delta(x)^2\,{C}_{4}^{(B)\,a_1a_2b_1b_2}
+\cdots
\label{w10-6}
\end{equation}
with 
\begin{equation}
\label{v42Bapp}
v_{4,2}^{(B)}=\Big(\frac{g_0^2}{8\pi^2}\Big)^2 \,3\,\zeta(3) (\pi x^2)^{2\epsilon}+\ldots
\end{equation}
in agreement with formula (\ref{v42is}) of the main text and with the findings of \cite{Billo:2017glv}.

This completes the calculation of the two-loop diagrams contributing to the chiral/anti-chiral 
correlators.

\section{Feynman integral on the sphere}
\label{app:feynsphere}
In this appendix, we evaluate the integral $I(x_1,x_2)$ that appears in the expression for 
the one-loop correlation function on the sphere, given in (\ref{WS1}) and (\ref{WS2}).

For convenience we first rewrite here the definition (\ref{Iis}) of the integral $I(x_1,x_2)$, namely
\begin{equation}
I(x_1,x_2)=\!\int \!
d^D x_3\,  d^2\bar\theta_3 \,d^D x_4\, d^2\theta_4 \,\Delta(x_{13})
\big(\rme^{-2\ii \theta_4 \partial_{x_{43}}\bar\theta_3}\Delta(x_{43})\big)^2
\Delta(x_{42})\,\big[\kappa(x_3)\,\kappa(x_4)\big]^{-\epsilon}~,
\label{Iisapp}
\end{equation}
where $\Delta(x)$ is the scalar propagator in $D-$dimension and $\kappa(x)$ is the scaling
factor (\ref{kappa}). We then observe that
\begin{equation}
\begin{aligned}
\rme^{-2\ii \theta_4 \partial_{x_{43}}\bar\theta_3}\Delta(x_{43})=
\Delta(x_{43})-2\ii \theta_4 \partial_{x_{43}}\bar\theta_3\,\Delta(x_{43})
-\theta_4^2\,\bar\theta_3^2\,\delta^{(D)}(x_{43})~,
\end{aligned}
\end{equation}
where the last term follows from $\square_x\,\Delta(x)=-\delta^{(D)}(x)$.
Using this relation and performing the Grassmann integrations over $\bar{\theta}_3$ and $\theta_4$, we get
\begin{equation}
\int\! d^2\bar\theta_3\, d^2\theta_4 \,
\big(\rme^{-2\ii \theta_4 \partial_{x_{43}}\bar\theta_3}\Delta(x_{43})\big)^2
= 2 \,\partial_{x_{43}}\Delta(x_{43})\cdot \partial_{x_{43}}\Delta(x_{43})
-2\Delta(x_{43})\,\delta^{(D)}(x_{43})~.
\label{int43}
\end{equation}
Inserting this expression in the integral (\ref{Iisapp}), we see that the term proportional to the
$\delta$-function yields a tadpole-like contribution, which vanishes in dimensional
regularization and thus can be discarded. We then remain with
\begin{equation}
\begin{aligned}
I(x_1,x_2)&=2\int \!
d^D x_3 \,d^D x_4 \,\Delta(x_{13})\,
\partial_{x_{43}}\Delta(x_{43})\cdot \partial_{x_{43}}\Delta(x_{43})\,
\Delta(x_{42})\,\big[\kappa(x_3)\,\kappa(x_4)\big]^{-\epsilon}\\
&=8\,\bigg(\frac{\Gamma(1-\epsilon)}{4\pi^{2-\epsilon}}\bigg)^4(1-\epsilon)^2
\int \! d^D x_3 \,d^D x_4 \,\frac{\big[\kappa(x_3)\,\kappa(x_4)\big]^{-\epsilon}}{
	(x_{13}^2)^{1-\epsilon}\,(x_{43}^2)^{3-2\epsilon}\,(x_{42}^2)^{1-\epsilon}}~,
\end{aligned}
\end{equation}
where in the second step we used the explicit expression (\ref{Delta}) of the scalar propagator.

To simplify the calculation, without any loss of generality, we set $R=1$ and choose the point
$\eta_2$ to be at the north pole on the sphere, namely $\eta_2=(1,0,\ldots,0)$. According to the stereographic projection (\ref{map}), this corresponds to sending $x_2\to \infty$. We therefore find
\begin{equation}
\big[\kappa(x_1)\,\kappa(x_2)\big]^{1-\epsilon}\,I(x_1,x_2) 
~\stackrel{x_2\to \infty}{\approx } ~ 2^{2+3\epsilon}
\,\bigg(\frac{\Gamma(1-\epsilon)}{4\pi^{2-\epsilon}}\bigg)^2
\,\bigg(\frac{x_1^2+1}{2}\bigg)^{1-\epsilon}\,Y(x_1^2)~,
\label{W-int2}
\end{equation}
where
\begin{equation}
Y(x_1^2)= \bigg(\frac{\Gamma(2-\epsilon)}{4\pi^{2-\epsilon}}\bigg)^2 
\int \! d^D x_3 \,d^D x_4 \,
\frac{1}{(x_{13}^2)^{1-\epsilon}\,(x_{43}^2)^{3-2\epsilon}\,(x_3^2+1)^{\epsilon}\,(x_4^2+1)^{\epsilon}}~.
\label{Yis}
\end{equation}
It is not difficult to realize that this function is regular for $x_1^2\to 0$ and
satisfies the following differential equation
\begin{equation}
\square_{x_1} Y(x_1^2)=-\frac{\Gamma(2-\epsilon)}{4\pi^{2-\epsilon}}\,(1-\epsilon) 
\,(x_1^2+1)^{-\epsilon}\int\!d^D x_4 \,\frac{1}{
	(x_{41}^2)^{3-2\epsilon}\,(x_4^2+1)^{\epsilon}}~.
\label{boxI}
\end{equation}
We rewrite the right hand side of (\ref{boxI}) using the Schwinger parametrization
\begin{equation}
\frac{1}{(x^2+a^2)^\alpha}=\frac{1}{\Gamma(\alpha)}\int_0^\infty\!ds\,s^{\alpha-1}
\,\rme^{-s(x^2+a^2)}~,
\label{Schwinger} 
\end{equation}
and, after computing the resulting Gaussian integral over $x_4$, we obtain
\begin{equation}
\begin{aligned}
\square_{x_1} Y(x_1^2)\,&\,=-\frac{\Gamma(2-\epsilon)}{8 \Gamma(2-2\epsilon)\,\Gamma(\epsilon)} 
\,(x_1^2+1)^{-\epsilon} 
\int_0^\infty\!ds_1\int_0^\infty\!ds_2\,\,\frac{s_1^{-1+\epsilon}\,s_2^{2-2\epsilon}}{(s_1+s_2)^{2-\epsilon}}\,\,\rme^{-s_1\frac{s_1+s_2(x_1^2+1)}{s_1+s_2}}\\[2mm]
&\,=-\frac{\Gamma(2-\epsilon)}{8\,\Gamma(2-2\epsilon)\,\Gamma(\epsilon)}\,(x_1^2+1)^{-\epsilon}
\int_0^\infty\!dt\,t^{-2+\epsilon}\,(1+t)^{-1+\epsilon}\,\frac{1}{t+x_1^2+1}~,
\end{aligned}
\label{boxI1}
\end{equation}
where the last step follows from changing the integration variable according to $s_1\to t \,s_2$ and
performing the resulting integral over $s_2$. With the further change of integration variable 
$t\to\frac{1-y}{y}$, we can rewrite the $t$-integral as
\begin{equation}
\int_0^1\!dy\,y^{2-2\epsilon}\,(1-y)^{-2+\epsilon}\,\frac{1}{1+x_1^2\,y}=\frac{\Gamma(3-2\epsilon)
	\,\Gamma(\epsilon-1)}{\Gamma(2-\epsilon)}\,{}_2F_1(1,3-2\epsilon,2-\epsilon;-x_1^2)~.
\label{hyper}
\end{equation}
Substituting this into (\ref{boxI1}), in the end we find
\begin{equation}
\begin{aligned}
\square_{x_1} Y(x_1^2) &= \frac{1}{4}\,(x_1^2+1)^{-\epsilon}\,{}_2F_1(1,3-2\epsilon,2-\epsilon;-x_1^2)
=\frac{x_1^2+2}{8(x_1^2+1)^2}+O(\epsilon)~.
\end{aligned}
\label{boxI2}
\end{equation}
The general solution to this differential equation which is regular for $x_1^2\to 0$ is
\begin{equation}
Y(x_1^2) = \frac{1}{32}\,\big(c_0+\ln(x_1^2+1)+O(\epsilon)\big)
\label{Yfin}
\end{equation}
with $c_0$ an arbitrary constant. To fix it, we examine $Y(x_1^2)$ for $x_1\to \infty$, corresponding 
to the short-distance limit on the sphere in which also $\eta_1$ is sent to the north pole. 
In this limit the leading contribution to (\ref{Yis}) comes from large $x_3^2$ and $x_4^2$, allowing 
us to replace the scaling factors $(1+x_i^2)^\epsilon$ with $(x_i^2)^\epsilon$. This leads to
\begin{align}
Y(x_1^2)
{}&
\stackrel{x_1^2\to \infty}{\simeq }  
\bigg(\frac{\Gamma(1-\epsilon)}{4\pi^{2-\epsilon}}\bigg)^2\,(1-\epsilon)^2
\int\! d^{4-2\epsilon} x_3\, \frac{1}{(x_{13}^2)^{1-\epsilon}  (x_3^2  )^{\epsilon}}
\,\int\! d^{4-2\epsilon} x_4 \, \frac{1}{(x_{34}^2)^{3-2\epsilon} (x_4^2)^{\epsilon}}
\notag\\[2mm]
&~~\,\simeq -  \frac{(x_1^2)^{-\epsilon}}{32\epsilon\,(1-2 \epsilon )}
\simeq
\frac{1}{32}\,\Big(\!-\frac{1}{\epsilon\,(1-2 \epsilon )}+\ln x_1^2+O(\epsilon)\Big)~.
\end{align}
Comparing with (\ref{Yfin}) in the limit $x_1^2\to\infty$, we deduce that
\begin{equation}
c_0=-\frac{1}{\epsilon\,(1-2 \epsilon )}~.
\end{equation}
Therefore, we can write
\begin{equation}
Y(x_1^2)=-\frac{(x_1^2+1)^{-\epsilon}}{32\epsilon\,(1-2\epsilon)}+O(\epsilon)~.
\label{Yfinal}
\end{equation}

The $x_2$-dependence can be easily restored by noticing that  
$\eta_{12}^2 \simeq 4/(x_1^2+1)$ at large $x_2$; this means that at finite $x_2$, the variable
$x_1^2$ must be replaced by
\begin{equation}
r_{12}^2 = \frac{4}{\eta_{12}^2} -1
\end{equation}
and the function $Y(x_1^2)$ by
\begin{equation}
Y(r_{12}^2)= -\frac{2^{-2\epsilon}\,(\eta_{12}^2)^\epsilon}{
	32\epsilon\,(1-2\epsilon)}+O(\epsilon)~.
\label{Yfinal1}
\end{equation}
We now use this information in (\ref{W-int2}) and find
\begin{equation}
\begin{aligned}
W_{1\,\cS}(\eta_{12})&\equiv
\big[\kappa(x_1)\,\kappa(x_2)\big]^{1-\epsilon}\,I(x_1,x_2) \\[1mm]
&= 2^{3+2\epsilon}\,
\frac{\Gamma(1-\epsilon)}{4\pi^{2-\epsilon}}\,\Delta_{\cS}(\eta_{12})\,Y(r_{12}^2)\\[1mm]
&= \frac{(\pi \eta_{12}^2)^\epsilon\,\Gamma(-\epsilon)}{(4\pi)^2\,(1-2\epsilon)}
\,\Delta_{\cS}(\eta_{12}) + O(\epsilon)~,
\end{aligned}
\label{finalresult}
\end{equation}
where we used (\ref{DeltaS}) in the second line, and (\ref{Yfinal1}) in the final step. 
This is the formula (\ref{WS3}) of the main text.
We have also computed the $O(\epsilon)$ terms, finding 
\begin{equation}
W_{1\,\cS}(\eta_{12})
= \frac{(\pi \eta_{12}^2)^\epsilon\,\Gamma(-\epsilon)}{(4\pi)^2\,(1-2\epsilon)}
\,\Delta_{\cS}(\eta_{12}) \,\Big(1-\epsilon^2\,\phi(r_{12}^2)+ O(\epsilon^3)\Big)
\label{finalresult1}
\end{equation}
with
\begin{equation}
\phi(x^2)=\mathrm{Li}_2(-x^2)+\frac{1}{2}\,\ln^2(x^2+1)+\frac{\ln(x^2+1)}{x^2}+\frac{\pi^2}{6}~.
\end{equation}
It is straightforward to verify that $\phi(x^2)$ vanishes at large $x^2$ and approaches a finite value
for $x^2\to 0$.

\providecommand{\href}[2]{#2}
\begingroup\raggedright
\endgroup

\end{document}